\documentclass[aps,prd,superscriptaddress,twocolumn,nofootinbib,10pt]{revtex4}
\usepackage[pdftex]{color}
\usepackage[sort&compress]{natbib}
\usepackage[colorlinks=true,linkcolor=blue,filecolor=blue,urlcolor=blue,citecolor=blue,pdftex,plainpages=false]{hyperref}
\usepackage{url}
\pdfoutput=1
\usepackage{ifpdf}
\usepackage{color,hhline,psfrag,rotating}
\usepackage{dcolumn}
\usepackage{verbatim}
\usepackage{bm}
\usepackage{graphicx}
\usepackage{amsfonts,amssymb,amsmath}
\usepackage{booktabs}  
\usepackage{array}

\graphicspath{{figs/}}

\newcommand{\Xsl}[1]{\raise.15ex\hbox{/}\kern-.57em #1}
\newcommand{\order}[1]{\mathcal{O}(#1)}

\newcommand\brachat[1]{\mathord{\mathop{#1}\limits^{\scriptscriptstyle(\wedge)}}}

\begin{document}
\title{$B_s\to D_s \ell\nu$ Form Factors for the full $q^2$ range from Lattice QCD with non-perturbatively normalized currents}

\author{E.~McLean}
\email[]{e.mclean.1@research.gla.ac.uk}
\affiliation{SUPA, School of Physics and Astronomy, University of Glasgow, Glasgow, G12 8QQ, UK}
\author{C.~T.~H.~Davies}
\email[]{christine.davies@glasgow.ac.uk}
\affiliation{SUPA, School of Physics and Astronomy, University of Glasgow, Glasgow, G12 8QQ, UK}
\author{J.~Koponen}
\affiliation{High Energy Accelerator Research Organisation (KEK), Tsukuba 305-0801, Japan}
\author{A.~T.~Lytle}
\affiliation{INFN, Sezione di Roma Tor Vergata, Via della Ricerca Scientifica 1, 00133 Roma RM, Italy}
\collaboration{HPQCD collaboration}
\homepage{http://www.physics.gla.ac.uk/HPQCD}
\noaffiliation

\date{\today}

\begin{abstract}
  We present a lattice QCD determination of the $B_s \to D_s \ell\nu$ scalar and vector 
form factors over the full physical range of momentum transfer. 
The result is derived from correlation functions computed using the Highly Improved Staggered 
Quark (HISQ) formalism, on the second generation MILC gluon ensembles accounting for up, down, 
strange and charm contributions from the sea. 
We calculate correlation functions for three lattice spacing values and an array of unphysically 
light $b$-quark masses, and extrapolate to the physical value. 
Using the HISQ formalism for all quarks means that the lattice current coupling to the $W$ 
can be renormalized non-perturbatively, 
giving a result free from perturbative matching errors for the first time. 
Our results are in agreement with, and more accurate than, 
previous determinations of these form factors. From the form factors we also 
determine the ratio of branching fractions that is sensitive to violation 
of lepton universality: 
$R(D_s) = \mathcal{B}(B_s\to D_s \tau \nu_{\tau})/\mathcal{B}(B_s\to D_s \ell \nu_{l})$, where $\ell$ 
is an electron or a muon. We find $R(D_s) = 0.2987(46)$, which is also more accurate than 
previous lattice QCD results.  
Combined with a future measurement of $R(D_s)$, this could supply a new test 
of the Standard Model. We also compare the dependence on heavy quark mass of our form factors 
to expectations from Heavy Quark Effective Theory.  
\end{abstract}

\maketitle
\section{Introduction}
\label{sec:intro}

The weak decay processes of mesons such as the $B$ and $B_s$, 
containing $b$ quarks, are a key potential 
source of insights into physics beyond the Standard Model (SM). Flavour-changing $B$ decays 
have gained a lot of interest because of a number of related tensions between 
experimental measurements and SM predictions~\cite{Wei:2009zv,Lees:2012xj,Lees:2012tva,Lees:2013uzd,Aaij:2014pli,Aaij:2014ora,Huschle:2015rga,Aaij:2015oid,Aaij:2015yra,Aaij:2015xza,Aaij:2016flj,Wehle:2016yoi,Sato:2016svk,Hirose:2017dxl,Aaij:2017deq,Aaij:2017uff,Aaij:2017vbb,Sirunyan:2017dhj,Aaboud:2018krd}.
These tensions drive the need for improved theoretical calculations in the SM using methods and 
studying processes where we have good control of the uncertainties.  

Lattice QCD is the method of choice for providing the hadronic input known as form factors 
that determine, up to a normalisation factor, the differential branching fraction for exclusive 
decay processes such $B \to D \ell \nu$ (we suppress all electric 
charge and particle-antiparticle labels here in 
referring to decay processes). The normalisation factor that can then be extracted by comparison 
of theory with experiment is the Cabibbo-Kobayashi-Maskawa matrix element, in this 
case $|V_{cb}|$~\cite{Aubert:2008yv,Aubert:2009ac,Lattice:2015rga,Na:2015kha,Glattauer:2015teq}. 
Determination of $|V_{cb}|$ then feeds into constraints on new 
physics through, for example, tests of the unitarity triangle.  

There has been a long-standing tension in determinations of $|V_{cb}|$ between {\it{exclusive}} 
(from $B\to D \ell \nu$ and $B\to D^* \ell \nu$ decays), and {\it{inclusive}} 
(from $B \to X_c \ell \nu$, where $X_c$ is any charmed hadronic state) processes.
The most accurate exclusive results came from studies of the $B \to D^* \ell \nu$ decay at zero recoil.
It now seems likely that the uncertainties there were being underestimated because 
of the use of a very constrained parameterisation in the extrapolation of the 
experimental $B \to D^*$ data to the zero recoil 
limit~\cite{Bernlochner:2017jka,Bigi:2017njr,Grinstein:2017nlq,Tanabashi:2018oca}, 
but see also~\cite{Dey:2019bgc, Gambino:2019sif}. 
This underlines the importance in future of comparing theory and experiment across 
the full range of squared 4-momentum transfer ($q^2$)  
(a point emphasised 
for $D \to K$ in~\cite{Koponen:2013tua}). It also demonstrates the need for comparison 
of accurate results from multiple decay processes for a more complete picture.  
Improved methods for producing the theoretical input to $|V_{cb}|_{\text{excl}}$, 
namely lattice QCD determinations of form factors, are clearly necessary. 

Here we provide improved accuracy for the form factors 
for the $B_s \to D_s \ell \nu$ decay using a new lattice QCD method 
that covers the full $q^2$ range of the decay for the 
first time. Preliminary results appeared in~\cite{McLean:2019jll}. 
The $B_s \to D_s$ form factors are more attractive 
than $B \to D$ for a first calculation to test methodology. 
They are numerically faster 
to compute and have higher statistical accuracy and smaller finite-volume 
effects because no valence 
$u/d$ quarks are present. Chiral perturbation theory~\cite{Laiho:2005ue} 
expects that the $B \to D$ form factors should be relatively insensitive to the 
spectator quark mass and hence should be very similar between 
$B_s \to D_s$ and $B \to D$. This is confirmed at the 5\% level by lattice 
QCD calculations~\cite{Bailey:2012rr, Monahan:2017uby}. Hence improved calculations 
of $B_s \to D_s$ form factors can also offer information on $B \to D$.     

Given an experimental determination, 
the $B_s\to D_s \ell\nu$ decay can supply a new method for precisely determining 
the Cabibbo-Kobayashi-Maskawa (CKM) element $|V_{cb}|$. 
It can also supply a new test of the SM through quantities sensitive 
to lepton universality violation. We give the SM result for 
$R(D_s) = \mathcal{B}(B_s\to D_s \tau \nu_{\tau}) / \mathcal{B}(B_s\to D_s l \nu_l)$, where $l=e$ or $\mu$. 
An experimental value for comparison to this 
would help to clarify the tension
found between the SM and experiment 
in the related ratios 
$R(D^{(*)}) = \mathcal{B}(B\to D^{(*)} \tau \nu_{\tau}) / \mathcal{B}(B\to D^{(*)} l \nu_l)$~\cite{Amhis:2016xyh} (see also a preliminary new analysis by Belle~\cite{CariaMoriond}). 

Three lattice QCD calculations of $B_{(s)} \to D_{(s)}$ form factors 
have already been performed. The FNAL/MILC collaboration~\cite{Bailey:2012rr, Lattice:2015rga} 
used the Fermilab action for the $b$ and $c$ 
quarks and the asqtad action for the light quarks on MILC gluon field ensembles that include 
$2+1$ flavours of asqtad sea quarks. On the same gluon field ensembles the HPQCD collaboration 
has calculated the form factors using Nonrelativistic QCD (NRQCD) for the valence $b$ and
the Highly Improved Staggered Quark (HISQ) action for the 
other valence quarks~\cite{Na:2015kha, Monahan:2017uby}. A further 
calculation has been done using maximally twisted Wilson quarks on $n_f=2$ gluon field 
ensembles~\cite{Atoui:2013zza}. Preliminary results using domain-wall 
quarks are given in~\cite{Flynn:2019any}. 

A considerable limitation in the FNAL/MILC and HPQCD/NRQCD studies is the 
requirement for normalisation of the lattice QCD $b\to {c}$ current. 
The matching between this current and that of continuum QCD is done in lattice 
QCD perturbation theory through $\mathcal{O}(\alpha_s)$, giving a $\mathcal{O}(\alpha_s^2)$ 
systematic error which can be sizeable. 
Systematic errors coming from the truncation of the 
nonrelativistic expansion of the current are also a problem. 
In the Fermilab formalism the missing terms become $\mathcal{O}(\alpha_s a)$ 
discretisation effects on fine enough 
lattices; in the NRQCD formalism they mix discretisation effects and $\mathcal{O}(\alpha_s/m_b)$ 
(where $m_b$ is the $b$ quark mass) relativistic corrections.   
Here we dispense with both of these problems by using a relativistic formalism 
with absolutely normalised lattice QCD currents. 

Another limitation present in each of the previous studies is 
that the lattice QCD results are limited to a region of high $q^2$, close to zero recoil. 
The reason for this is mainly to avoid large statistical errors. The signal/noise 
degrades exponentially as 
the spatial momentum of the meson in the final state grows. For $b$ quark decays 
the maximum spatial momentum of the final state meson can be large (tending to $m_B/2$ for light 
mesons, where $m_B$ is the $B$ meson mass).  
Systematic errors from missing discretisation (and relativistic) corrections also grow 
away from zero recoil. This is particularly problematic if discretisation effects are 
$\mathcal{O}(a)$ as above and relatively coarse lattices are used (to reduce numerical 
cost). 
Working close to zero recoil means 
that the lattice results then have to be extrapolated from the 
high $q^2$ region into the rest of the physical $q^2$ range. 
Here we also overcome this problem by working with a highly improved quark 
action in which even $\mathcal{O}(a^2)$ errors have been eliminated 
at tree-level~\cite{Follana:2006rc}. We cover a range of values of the lattice 
spacing that includes very fine lattices and include results from lighter 
than physical $b$ quarks and this enables us to cover the full $q^2$ range 
in our lattice calculation. 

We perform our calculation on the second-generation MILC gluon ensembles~\cite{Bazavov:2012xda}, 
including effects from 2+1+1 flavours in the sea using the HISQ action~\cite{Follana:2006rc}.  
We also use the HISQ action for all valence quarks. 
Our calculation employs HPQCD's {\it{heavy-HISQ}} approach. 
In this we obtain lattice results at a number of unphysically light masses for 
the $b$ (we refer to this generically as the {\it{heavy quark}} $h$), reaching the $b$ quark 
mass on the finest lattices.  
This allows us to perform a combined fit in $m_h$ and lattice spacing that we can evaluate in the 
continuum limit at $m_h=m_b$ (and as a function of $m_h$ to compare, for example, to 
expectations from Heavy Quark Effective Theory (HQET)). 
By using only HISQ quarks, we can normalise all the lattice currents fully 
non-perturbatively and avoid systematic errors from current matching. 

This calculation adds to a growing number of successful demonstrations 
of the heavy-HISQ approach. The method was developed for determination of the $b$ quark mass, 
$B$ meson masses and decay constants~\cite{McNeile:2010ji,McNeile:2011ng,McNeile:2012qf} and 
is now also being used by other groups for these 
calculations~\cite{Bazavov:2017lyh, Petreczky:2019ozv}. 
A proof-of-principle application of heavy-HISQ to form factors was given for 
$B_c \to \eta_c$ and $B_c \to J/\psi$ in~\cite{Lytle:2016ixw,Colquhoun:2016osw}, 
covering the full $q^2$ range for these decays and this work builds on those results.  
The $B_s\to D_s^*$ axial form factor at zero recoil was calculated using heavy-HISQ in~\cite{McLean:2019sds}. 

This article is structured in the following way: Section~\ref{sec:lattice} 
lays out our lattice QCD approach to calculating the form factors and then 
Section~\ref{sec:results} presents our results, along with several consistency checks 
and our determination of $R(D_s)$. We also give curves showing the heavy-quark mass dependence of 
some features of the form factors that can be compared to HQET. 
For those simply hoping to use our calculated $B_s \to D_s$ form factors, 
Appendix~\ref{sec:reconstructing_formfactors} gives the parameters and covariance matrix 
required to reconstruct them.

\section{Calculation Details}
\label{sec:lattice}

\subsection{Form Factors}
\label{sec:formfactors}

In this section we specify our notation for the form factors and matrix elements. 
The differential decay rate for $B_s\to D_s l \nu$ decays is given in the SM by
\begin{align}
\label{eq:diffrate}
  &{d\Gamma\over dq^2} = \eta_{\text{EW}} { G_F^2 |V_{cb}|^2\over 24 \pi^3 M_{B_s}^2 } \left( 1 - {m_\ell^2\over q^2}\right)^2 |{\bf{p}}_{D_s}| \,\,\times \\
  \label{eq:branchingfraction}
  &\left[ \left( 1 + {m_\ell^2\over 2q^2}\right) M_{B_s}^2 |{\bf{p}}_{D_s}|^2 f_+^{s\,2}(q^2) + \right. \nonumber \\
&\hspace{10em}\left. {3m_\ell^2\over 8q^2} (M_{B_s}^2-M_{D_s}^2)^2 f_0^{s\,2}(q^2) \right] \nonumber
\end{align}
where $m_\ell$ is the mass of the lepton, $\eta_{\text{EW}}$ is the electroweak 
correction, $q^2 = (p_{B_s} - p_{D_s})^2$ is the momentum transfer 
and $f_0^s(q^2)$, $f_+^s(q^2)$ are the scalar and vector form factors that parameterize 
the fact that the decay process involves hadrons. We use superscript `$s$' to denote 
the strange spectator valence quark.  
The allowed range of $q^2$ values if the final states are on-shell is
\begin{align}
  m_\ell^2 \leq q^2 \leq (M_{B_s}-M_{D_s})^2.
\end{align}
The form factors are determined from matrix elements of the electroweak current 
between $B_s$ and $D_s$ states, $\langle D_s | (V-A)^{\mu} | B_s \rangle$ 
where $V^{\mu}=\bar{b}\gamma^{\mu}c$ is the vector component 
and $A^{\mu}=\bar{b}\gamma^5\gamma^{\mu} c$ is the axial vector component. 
In a pseudoscalar-to-pseudoscalar amplitude, only $V^{\mu}$ contributes, 
since $\langle D_s | A^{\mu} | B_s \rangle$ does not satisfy the parity invariance 
of QCD. In terms of form factors, the vector current matrix element is given by
\begin{align}
  \langle D_s | V^{\mu} | B_s \rangle &= f_+^s(q^2) \left[ p_{B_s}^{\mu} + p_{D_s}^{\mu} - {M_{B_s}^2 - M_{D_s}^2 \over q^2} q^{\mu} \right] \nonumber \\
  &+ f_0^s(q^2) { M_{B_s}^2 - M_{D_s}^2 \over q^2} q^{\mu}.
  \label{eq:vectorcurrent}
\end{align}
Analyticity of this matrix element demands that
\begin{align}
  f_+^s(0) = f_0^s(0).
  \label{eq:q20constraint}
\end{align}

Via the partially conserved vector current relation (PCVC), the form factor $f_0^s(q^2)$ is 
also directly related to the matrix element of the scalar current $S=\bar{b}c$;
\begin{align}
  (m_b-m_c)\langle D_s | S | B_s \rangle = (M^2_{B_s} - M^2_{D_s}) f_0^s(q^2).
    \label{eq:scalarcurrent}
\end{align}
In our calculation we determine the form factors by computing matrix elements of 
the temporal vector current $V^0$ and the scalar current $S$. The form factors can be 
extracted from this combination using expressions derived from Eqs.~\eqref{eq:vectorcurrent} 
and~\eqref{eq:scalarcurrent} (once the currents have the correct continuum 
normalisation - see Section~\ref{subsec:norm}):
\begin{align}
\label{eq:fmat}
  f_0^s(q^2) &= {m_b - m_c\over M_{B_s}^2 - M_{D_s}^2 } \langle D_s | S | B_s \rangle, \nonumber \\
  f_+^s(q^2) &= {1\over 2M_{B_s}} { \delta^M \langle D_s | S | B_s \rangle - q^2 \langle D_s | V^0 | B_s \rangle \over {\bf{p}}^2_{D_s}}, \\
    &( \, \delta^M = (m_b - m_c)(M_{B_s}-E_{D_s}) \, ). \nonumber
\end{align}

Our goal is to compute $f_0^s(q^2)$ and $f_+^s(q^2)$ throughout the range 
of $q^2$ values $0 \leq q^2 \leq (M_{B_s}-M_{D_s})^2 \equiv q^2_{\text{max}}$. 
We extend the range to $q^2=0$ in order to take advantage of the constraint 
from Equation~\eqref{eq:q20constraint}.

\begin{table}[t]
  \begin{center}
    \begin{tabular}{c c c c c c c c c c}
      \hline
      set & handle & $w_0/a$ & $N_x^3\times N_t$ & $am_{l0}$ & $am_{s0}$ & $am_{c0}$  \\ [0.5ex]
      \hline
      1 & \bf{fine} & 1.9006(20) & $32^3\times96$ & 0.0074 & 0.037 & 0.440 \\ [1ex]
      2 & \bf{fine-} & 1.9518(7) & $64^3\times96$ & 0.0012 & 0.0363 & 0.432 \\ 
       & \bf{physical} &  &  &  &  &  \\ [1ex]
      3 & \bf{superfine} & 2.896(6) & $48^3\times144$ & 0.0048 & 0.024 & 0.286 \\ [1ex]
      4 & \bf{ultrafine} & 3.892(12) &  $64^3\times192$ & 0.00316 & 0.0158 & 0.188  \\ [1ex]
      \hline
    \end{tabular}
  \end{center}
  \caption{Parameters for gluon field ensembles~\cite{Bazavov:2010ru,Bazavov:2012xda}. 
$a$ is the lattice spacing, determined from the Wilson flow parameter $w_0$~\cite{Borsanyi:2012zs}. 
Values for $w_0/a$ are from: set 1~\cite{Chakraborty:2016mwy}, sets 2 and 3~\cite{Chakraborty:2014aca}, 
set 4~\cite{mcneile:private}. The physical value of $w_0$ was determined 
to be $0.1715(9)$fm in~\cite{Dowdall:2013rya}. $N_x$ is the spatial extent and $N_t$ the 
temporal extent of the lattice in lattice units. Light ($m_u=m_d$), strange and charm quarks are 
included in the sea, their masses are given in columns 5-7. \label{tab:ensembles}}
\end{table}

\begin{table*}[t]
  \begin{center}
    \begin{tabular}{c c c c c c c c c c c}
      \hline
      set & $am_{s0}^{\text{val}}$ & $am_{c0}^{\text{val}}$ & $am^{\text{val}}_{h0}$ & $|a{\bf{p}}_{D_s}|$ & $n_{\text{cfg}}\times n_{\text{src}}$ & $T$ \\ [0.5ex]
      \hline
      1 & 0.0376 & 0.45
      & 0.5 & 0, 0.056 & $986\times 8$ & 14, 17, 20 \\ [1ex]
      & & & 0.65 & 0, 0.142, 0.201 & &   \\ [1ex]
      & & & 0.8 & 0, 0.227, 0.323 & &  \\ [1ex]

      \hline
      2 & 0.036 & 0.433
      & 0.5 & 0, 0.0279 & $286\times 4$ & 14, 17, 20  \\ [1ex]
      & & & 0.8 & 0, 0.162 & & \\ [1ex]

      \hline
      3 & 0.0234 & 
      0.274 & 0.427 & 0, 0.113, 0.161 & $250\times 8$ & 22, 25, 28 \\ [1ex]
      & & & 0.525 & 0, 0.161, 0.244 & & \\ [1ex]
      & & & 0.65 & 0, 0.244, 0.338 & & \\ [1ex]
      & & & 0.8 & 0, 0.338, 0.438 & &  \\ [1ex]

      \hline
      4 & 0.0165 
      & 0.194 & 0.5 & 0, 0.202, 0.281 & $237\times 4$ & 31, 36, 41 \\ [1ex]
      & & & 0.65 & 0, 0.202, 0.281, 0.382 & & \\ [1ex]
      & & & 0.8 & 0, 0.281, 0.382, 0.473 & & \\ [1ex]
      \hline
    \end{tabular}
  \end{center}
  \caption{Calculational details. 
Columns 2 and 3 give the $s$ and $c$ valence quark masses in lattice units, 
which were tuned in \cite{Chakraborty:2014aca}. 
In column 4 we give the heavy quark masses that we used in lattice units. 
We use a number of heavy quark masses to enable the heavy-quark mass dependence to be determined 
in our fit. Column 5 gives the absolute value of the spatial momentum (in lattice units) given 
to the $D_s$ meson using a momentum twist on the charm quark propagator. 
These values are chosen with the following rationale: when only two values are given, 
these correspond to the $q^2=0$ and $q^2_{\text{max}}$ points (except on the fine-physical 
ensemble, where we use the points $q^2_{\text{max}}$ and $q^2_{\text{max}}/2$); when three 
values are given, the momenta correspond to $q^2=0$, $q^2=q^2_{\text{max}}/2$, and $q^2_{\text{max}}$; 
when four values are given, these are points corresponding to $q^2_{\text{max}}$, 
$3q^2_{\text{max}}/4$, $q^2_{\text{max}}/2$, $q^2_{\text{max}}/4$ and $q^2=0$. 
We used twisted momenta in the $(1,1,1)$ direction to minimise discretisation effects. 
Column 6 gives the number of gluon field configurations used for that ensemble, $n_{\text{cfg}}$, 
and the number of different time sources used per configuration to increase statistics, $n_{\text{src}}$. 
Column 7 gives the temporal separations between source and sink, $T$, of the three-point 
correlation functions computed on each ensemble.}
  \label{tab:simulation}
\end{table*}

\subsection{Lattice Calculation}
\label{sec:correlationfunctions}

This calculation closely follows the approach employed in our calculation 
of the $B_s\to D_s^*$ axial form factor at zero recoil~\cite{McLean:2019sds}. 
Here, however, we must give spatial momentum to the charm quark in the final 
state so that we can cover the full $q^2$ range of the decay. 

The gluon field configurations used in this calculation were generated by the MILC 
collaboration~\cite{Bazavov:2010ru,Bazavov:2012xda}. 
The relevant parameters for the specific ensembles we use are given in Table~\ref{tab:ensembles}. 
The gluon fields are generated using a Symanzik-improved gluon action with 
coefficients matched to continuum QCD through $\order{\alpha_s a^2,n_f\alpha_s a^2}$~\cite{Hart:2008sq}. 
The gluon fields include the effect of 2+1+1 flavours of quarks in the sea ($u$, $d$, $s$, $c$, 
where $m_{u0}=m_{d0}\equiv m_{l0}$) using the HISQ action~\cite{Follana:2006rc}. 
In three of the four ensembles (sets 1, 3 and 4), the bare light quark mass is 
set to $m_{l0}/m_{s0} = 0.2$. The fact that the $m_{l0}$ value is unphysically high 
is expected to have only a small effect on the form factors here, since we have no 
valence light quarks. We quantify this small effect by including a fourth 
ensemble (set 2) with roughly physical $m_{l0}$. 

We use a number of different masses for the valence heavy 
quark $am_{h0}^{\text{val}}$. This allows us to resolve the dependence of the form 
factors on the heavy quark mass, so that a fit in $m_h$ can be performed and the results 
of the fit evaluated at $m_h=m_b$. With a heavy quark mass varying both on a given ensemble 
and between ensembles, we can resolve both the discretisation effects that grow with 
large ($am^{\text{val}}_{h0} \lesssim 1$) masses and the physical dependence of 
the continuum form factors on $m_h$. Using unphysically light $h$-quarks also reduces 
the $q^2$ range, meaning that we can obtain lattice results across the full range while 
the statistical noise remains under control. 

Staggered quarks have no spin degrees of freedom. Spin-parity quantum numbers are accounted for 
by construction of appropriate fermion bilinears and including an appropriate 
space-time dependent phase with each operator in the path integral. 
We categorize these phases according to the standard {\it{spin-taste}} 
notation, $(\gamma_n\otimes \gamma_s)$, where $\gamma_n$ is the spin structure of 
the operator in the continuum limit, and $\gamma_s$ is the `taste' structure which accounts for 
the multiple possible copies of the operator constructed from staggered quark fields.

We have designed this calculation to use only local operators (combining fields at the same 
space-time point and having $(\gamma_n\otimes\gamma_n)$ spin-taste) 
for the calculation of the current matrix elements that we require. 
This is an advantage since point-split operators can lead to noisier correlation functions. 
The spin-taste operators we use are: scalar $(1\otimes1)$, 
pseudoscalar $(\gamma^5\otimes \gamma^5)$, 
vector $(\gamma^{\mu}\otimes \gamma^{\mu})$, 
and temporal axial-vector $(\gamma^{0}\gamma^5\otimes \gamma^{0}\gamma^5)$.

We compute a number of correlation functions on the ensembles 
detailed in Table~\ref{tab:ensembles}. Valence quark masses, 
momenta and other inputs to the calculation are given in Table~\ref{tab:simulation}. 
We use random wall sources to generate all staggered propagators from the source since this 
gives improved statistical errors~\cite{Davies:2010ip}. 
First we compute two-point correlation functions between meson eigenstates of 
momentum $a{\bf{p}}$, 
\begin{align}
  C^{a{\bf{p}}}_{M}(t) &= \frac{1}{N_{\text{taste}}}\langle \tilde{\Phi}_M ({\bf{p}},t) \tilde{\Phi}_M^{\dagger}({\bf{p}},0) \rangle, \\ 
  \tilde{\Phi}_M({\bf{p}},t) &= \sum_{{\bf{x}}} e^{-i\bf{p}\cdot \bf{x}} \bar{q}({\bf{x}},t) \Gamma q'({\bf{x}},t), \nonumber
\end{align}
where $\langle \rangle$ represents a functional integral over all fields, $q,q'$ are valence 
quark fields of the flavours the $M$ meson is charged under, $\Gamma$ is the 
spin-taste structure of $M$ and the division by the number of tastes is required 
to normalise closed loops made from staggered quarks~\cite{Follana:2006rc}. 
We compute these correlation functions for all $t$ values, i.e. $0\leq t \leq N_t$.

\begin{figure*}
  \hspace*{-30pt}
  \includegraphics[width=0.9\textwidth]{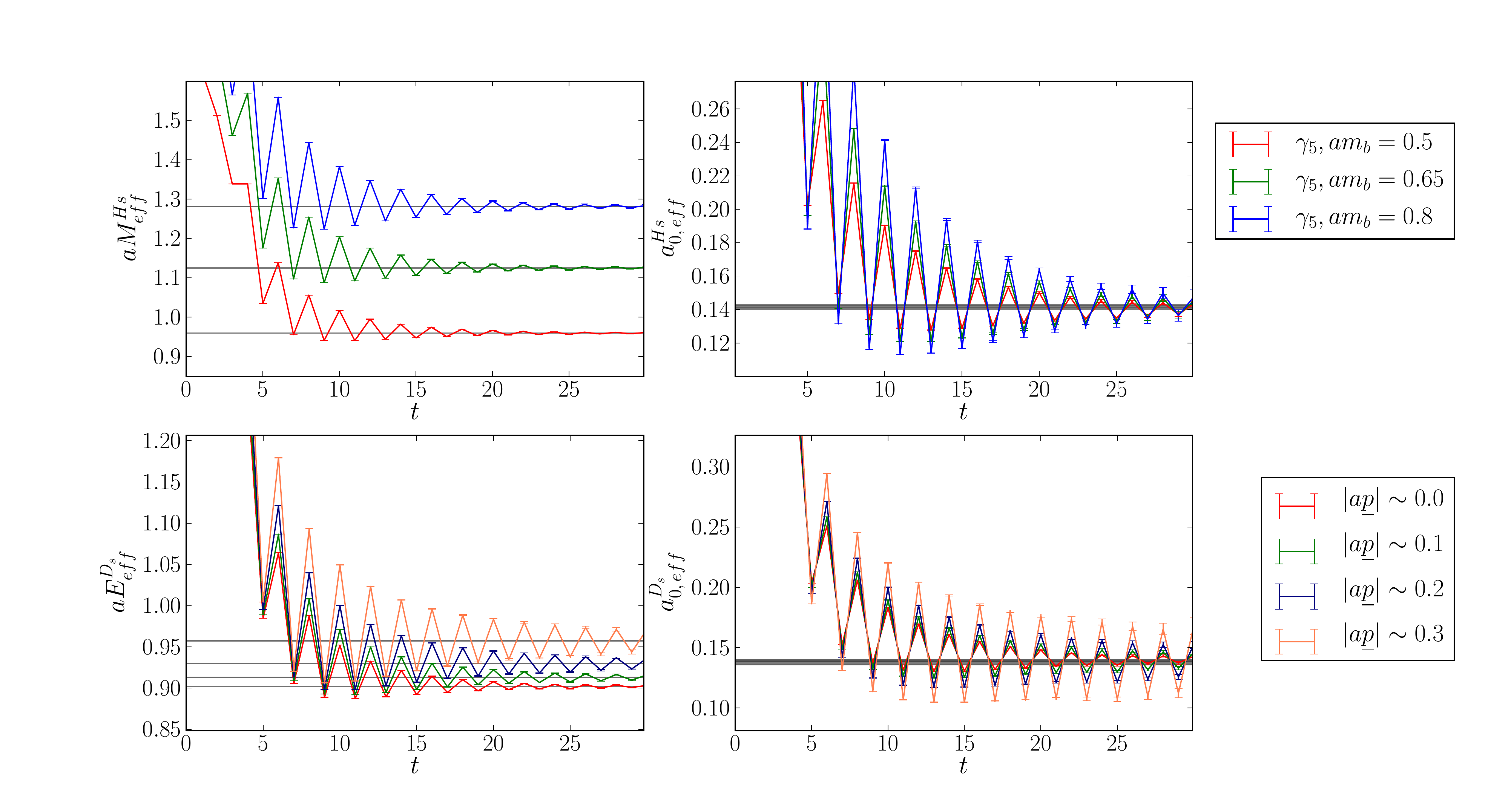}
  \caption{Effective energies and amplitudes, 
defined from Eqs.~\eqref{eq:effmass} and~\eqref{eq:effamp} on the 
fine ensemble. Grey lines 
show the fit result for the ground-state energies and amplitudes. \label{fig:2ptplots}}
\end{figure*}

\begin{table*}
\begin{center}
\begin{tabular}{ c c c c c c c c c }
\hline
Set & $am_h^{\text{val}}$ & $aM_{H_s}$& $aM_{D_s}$& $aM_{H_c}$& $af_{H_c}$& $aM_{\eta_h}$& $aM_{\eta_c}$& $aM_{\eta_s}$\\ [0.5ex]
\hline
1 & 0.5 & 0.95971(12) & 0.90217(11) & 1.419515(41) & 0.186299(70) & 1.471675(38) & 1.367014(40) & 0.313886(75)\\ [0.5ex] 
 & 0.65 & 1.12507(18) &  & 1.573302(40) & 0.197220(77) & 1.775155(34) & &  \\ [0.5ex] 
 & 0.8 & 1.28129(19) &  & 1.721226(39) & 0.207068(78) & 2.064153(30) & & \\ [0.5ex] 
\hline
2 & 0.5 & 0.95446(13) & 0.87715(11) & 1.400025(26) & 0.183482(46) & 1.470095(25) & 1.329291(27) & 0.304826(52)\\ [0.5ex] 
 & 0.8 & 1.27560(24) &  & 1.702438(24) & 0.203382(50) & 2.062957(19) & & \\ [0.5ex] 
\hline
3 & 0.427 & 0.77443(17) & 0.59151(11) & 1.067224(46) & 0.126564(70) & 1.233585(41) & 0.896806(48) & 0.207073(96)\\ [0.5ex] 
 & 0.525 & 0.88470(21) &  & 1.172556(46) & 0.130182(72) & 1.439515(37) & & \\ [0.5ex] 
 & 0.65 & 1.01973(28) &  & 1.303144(46) & 0.133684(75) & 1.693895(33) & &  \\ [0.5ex] 
 & 0.8 & 1.17436(40) &  & 1.454205(46) & 0.137277(79) & 1.987540(30) & & \\ [0.5ex] 
\hline
4 & 0.5 & 0.80235(19) & 0.439899(86) & 1.011679(25) & 0.099031(45) & 1.342747(27) & 0.666754(39) & 0.153827(77)\\ [0.5ex] 
 & 0.65 & 0.96344(27) & & 1.169780(26) & 0.100598(49) & 1.650264(23) & & \\ [0.5ex] 
 & 0.8 & 1.11728(35) &  & 1.321660(28) & 0.101765(54) & 1.945763(21) & & \\ [0.5ex] 
\hline
\end{tabular}
\caption{
Parameters determined from our correlation function fits.  
The decay constant $af_{H_c}$ is extracted from the amplitude obtained via Eqs.~\eqref{eq:decayampfit} 
and~\eqref{eq:decayconstantfit}. $D_s$ energies at non-zero spatial momentum are given in Table~\ref{tab:kinetic}. \label{tab:masses}}
\end{center}
\end{table*}

\begin{table*}
\begin{center}
\begin{tabular}{ c c c c c c c c }
\hline
Set & $am_h^{\text{val}}$ & $q^2$[GeV$^2$]& $aE_{D_s}$& $f^s_0(q^2)$& $f^s_+(q^2)$& $f^s_0(q^2)/f_{H_c}\sqrt{M_{H_c}}$& $f^s_+(q^2)/f_{H_c}\sqrt{M_{H_c}}$\\ [0.5ex]
\hline\hline
1 & 0.5 & 0.01584(17) & 0.90217(11) & 1.0009(14) &  & 1.394(11) & \\ [0.5ex] 
 &  & 0.00026(50) & 0.90386(11) & 0.9997(14) & 0.9997(15) & 1.393(11) & 1.393(11)\\ [0.5ex]
 \hline
 & 0.65 & 0.2376(26) & 0.90217(11) & 1.0047(28) &  & 1.256(11) & \\ [0.5ex] 
 &  & 0.1201(15) & 0.91308(13) & 0.9956(31) & 1.0014(77) & 1.245(11) & 1.252(14)\\ [0.5ex] 
 &  & 0.0027(12) & 0.92399(15) & 0.9878(31) & 0.9880(32) & 1.235(11) & 1.235(11)\\ [0.5ex]
 \hline
 & 0.8 & 0.6874(74) & 0.90217(11) & 1.0092(17) &  & 1.1488(94) & \\ [0.5ex] 
 &  & 0.3473(40) & 0.92992(17) & 0.9898(18) & 1.0079(56) & 1.1267(93) & 1.147(11)\\ [0.5ex] 
 &  & 0.0082(29) & 0.95759(27) & 0.9710(18) & 0.9714(19) & 1.1053(91) & 1.1057(91)\\ [0.5ex] 
\hline\hline
2 & 0.5 & 0.03014(32) & 0.87715(11) & 1.0004(15) &  & 1.369(11) & \\ [0.5ex] 
 &  & 0.02593(49) & 0.87759(11) & 1.0001(15) & 1.002(19) & 1.369(11) & 1.371(28)\\ [0.5ex]
 \hline
 & 0.8 & 0.8007(84) & 0.87715(11) & 1.0054(18) &  & 1.1258(91) & \\ [0.5ex] 
 &  & 0.6126(66) & 0.89178(15) & 0.9948(22) & 1.030(18) & 1.1139(91) & 1.154(22)\\ [0.5ex] 
\hline\hline
3 & 0.427 & 0.3715(42) & 0.59151(11) & 0.9942(24) &  & 1.250(11) & \\ [0.5ex] 
 &  & 0.1877(26) & 0.60220(14) & 0.9807(26) & 0.9928(61) & 1.233(11) & 1.248(13)\\ [0.5ex] 
 &  & 0.0053(23) & 0.61281(17) & 0.9685(26) & 0.9688(26) & 1.218(11) & 1.218(11)\\ [0.5ex]
 \hline
 & 0.525 & 0.954(11) & 0.59151(11) & 0.9876(25) &  & 1.152(10) & \\ [0.5ex] 
 &  & 0.5361(66) & 0.61281(17) & 0.9614(27) & 0.9901(80) & 1.121(10) & 1.155(13)\\ [0.5ex] 
 &  & 0.0124(54) & 0.63946(31) & 0.9320(31) & 0.9326(31) & 1.0870(99) & 1.0876(99)\\ [0.5ex]
 \hline
 & 0.65 & 2.036(23) & 0.59151(11) & 0.9791(28) &  & 1.0548(95) & \\ [0.5ex] 
 &  & 0.950(12) & 0.63946(31) & 0.9227(37) & 0.9611(81) & 0.9940(93) & 1.035(12)\\ [0.5ex] 
 &  & 0.007(13) & 0.68113(58) & 0.8821(42) & 0.8823(41) & 0.9503(92) & 0.9505(92)\\ [0.5ex]
 \hline
 & 0.8 & 3.772(43) & 0.59151(11) & 0.9709(37) &  & 0.9643(90) & \\ [0.5ex]
 &  & 1.435(22) & 0.68113(58) & 0.8731(53) & 0.9138(93) & 0.8671(90) & 0.908(12)\\ [0.5ex] 
 &  & -0.030(44) & 0.7373(17) & 0.825(10) & 0.8242(97) & 0.819(12) & 0.819(12)\\ [0.5ex] 
\hline\hline
4 & 0.5 & 2.634(32) & 0.439899(86) & 0.9741(28) &  & 1.0319(99) & \\ [0.5ex] 
 &  & 1.179(16) & 0.48514(23) & 0.9134(32) & 0.9522(63) & 0.9676(95) & 1.009(11)\\ [0.5ex] 
 &  & -0.027(18) & 0.52261(54) & 0.8666(45) & 0.8657(44) & 0.9181(96) & 0.9172(96)\\ [0.5ex]
 \hline
 & 0.65 & 5.497(67) & 0.439899(86) & 0.9575(34) &  & 0.9287(91) & \\ [0.5ex] 
 &  & 3.748(47) & 0.48514(23) & 0.8985(39) & 1.004(18) & 0.8714(88) & 0.974(19)\\ [0.5ex] 
 &  & 2.301(35) & 0.52261(54) & 0.8529(55) & 0.913(11) & 0.8272(92) & 0.886(13)\\ [0.5ex] 
 &  & 0.198(93) & 0.5770(24) & 0.770(15) & 0.775(14) & 0.746(16) & 0.751(15)\\ [0.5ex]
 \hline
 & 0.8 & 9.20(11) & 0.439899(86) & 0.9433(40) &  & 0.8508(86) & \\ [0.5ex] 
 &  & 5.495(72) & 0.52261(54) & 0.8438(67) & 0.974(23) & 0.7611(92) & 0.878(22)\\ [0.5ex] 
 &  & 3.06(11) & 0.5770(24) & 0.768(16) & 0.842(23) & 0.693(15) & 0.759(22)\\ [0.5ex] 
 &  & 0.19(26) & 0.6410(57) & 0.721(27) & 0.724(25) & 0.650(25) & 0.653(23)\\ [0.5ex] 
\hline\hline
\end{tabular}
\caption{Parameters from our correlation function fits at varying $q^2$ points. 
$f_{0,+}^s(q^2)$ are extracted via~\eqref{eq:currentfit}
and~\eqref{eq:normalizations}, and $R^s_{0,+}(q^2)$ is defined in~\eqref{eq:ratio}. 
We show statistical/fit errors on each of the quantities, including the value 
of $q^2$ that is derived from $M_{H_s}$ and $E_{D_s}$. $q^2$ and $R^s$ uncertainties 
also include those from the determination of the lattice spacing.   
\label{tab:kinetic}}
\end{center}
\end{table*}

We compute correlation functions for a heavy-strange pseudoscalar, $H_s$, 
with spin-taste structure $(\gamma^5\otimes \gamma^5)$, at rest. 
In terms of staggered quark propagators this takes the form
\begin{align}
  C_{H_s}(t) = \frac{1}{4}\sum_{\bf{x},\bf{y}} \langle \text{Tr}\left[ g_h(x,y) g_s^{\dagger}(x,y) \right] \rangle,
  \label{eq:pseudoscalar_corrs}
\end{align}
where $g_q(x,y)$ is a staggered propagator for flavour $q$, and the trace is over color. 
Here $x_0=0$ and $y_0=t$, and the sum is over spatial sites labelled {$\bf{x}$, $\bf{y}$}. 
We also compute correlators for a charm-strange pseudoscalar meson $D_s$, with 
structure $(\gamma^{5}\otimes \gamma^{5})$. For these correlators we need both zero and 
non-zero spatial momentum. 
Non-zero spatial momentum is given to the $D_s$ by imposing twisted 
boundary conditions on the gluon fields when computing the charm quark propagators~\cite{Guadagnoli:2005be}. 
Then
\begin{align}
\label{eq:dscorrs}
  C^{a\bf{p}}_{D_s}(t) = \frac{1}{4}\sum_{\bf{x},\bf{y}} \langle \text{Tr}\left[ g^{\theta}_c(x,y) g_s^{\dagger}(x,y) \right] \rangle,
\end{align}
where $g_q^{\theta}(x,y)$ denotes a propagator with momentum twist $\theta$. 
We compute these correlation functions using several different twists to produce 
the range of momenta given in Table~\ref{tab:simulation}. 
We design the $c$ propagators to have momentum $a{\bf{p}} = |a{\bf{p}}|(1,1,1)$, 
by imposing a twist $\theta = N_x |a{\bf{p}}| / \pi \sqrt{3}$ in each spatial direction.

Necessary for extracting the vector current matrix element, 
we also compute correlation functions for a non-goldstone 
pseudoscalar heavy-strange mesons at rest, denoted $\hat{H}_s$. 
This has spin-taste structure $(\gamma^0\gamma^5\otimes \gamma^0\gamma^5)$. 
$\hat{H}_s$ correlators are computed using
\begin{align}
  C_{\hat{H}_s}(t) = \frac{1}{4}\sum_{\bf{x},\bf{y}}\langle (-1)^{\bar{x}_0+\bar{y}_0} \text{Tr}\left[ g_h(x,y) g^{\dagger}_s(x,y) \right] \rangle,
\end{align}
where we use the notation $\bar{z}_{\mu} = \sum_{\nu\neq\mu} z_{\nu}$.

We also compute correlators for $H_c$ mesons, heavy-charmed pseudoscalars, using 
the same form as those for $H_s$, Equation~\eqref{eq:pseudoscalar_corrs}. 
These are used to find $H_c$ decay constants, which are useful in some of our continuum and $m_h$ fits.
In our fits to heavy-quark mass dependence we will use the mass of the heavy-heavy pseudoscalar meson, 
$\eta_h$ as a physical proxy for the quark mass. 
To quantify mistuning of the charm and strange quark masses, we also require masses for 
$\eta_c$ and $\eta_s$ mesons, identical to $\eta_h$ with $h$ replaced $c$ and $s$ quarks 
respectively. We compute correlators for each of these at rest, using a spin-taste 
structure $(\gamma^5\otimes \gamma^5)$, taking the same form as those of the 
$H_s$, Equation~\eqref{eq:pseudoscalar_corrs}. Note that all of the $\eta$ mesons discussed 
here are artificially forbidden to annihilate in our lattice QCD calculation. We expect this 
to have negligible effect, for the purposes of this calculation, 
on the masses of the $\eta_c$ and the $\eta_b$~\cite{Chakraborty:2014aca}; 
the $\eta_s$ is an unphysical meson that can be defined in this limit in a lattice QCD 
calculation and is convenient for tuning the $s$ quark mass~\cite{Davies:2009tsa, Dowdall:2013rya}. 

Three-point correlation functions are needed to allow determination of the current 
matrix elements for $B_s \to D_s$ decay. 
We require two sets of such correlation functions, one with a scalar and one with a 
temporal vector current insertion. The first takes the form
\begin{align}
  C^{a{\bf{p}}_{D_s}}_S(t,T) &= \frac{1}{N_{\text{taste}}}\sum_{{\bf{y}}} \langle \tilde{\Phi}_{D_s}({\bf{p}},T)\, S({\bf{y}},t) \,\tilde{\Phi}_{H_s}({\bf{0}},0) \rangle, 
\\ S({\bf{y}},t) &= \bar{c}({\bf{y}},t) h({\bf{y}},t). \nonumber
\end{align}
In terms of the staggered quark formalism, both the $H_s$ source and $D_s$ sink are given 
structure $(\gamma^5\otimes \gamma^5)$, and the current insertion $(1\otimes1)$. 
We combine staggered propagators to construct these correlation functions as: 
\begin{align}
  C^{a{\bf{p}}_{D_s}}_S(t,T) =& \frac{1}{4}\sum_{{\bf{x},\bf{y},\bf{z}}} \langle \text{Tr}\left[ g_h(x,y)g^{\theta}_c(y,z) g^{\dagger}_s(x,z) \right] \rangle,
\end{align}
where we fix $x_0 = 0$, $y_0=t$ and $z_0=T$, and once again the charm propagator is 
given the appropriate twist $\theta$. We compute these correlation functions 
for all $t$ values within $0\leq t\leq T$, using 3 $T$ values to make sure that excited state effects 
are accounted for. The $T$ values vary with lattice spacing to give approximately the same physical range 
and always include both even and odd values. 
The values are given in Table~\ref{tab:simulation}.

The three-point correlation function with temporal vector current insertion is given by
\begin{align}
  C^{a{\bf{p}}_{D_s}}_{V^0}(t,T) &= \frac{1}{N_{\text{taste}}}\sum_{{\bf{y}}} \langle \tilde{\Phi}_{D_s}({\bf{p}},T)\, V^0({\bf{y}},t) \,\tilde{\Phi}_{\hat{H}_s}({\bf{0}},0) \rangle, \nonumber \\ 
V^0({\bf{y}},t) &= \bar{c}({\bf{y}},t) \gamma^0 h({\bf{y}},t).
\end{align}
This is generated using spin-taste $(\gamma^0\gamma^5\otimes \gamma^0\gamma^5)$ at 
the $\hat{H}_s$ source, $(\gamma^5\otimes \gamma^5)$ at the $D_s$ sink, 
and $(\gamma^0\otimes \gamma^0)$ at the current insertion. To achieve this we compute
\begin{align}
  C^{a{\bf{p}}_{D_s}}_{V_0}(t,T) =& \frac{1}{4}\sum_{{\bf{x},\bf{y},\bf{z}}} \langle (-1)^{\bar{x}_0+\bar{y}_0} \nonumber \\ &\times \text{Tr}\left[ g_h(x,y)g^{\theta}_c(y,z) g^{\dagger}_s(x,z) \right]\rangle.
\end{align}
The non-goldstone $\hat{H}_s$ is required here to ensure that taste cancels in the correlation function. 
The difference between the $\hat{H}_s$ and the $H_s$, for example in their masses, is generated by 
taste-exchange discretisation effects. In practice it is very small for heavy mesons~\cite{Follana:2006rc}, 
being suppressed by the heavy meson mass.  

\subsection{Analysis of Correlation Functions}
\label{sec:analysiscorrs}

\begin{figure*}
  \hspace*{-30pt}
  \includegraphics[width=0.98\textwidth]{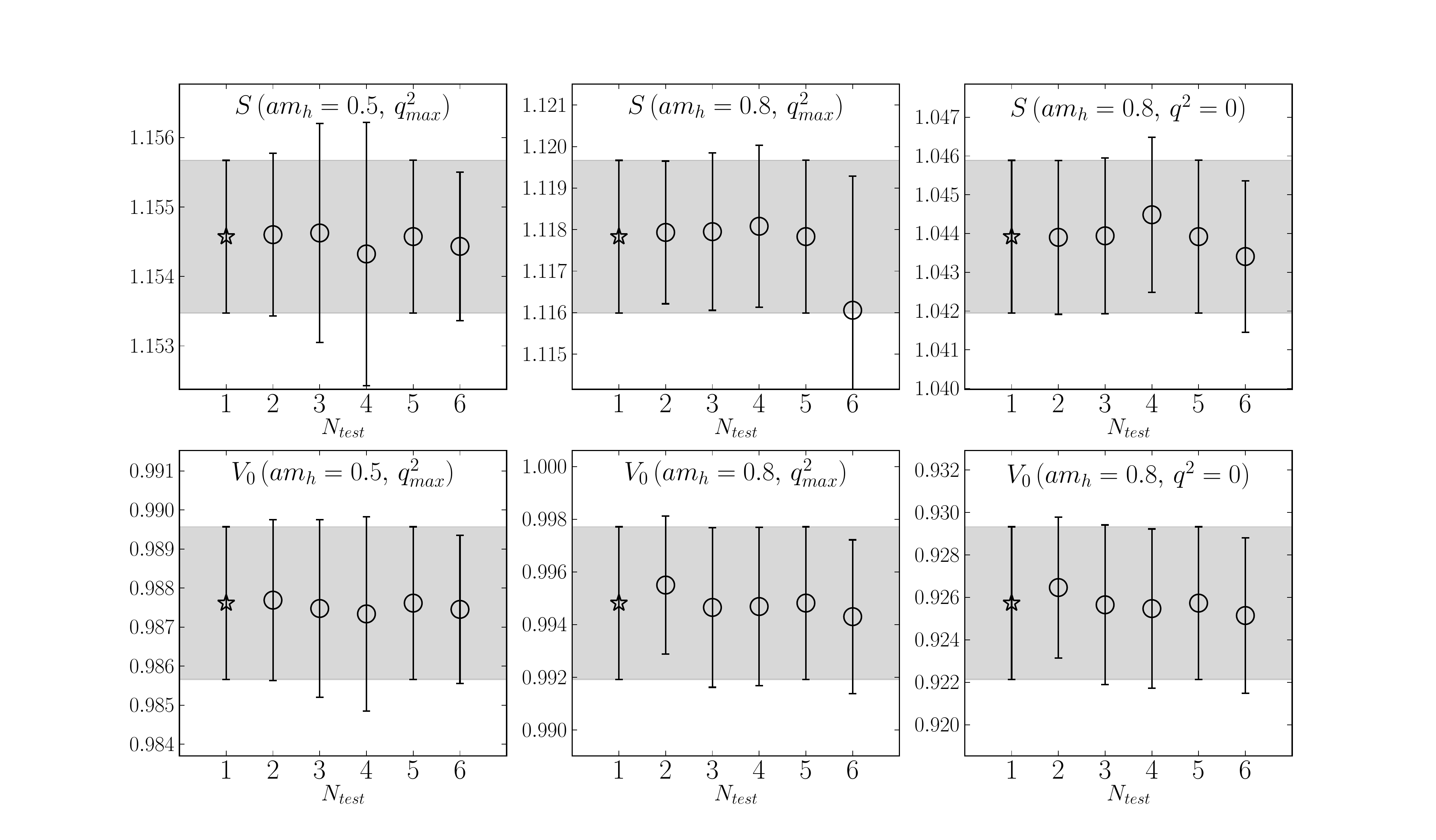}
  \caption{Tests on the correlator fits on the fine ensemble. The y-axis shows the best fit result for $J_{00}^{nn}$, with the appropriate current, heavy mass and $q^2$ specified. At $N_{\text{test}}=1$ we give our final 
result, reproduced by the light grey band for ease of comparison. 
$N_{\text{test}}=2$ and $3$ give the results of setting $N_{\text{exp}}=4$ and $6$ 
respectively.  $N_{\text{test}}=4$ gives the result of setting $t_{\text{cut}}=3$ for all 2-point correlators (in the final fit $t_{\text{cut}}=2$ for all correlators). $N_{\text{test}}=5$ gives the value when the prior width on the $J_{00}^{nn}$ parameters 
is doubled. $N_{\text{test}}=6$ gives the results from the output
from a fit to the appropriate correlators from that heavy 
mass and $q^2$ value only, and therefore not including correlation with results from other 
masses and momentum values. \label{fig:corr_tests}}
\end{figure*}

\begin{figure}
  \includegraphics[width=0.45\textwidth]{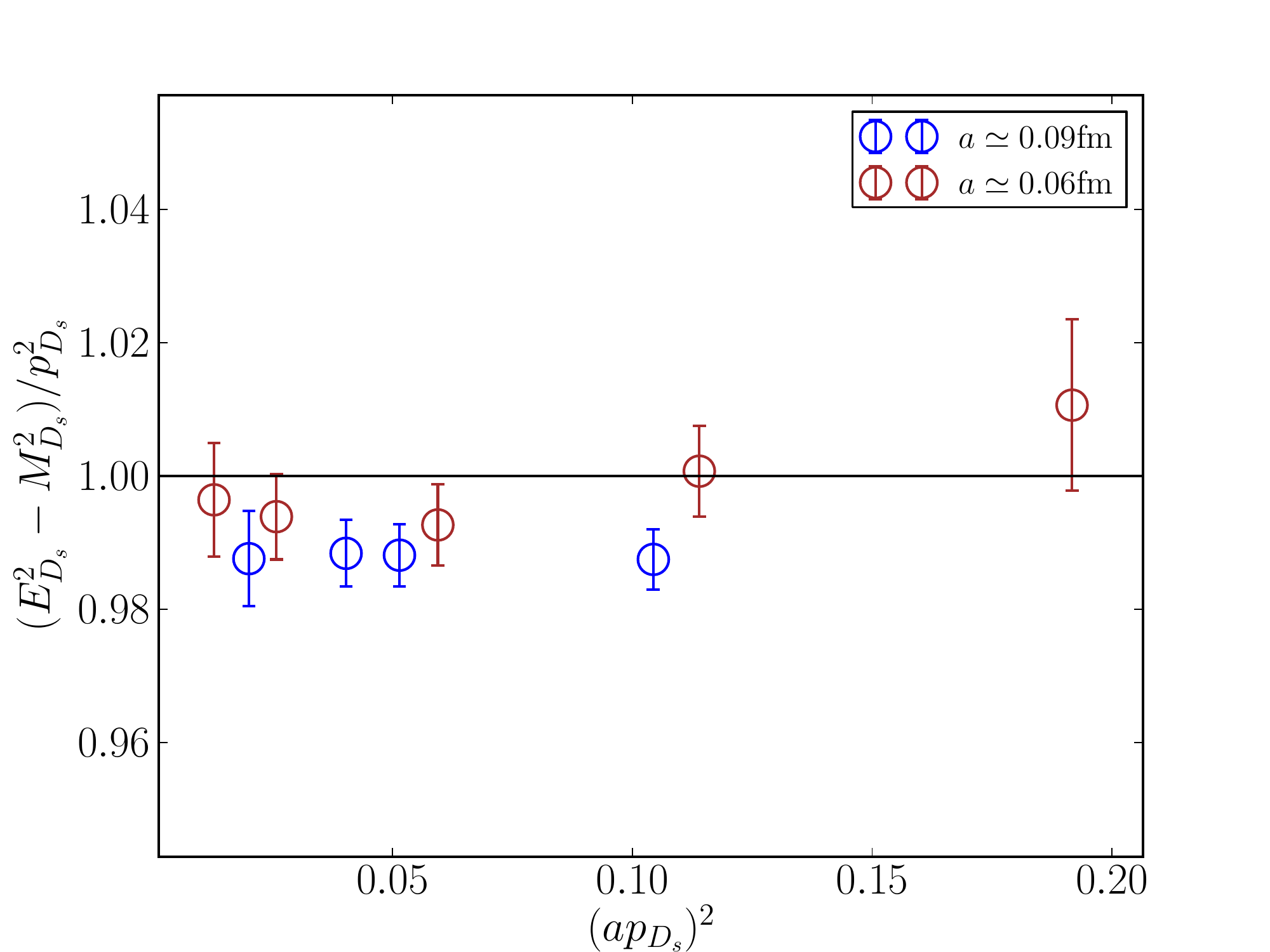}
  \caption{A comparison of the relativistic dispersion relation 
for our $D_s$ mesons on gluon field ensemble sets 1 and 3. 
We plot the square of the `speed of light' against the square 
of the spatial momentum of the $D_s$ in lattice units. Values 
on the coarser lattices, set 1, show a small deviation from 1 that is reduced
on the finer lattices. 
\label{fig:csq}}
\end{figure}

We now describe our simultaneous multi-exponential fits to the correlation functions
using a standard Bayesian apporach~\cite{Lepage:2001ym,corrfitter}.
The parameters that we wish to determine are ground-state energies, two-point amplitudes and 
ground-state to ground-state matrix elements. Our correlation functions, however, are contaminated
by contributions from excited states. These excited states must be included in our fits so that 
the systematic error on the ground-state parameters from the presence of the excited states is 
fully taken into account. Multi-exponential fits are then mandatory, guided by Bayesian priors 
for the parameters, discussed below. To reduce the number of exponentials needed by the fits, 
we drop values of the correlation functions when they are within $t_{\text{cut}}$ of the 
end-points (where excited states contribute most). 
We use values of $t_{\text{cut}}$ varying from 2 to 10 throughout the correlator fits.
We take results from fits using 5 exponentials ($N_{\text{exp}}=5$ in the 
fit forms below), where good $\chi^2$ values are 
obtained and the ground-state parameters and their uncertainties have stabilised. 

Two-point correlation functions are fit to the form
\begin{align}
\label{eq:twoptfit}
  C_M(t)|_{\text{fit}} = &\sum_{n}^{N_{\text{exp}}} \Big( | a^M_{n} |^2 f(E^M_n,t) \nonumber \\ 
  &- (-1)^t | a^{M,o}_n |^2 f(E_n^{M,o},t) \Big), 
\end{align}
where
\begin{equation}
  f(E,t) = \left( e^{-E t} + e^{-E (N_t-t)} \right),
\end{equation}
and $E^{M,(o)}_n$,$a^{M,(o)}_n$ are fit parameters. 
The second term in Equation~\eqref{eq:twoptfit} accounts for the 
opposite-parity states that arise from 
the staggered quark time doublers and are known 
as {\it{oscillating states}} (see Appendix G of \cite{Follana:2006rc}). 
These oscillating states do not appear when $M$ is a Goldstone-taste pseudoscalar with 
a quark and antiquark of the same mass, so in the $M=\eta_h,\eta_c,$ and $\eta_s$ cases 
the second term is not required. 

Figure~\ref{fig:2ptplots} shows the quality of our results. We plot effective 
energies and amplitudes for the $D_s$ and $H_s$ mesons on the 
fine ensemble. The effective energy is defined 
as 
\begin{equation}
\label{eq:effmass}
E_{\text{eff}}(t) = -\log \left( \frac{\tilde{C}(t)}{\tilde{C}(t-1)} \right).
\end{equation}
Here $\tilde{C}$ is a time-smeared mean correlator to reduce the impact of the oscillating 
states: 
\begin{equation}
\label{eq:smear}
\tilde{C}(t) = C(t-1) + 2C(t) + C(t+1) ,
\end{equation}
where $C(t)$ is defined in Equations~\eqref{eq:pseudoscalar_corrs} and~\eqref{eq:dscorrs}.  
At large $t$ we expect $E_{\text{eff}}(t)$ to stabilise at the ground-state energy 
and we see that happening in Figure~\ref{fig:2ptplots}. An effective amplitude can 
then be defined from 
\begin{equation}
\label{eq:effamp}
a_{\text{eff}}(t) = \sqrt{\tilde{C}(t)e^{E_{\text{eff}}(t)t}}/[2\cosh (E_{\text{eff}}/2)]
\end{equation}
This again converges to the ground-state amplitude from the fit. 

For the three-point correlation functions we use the fit form
\begin{align}
  C_{J}&(t,T)|_{\text{fit}} = \\
&\sum_{k,j=0}^{N_{\text{exp}},N_{\text{exp}}} \Big(\,
  a^{\brachat{H}_s}_j J^{nn}_{jk} a^{D_s}_k f(E^{H_s},t) f(E^{D_s}_n,T-t) \nonumber
  \\ 
  &+a^{\brachat{H}_s,o}_j J^{on}_{jk} a^{D_s}_k (-1)^t f(E^{\brachat{H}_s,o}_n,t) f(E^{D_s},T-t) \nonumber
  \\ 
  &+a^{\brachat{H}_s}_j J^{no}_{jk} a^{D_s,o}_k (-1)^{T-t} f(E^{\brachat{H}_s},t) f(E^{D_s^*,o}_n,T-t) \nonumber
  \\ 
  &+a^{\brachat{H}_s,o}_j J^{oo}_{jk} a^{D_s,o}_k (-1)^T f(E^{\brachat{H}_s,o}_n,t) f(E^{D_s,o},T-t) \,\Big). \nonumber
\end{align}
This includes fit parameters common to the fits of $H_s$ (when $J=S$), $\hat{H}_s$ (when $J=V_0$) 
and $D_s$ two-point correlators, 
along with new fit parameters $J_{jk}$ that are related to the current matrix elements. 
We perform a single simultaneous fit containing each correlator computed 
($C_{H_s},C_{\hat{H}_s},C_{D_s},C_{\eta_h},C_{\eta_c},C_{\eta_s},C_{H_c},C_S,C_{V_0}$) at 
every $am_h$ and every $|a{\bf{p}}_{D_s}|$, for each ensemble.

These simultaneous fits are very large and this causes problems for the covariance matrix which must 
be inverted to determine $\chi^2$. 
We take two steps toward mitigating this. The first is to impose an svd (singular value decomposition) 
cut $c_{\text{svd}}$. 
This replaces any eigenvalue of the covariance matrix smaller than $c_{\text{svd}}x$ 
with $c_{\text{svd}}x$, where $x$ is the largest eigenvalue of the matrix. 
The small eigenvalues are driven to zero if the statistics available are 
not high enough. The application of the svd cut makes the matrix less 
singular, and can be considered a conservative move since 
the only possible effect on the error of the final results is to inflate them.
An appropriate value for $c_{\text{svd}}$ is found by comparing estimates 
of covariance matrix eigenvalues between different bootstrap samples of the data 
using the Corrfitter package~\cite{corrfitter}. The resulting $c_{\text{svd}}$ varies 
between ensembles since it depends on the statistical quality of the dataset, but we find them 
to be of order $10^{-3}$.

The other step we take towards a stable fit is employing a 
chained-fitting approach. 
We first perform an array of smaller fits, each fitting the correlators relevant only to 
one $m_h$ and one $|a{\bf{p}}_{D_s}|$ value. In the case of set 4, for example, this 
results in 11 separate fits. Then, a full simultaneous fit of all of the correlators is 
carried out, using as priors the results of the smaller fits. This both speeds up the full fit and improves stability of the results. 

The priors for the fits 
were set up as follows. We set gaussian priors 
for the parameters $J_{jk}$, and log-normal priors for amplitudes $a_i^M$, ground-state 
energies $E_0^M$, and excited-state energy differences $E_{i+1}^M-E_{i}^M$. 
Using log-normal distributions forbids ground-state energies, excited state energy differences 
and amplitudes moving too close to zero or becoming negative, improving stability of the fit.

Priors for ground state energies $E_0^M$ and amplitudes $a_0^M$ are set 
according to an empirical-Bayes approach, plots of the effective amplitude of 
the correlation functions are inspected to deduce reasonable priors. 
The ground-state oscillating parameters $a_0^{M,o}$, $E_0^{M,o}$, are given the 
same priors as the non-oscillating states, with uncertainties inflated by 50\%. 
The resulting priors always have a standard deviation at least 10 times that of the final result. 
The logs of the excited-state energy differences are given prior 
values $2a\Lambda_{\text{QCD}}\pm a\Lambda_{\text{QCD}}$ where $\Lambda_{\text{QCD}}$ was taken 
as 0.5 GeV. The log of oscillating and non-oscillating excited state amplitudes are 
given priors of $-1.9\pm 3.3$. The ground-state non-oscillating to non-oscillating 
three-point parameter, $J_{00}^{nn}$ is given a prior of $1\pm 0.5$, and the 
rest of the three-point parameters $J_{jk}^{nn}$ are given $0\pm 1$.

The physical quantities that we need here 
are extracted from the ground-state fit parameters and given in Tables~\ref{tab:masses} 
and~\ref{tab:kinetic}.
$E_0^M$ are the ground-state meson energies in lattice units. 
For mesons at rest, this corresponds to the mass of the meson, i.e. $E_0^M = aM_M$. 
The annihilation amplitude for an $M$-meson at rest is given in lattice units by
\begin{align}
  \langle 0 | \tilde{\Phi}_M | M \rangle |_{\text{lat}} = \sqrt{2M_M} a_0^M .
  \label{eq:decayampfit}
\end{align}
If $\tilde{\Phi}_M$ is a $\gamma_5 \otimes \gamma_5$ pseudoscalar operator $P$, 
the decay constant can be found from this via
\begin{align}
  f_{M} = { m^{\text{val}}_{q0} + m^{\text{val}}_{q'0} \over M_{M}^2 } \langle \Omega | P | M \rangle |_{\text{lat}} ,
  \label{eq:decayconstantfit}
\end{align}
where $q$,$q'$ are the quark flavours that $M$ is charged under. We use this to 
determine the $H_c$ meson decay constant in Table~\ref{tab:masses}.  
The current matrix elements that we are focussed on here can be extracted from the fit parameters via
\begin{align}
  \langle D_s| J | H_s \rangle |_{\text{lat}} = 2 \sqrt{M_{H_s}E_{D_s}} J^{nn}_{00}.
  \label{eq:currentfit}
\end{align}
These can be converted into values for the form factors once the currents 
have been normalised (Section~\ref{subsec:norm}). 

Figure~\ref{fig:corr_tests} shows the results of a number of tests we performed on 
the fits to correlators on the fine ensemble. Each test modifies one of the features of the fits and 
we then plot the resultant value of the key output parameter $J_{00}^{nn}$. 
The robustness of the fits can be gauged by the effect of these changes, which are all small. 

Figure~\ref{fig:csq} shows a comparison of the $D_s$ meson dispersion relation 
on the fine (set 1) and superfine (set 3) lattices. The dispersion relation 
is sensitive to discretisation effects in our quark action. The figure 
shows them to be small (see~\cite{Donald:2012ga} for more discussion of 
discretisation effects in dispersion relations for mesons using HISQ quarks).  

\begin{table}
\begin{center}
\begin{tabular}{ c c c c }
\hline
Set & $am_{h0}^{\text{val}}$ & $Z_V$& $Z_{\text{disc}}$\\ [0.5ex]
\hline
1 & 0.5 & 1.0155(23) & 0.99819\\ [0.5ex] 
 & 0.65 & 1.0254(35) & 0.99635\\ [0.5ex] 
 & 0.8 & 1.0372(32) & 0.99305\\ [0.5ex] 
\hline
2 & 0.5 & 1.0134(24) & 0.99829\\ [0.5ex] 
 & 0.8 & 1.0348(29) & 0.99315\\ [0.5ex] 
\hline
3 & 0.427 & 1.0025(31) & 0.99931\\ [0.5ex] 
 & 0.525 & 1.0059(33) & 0.99859\\ [0.5ex] 
 & 0.65 & 1.0116(37) & 0.99697\\ [0.5ex] 
 & 0.8 & 1.0204(46) & 0.99367\\ [0.5ex] 
\hline
4 & 0.5 & 1.0029(38) & 0.99889\\ [0.5ex] 
 & 0.65 & 1.0081(43) & 0.99704\\ [0.5ex] 
 & 0.8 & 1.0150(49) & 0.99375\\ [0.5ex] 
\hline
\end{tabular}
    \caption{Normalization constants applied to the lattice currents 
in Equation~\eqref{eq:normalizations}. $Z_V$ is found from Equation~\eqref{eq:ward} and 
$Z_{\text{disc}}$ from~\cite{McLean:2019sds}. \label{tab:norms}}
\end{center}
\end{table}

\subsection{Current Normalization}
\label{subsec:norm}

In the HISQ formalism, the local scalar current $(1\otimes 1)$ 
(multiplied by the mass difference of flavours it is charged under) is conserved, 
and hence requires no renormalization. 
This is not the case for the local temporal vector current $(\gamma^0\otimes \gamma^0)$. 
We use this instead of the conserved vector current because it is much simpler, but we 
then require a renormalisation factor to match to the continuum current. 
This is simple to obtain fully non-perturbatively within this calculation~\cite{Koponen:2013tua,Donald:2013pea}, 
at no additional cost. 

When both meson states in the matrix elements are at rest (the zero recoil point), 
the scalar and local vector matrix elements are related via the PCVC relation:
\begin{align}
  ( M_{H_s}& - M_{D_s} ) Z_V \langle D_s | V^0 | \hat{H}_s \rangle|_{\text{lat}} \nonumber \\ &= (m^{\text{val}}_{h0} - m^{\text{val}}_{c0}) \langle D_s | S | H_s \rangle|_{\text{lat}}.
  \label{eq:ward}
\end{align}
$Z_V$ can be extracted from this relation using the matrix elements we have computed. 
The $Z_V$ values found on each ensemble and for each $am^{\text{val}}_{h0}$ are given in Table~\ref{tab:norms}.

We also remove $\mathcal{O}(am_h^4)$ tree-level mass-dependent 
discretisation effects from the current 
using a normalization constant, $Z_{\text{disc}}$ derived in~\cite{Monahan:2012dq} and 
discussed in detail in~\cite{McLean:2019sds}. $Z_{\text{disc}}$ values are 
also tabulated in Table~\ref{tab:norms}; 
they have only a very small effect.  

Combining these normalizations with the lattice current from the 
simultaneous correlation function fits, we find values for the form factors 
at a given heavy mass, lattice spacing, and $q^2$:
\begin{align}
  \nonumber
  f_0^s&(q^2) = {m^{\text{val}}_{h0} - m^{\text{val}}_{c0}\over M_{H_s}^2 - M_{D_s}^2 } Z_{\text{disc}} \langle D_s | S | H_s \rangle|_{\text{lat}}(q^2) \\
  \label{eq:normalizations}
  f_+^s&(q^2) = {Z_{\text{disc}} \over 2M_{H_s}} \times \\ \nonumber &{ \delta^M \langle D_s | S | B_s \rangle|_{\text{lat}}(q^2) - q^2 Z_V \langle D_s | V^0 | B_s \rangle|_{\text{lat}}(q^2) \over {\bf{p}}^2_{D_s}}.
\end{align}
where $\delta^M$ is defined in Equation~(\ref{eq:fmat}) and 
we have made the dependence of the matrix elements on $q^2$ explicit. 

\subsection{Obtaining a result at the Physical Point}
\label{sec:extrapolation}

We now discuss how we fit our results for $f_{0}^s(q^2)$ and $f_{+}^s(q^2)$ 
as a function of valence heavy quark mass, sea light quark mass and lattice spacing. 
Evaluating these fits at the mass of the $b$, with physical $l,s$ and $c$ masses and 
zero lattice spacing will then give us the physical form factor curves from 
which to determine the differential decay rate, using Equation~\eqref{eq:branchingfraction}. 

Following~\cite{McLean:2019sds} we use two methods; one a direct approach to 
fitting the form factors and the other in which we fit the ratio
\begin{align}
  R^s_{0,+}(q^2) \equiv {f^s_{0,+}(q^2)\over f_{H_c}\sqrt{M_{H_c}}}\,,
  \label{eq:ratio}
\end{align}
in which 
discretisation effects are somewhat reduced. We will take our final result 
from the direct approach and we describe that here. We use the ratio approach as a test of uncertainties and we describe that in more detail in Appendix~\ref{appendix:ratio}.

We use identical fit functions for both approaches. 
We feed into the fit our results from Tables~\ref{tab:masses} and~\ref{tab:kinetic}, 
retaining the correlations (not shown in the Tables) 
between values for different heavy quark masses and 
$q^2$ values on a given gluon field ensemble that we are able to capture in our 
simultaneous fits (Section~\ref{sec:analysiscorrs}). We also include, where 
needed, correlated lattice spacing uncertainties.  

\subsubsection{Kinematic Behaviour}
\label{eq:kin}

Our fit form is a modified version of the 
Bourrely-Caprini-Lellouch (BCL) parameterisation for 
pseudoscalar-to-pseudoscalar form factors~\cite{Bourrely:2008za}:
\begin{align}
  \label{eq:BCL}
  f^s_0(q^2)|_{\text{fit}} &= {1\over 1 - {q^2\over M_{H_{c0}}^2} } \sum_{n=0}^{N-1} a_n^0 z^n(q^2), \\
  f^s_+(q^2)|_{\text{fit}} &= {1\over 1 - {q^2\over M_{H_c^*}^2} } \times \nonumber \\
&\sum_{n=0}^{N-1} a_n^+ \Big( z^n(q^2) - {n\over N} (-1)^{n-N} z^N(q^2) \Big). \nonumber
\end{align}
The functon $z(q^2)$ maps $q^2$ to a small region inside the unit circle on the complex $q^2$ plane, defined by
\begin{align}
  z(q^2) = {\sqrt{t_+ - q^2} - \sqrt{t_+ - t_0} \over \sqrt{t_+ - q^2} + \sqrt{t_+ - t_0} }\,.
\end{align}
Here $t_+ = (M_{H_s}+M_{D_s})^2$ and we choose $t_0$ to be $t_0 = 0$. 
This $t_0$ choice means that $q^2=0$ maps to $z=0$ and 
the fit functions simplify to $f_{0,+}^s(0) = a_0^{0,+}$. 
For the physical range of $q^2$ for $B_s$ to $D_s$ decay, 
the range covered by $z$ is 
$|z|<0.06$, resulting in a rapidly converging series in 
powers of $z$. We truncate at $N=3$; adding further powers 
of $z^n$ does not effect the results of the fit.

The factors in front of the sums in the BCL parameterisation 
account for lowest mass pole expected in the full $q^2$ plane 
for each form factor coming from
the production of on-shell $H_{c0}$ and $H_c^*$ states in 
the crossed channel of the semileptonic decay. Note that these 
poles, even though they are below the cut (for $H_s + D_s$ production) 
that begins at $t_+$, are at much higher $q^2$ values than those 
covered by the semileptonic decay here (with maximum $q^2$ given by 
$(M_{H_s}-M_{D_s})^2$ ).  

We must estimate $M_{H_{c0}}$, the scalar heavy-charm 
meson mass, at each of the heavy masses we use. 
For this we use the fact that the 
splitting $\Delta_0 = M_{H_{c0}} - M_{H_c}$ is an orbital excitation 
and therefore largely independent of the heavy quark mass. 
The splitting has been calculated in~\cite{Dowdall:2012ab} 
to be $\Delta_0 = 0.429(13)$ GeV at the $b$ quark mass. 
Combined with an $H_c$ mass from our lattice results, 
we construct the $H_{c0}$ mass as $M_{H_{c0}} = M_{H_c} + \Delta_0$. 
We do not include the uncertainty on $\Delta_0$ in the fit, 
since any shift in the precise position of the pole 
will be absorbed into the other fit parameters.

To estimate $M_{H^*_c}$, the vector heavy-charm meson mass, 
we use the fact that the hyperfine splitting $M_{H_c^*} - M_{H_c}$ 
should vanish in the infinite $m_h$ limit. 
$M_{H_c^*}$ then takes the approximate form 
$M_{H_c^*} \simeq M_{H_c} + \order{1/m_h}$. 
To reproduce this behaviour we use the ansatz 
$M_{H_c^*}=M_{H_c} + x/M_{\eta_h}$, and fix $x$ at the $b$ quark mass 
using the value of $M_{H_c^*}-M_{H_c}$ from~\cite{Dowdall:2012ab}. 
This gives $x=0.507\,\text{GeV}^2$.

\subsubsection{Heavy Quark Mass and Discretisation Effects}

To account for dependence on the heavy quark mass and 
discretisation effects in a general way, we use the following 
form for each of the $a_n^{0,+}$ coefficients:
\begin{align}
  \nonumber  a_n^{0,+} = & \,\,\left( 1 + \rho^{0,+}_n \text{log}\left(M_{\eta_h}\over M_{\eta_c}\right)\right) \times \\ \nonumber
  &\sum_{i,j,k = 0}^{2,2,2} d^{0,+}_{ijkn} \left({2\Lambda_{\text{QCD}}\over M_{\eta_h} }\right)^{i} \left({ am^{\text{val}}_{h0} \over \pi }\right)^{2j} \left({ am^{\text{val}}_{c0} \over \pi }\right)^{2k} \\ & \times \left(\, 1 + \mathcal{N}_{\text{mistuning},n}^{0,+} \,\right)\,.
      \label{eq:mhfitfun}
\end{align}

To understand this form, focus first on the terms inside the sum. 
Powers of $(2\Lambda_{\text{QCD}}/M_{\eta_h})$ allow for variation of 
the coefficients as the heavy quark mass changes, using an HQET-inspired form 
since this is a heavy-light to heavy-light meson transition.
$M_{\eta_h}/2$ is proportional to $m_h$ at leading order in HQET, 
so is a suitable physical proxy for the heavy quark mass. 
We take $\Lambda_{\text{QCD}}$ here to be 0.5GeV.
The other two terms in the sum allow for discretisation effects. 
These can be set by two scales. 
One is the variable heavy quark mass $am_{h0}^{\text{val}}$
and the other is the charm quark mass, $am_{c0}^{\text{val}}$, constant 
on a given ensemble. 
Adding further discretisation effects set by smaller scales such 
as $a\Lambda_{\text{QCD}}$ had no impact on the results since  
such effects 
are subsumed into the larger $am_{c0}^{\text{val}}$ terms. 

The coefficients $d^{0,+}_{ijkn}$ are fit parameters given 
Gaussian prior distributions of $0\pm 2$. 

To account for any possible logarithmic dependence on $m_h$, 
arising from, for example, an ultraviolate matching between HQET 
and QCD, we include a log term in front of the sum. 
$\rho^{0,+}_n$ are fit parameters with prior distribution $0\pm 1$.

The fact that $f^s_+(0) = f^s_0(0) \,(\,\Rightarrow a^+_0 = a^0_0)$ is 
a powerful constraint within the heavy-HISQ approach. Since this relation 
must be true at all $m_h$, it translates to constraints on the 
fit parameters; $d^+_{i000}=d^0_{i000} \,\forall \,i$ and $\rho^+_0 = \rho^0_0$.We impose these constraints in the fit. 

\subsubsection{Quark Mass Mistuning}

To account for any possible mistunings in the $c$, $s$ and $l$ quark 
masses, we include the terms $\mathcal{N}_{\text{mistuning},n}^{0,+}$ 
in each $a^{0,+}_n$ coefficient, defined by 
\begin{align}
  \mathcal{N}^{0,+}_{\text{mistuning},n} &= {c^{\text{val},\,0,+}_{s,n} \delta^{\text{val}}_s + c^{0,+}_{s,n} \delta_s + 2c^{0,+}_{l,n}\delta_l \over 10m_s^{\text{tuned}}}
    \label{eq:mistuning}
  \\ &+c_{c,n}^{0,+} \left( { M_{\eta_c} - M_{\eta_c}^{\text{phys}} \over M_{\eta_c}^{\text{phys}} } \right)\,.
  \nonumber
\end{align}
Here $c^{0,+}_{l,n}$, $c^{0,+}_{s,n}$ and $c^{(\text{val}),\,0,+}_{s,n}$ 
are fit parameters with prior distributions $0\pm 1$. 

We define $\delta^{\text{(val)}}_s = m^{\text{(val)}}_{s0} - m_s^{\text{tuned}}$~\cite{Chakraborty:2014aca}, where $m_s^{\text{tuned}}$ is given by
\begin{align}
  m_s^{\text{tuned}} = m_{s0} \left({ M^{\text{phys}}_{\eta_s} \over M_{\eta_s} }\right)^2.
\end{align}
$M_{\eta_s}^{\text{phys}}$ is the mass of an unphysical $s\overline{s}$ 
meson but its mass can be determined in a lattice QCD calculation from 
the masses of the pion and kaon~\cite{Dowdall:2013rya}.

We similarly account for (sea) light quark mass mistuning by 
defining $\delta_l = m_{l0} - m_l^{\text{tuned}}$. We 
find $m_l^{\text{tuned}}$ from $m_s^{\text{tuned}}$, using the fact 
that the ratio of quark masses is regularization independent, and 
was determined in~\cite{Bazavov:2017lyh}:
\begin{align}
  \left.\frac{m_s}{m_l}\right\rvert_{\textrm{phys}} = 27.18(10).
\end{align}
We set $m_l^{\text{tuned}}$ to $m_s^{\text{tuned}}$ divided by this ratio.

All higher order contributions, such as $\delta_{s,l}^2$, $(M_{\eta_c}-M_{\eta_c}^{\text{phys}})^2$, or $(\Lambda_{\text{QCD}}/ M_{\eta_h})^2$ are too small to be resolved by our lattice data, so are not included in the fit.

In our lattice QCD calculation we set $m_u=m_d\equiv m_l$; 
this means that our results do not allow for strong-isospin breaking in
the sea quarks. By moving the $m_l^{\text{tuned}}$ value up and 
down by the PDG value for $m_d-m_u$~\cite{Tanabashi:2018oca}, 
we found that any impact of strong-isospin breaking on our results 
was negligible. 

\subsubsection{Finite-Volume and Topology Freezing Effects}

We expect finite-volume effects to be negligible in our calculation
and we do not include any associated error. Finite-volume corrections 
to the $B\to D\ell\nu$ form factors were calculated in chiral perturbation 
theory in~\cite{Laiho:2005ue} and found to be very small, less than 
one part in $10^4$, for typical lattice QCD calculations. For 
$B_s\to D_s \ell \nu$ form factors we expect finite-volume effects to be 
smaller than this because there are no valence $u/d$ quarks.  

The finest lattices that we use here have been shown 
to have only a slow variation of topological charge in Monte 
Carlo time. This means that averaging results over the ensemble could 
introduce a bias if the quantities we are studying are sensitive 
to topological charge. A study calculating the adjustment needed 
to allow for this gives only a 0.002\% effect on the $D_s$ 
decay constant~\cite{Bernard:2017npd}. For $B_s$ to $D_s$ form 
factors we might expect an effect of similar relative size. 
This is negligible compared to our other uncertainties. 

\subsubsection{Uncertainties in the Physical Point}

Once we have fit our lattice results as described above, 
we can determine the physical form factors by 
setting $a=0$, $M_{\eta_h}=M_{\eta_b}$. We also take $M_{H_c}=M_{B_c}$, 
$M_{\eta_c}=M_{\eta_c}^{\text{phys}}$, $m_{l0}=m_{l}^{\text{tuned}}$, 
and $m_{s0}=m_{s}^{\text{tuned}}$. We take the experimental value 
for $M_{\eta_b}$ but allow for an additional $\pm10$ MeV uncertainty 
beyond the experimental uncertainty, since our lattice QCD results 
do not allow for QED effects or for $\eta_b$ annihilation to 
gluons~\cite{McNeile:2012qf}. This additional uncertainty has 
no effect, however, because the heavy quark mass dependence is mild.

\begin{figure}[ht]
  \hspace{-10pt}
  \includegraphics[width=0.50\textwidth]{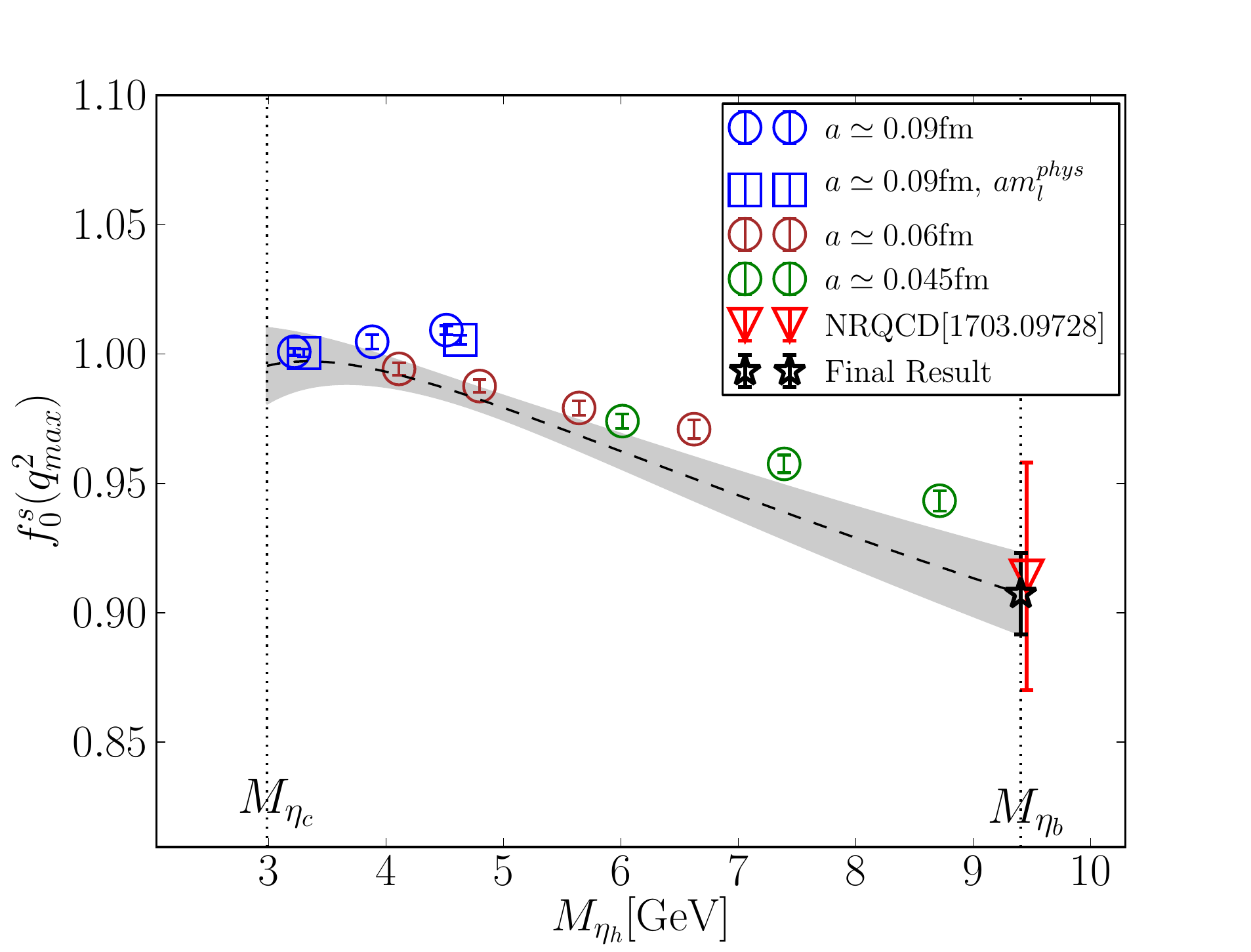}
  \caption{ $f_0^s(q^2_{\text{max}})$ against $M_{\eta_h}$ (a proxy for 
the heavy quark mass). The grey band shows the result of our fit
at $a=0$ and physical $l$, $s$ and $c$ masses. 
Notice that the $y$-axis scale does not begin at zero.
We also include the result from a previous lattice calculation, 
which used the NRQCD discretisation for the $b$ 
quark with a non-relativistic expansion of the current 
through $\mathcal{O}(\Lambda/m_b)$ and $\mathcal{O}(\alpha_s)$ 
matching to continuum QCD~\cite{Monahan:2017uby}. 
Sets of gluon field configurations listed in the legend follow 
the order of sets in table~\ref{tab:ensembles}. \label{fig:directq2max}}
\end{figure}

\begin{table}
  \begin{center}
    \begin{tabular}{c c}
      \hline
      Source & \% Fractional Error \\ [0.5ex]
      \hline
      Statistics & 1.11  \\ [1ex]
      $m_h \to m_b$ and $a\to 0$ & 1.20  \\ [1ex]
      Quark mistuning & 0.58 \\ [1ex]
      \hline
      Total & 1.73 \\ [1ex]
      \hline
    \end{tabular}
  \end{center}
  \caption{Error budget for $f_0^s(q^2_{\text{max}})$.\label{directq2max_budget}}
\end{table}

\begin{figure}[ht]
  \hspace{-20pt}
  \includegraphics[width=0.45\textwidth]{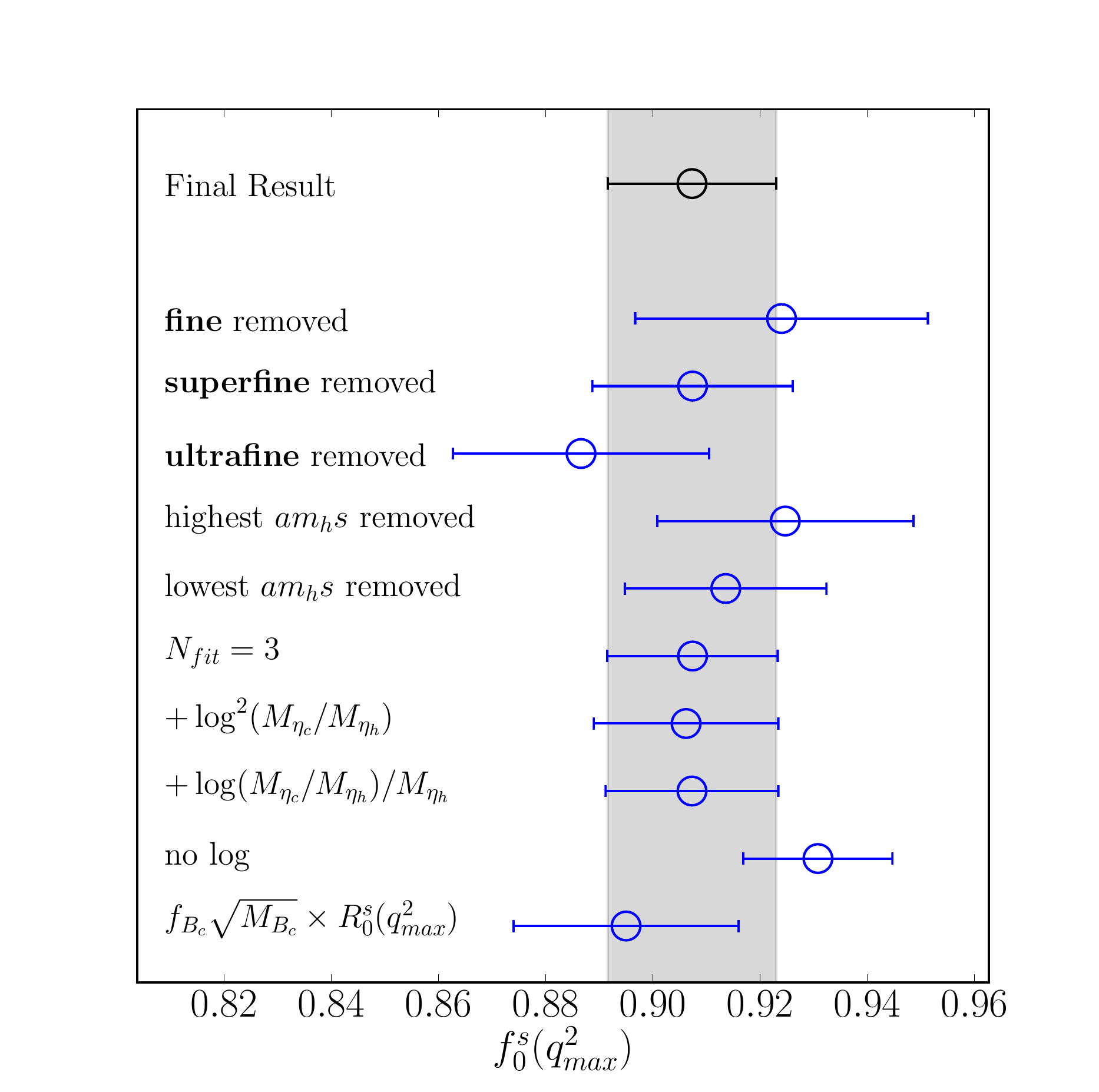}
  \caption{ Results of tests of the $f_0^s(q^2_{\text{max}})$ fit. The top three blue points 
show $f_0^s(q^2_{\text{max}})$ at continuum and
physical $b$ mass, if data from the fine, superfine or ultrafine ensembles are not 
used in the fit. The fourth and fifth blue points show the result if data at the 
highest/lowest $am_{h0}^{\text{val}}$ value on each ensemble are removed. 
The point labelled $N_{fit}=3$ is the result of extending the sum in Equation~\eqref{eq:mhfitfun} 
so that it truncates at 3 rather than 2 in each of the $i,j,k$ directions. 
The points labelled $+\text{log}^2(M_{\eta_h}/M_{\eta_c})$ represents the result of 
adding a $\rho_{2} \text{log}^2(M_{\eta_h}/M_{\eta_c})$ term in the first 
set of brackets in Equation~\eqref{eq:mhfitfun}, where $\rho_{2}$ is a new fit 
parameter with the same prior distribution as $\rho$. 
Similarly for the $+\text{log}(M_{\eta_h}/M_{\eta_c})/M_{\eta_h}$ point. 
The point labelled "no log" results from omitting the 
factor $(1+\rho \text{log}(M_{\eta_c}/M_{\eta_h}))$. The lowest point shows the value
from the fit result for $R_0^s(q^2_{\text{max}})$, multiplied by the experimental value 
for $\sqrt{M_{B_c}}$ \cite{Tanabashi:2018oca} and the result of our determination 
of $f_{B_c}$ at the physical point detailed in Appendix A of~\cite{McLean:2019sds}.
\label{fig:directtests}}
\end{figure}

\section{Results and Discussion}
\label{sec:results}

In tables~\ref{tab:masses} and~\ref{tab:kinetic}, we give our results 
for the form factors on each ensemble along with the meson masses needed 
for the fits of the form factors as a function of $m_h$ and $a$ discussed 
in Section~\ref{sec:extrapolation}. 

We first show results from simplified fits to zero recoil 
data to find $f_{0,+}^s(q^2_{\text{max}})$. This allows us to 
test the behaviour in $m_h$. 
We then perform the larger fit, described in 
Section~\ref{sec:extrapolation}, taking into account all the lattice 
data throughout the 
$q^2$ range. 

\subsubsection{Zero Recoil}
\label{sec:directq2max}

We performed a fit to $f^s_0(q^2_{\text{max}})$ as a function 
of $m_h$ and $a$ using the fit form 
\begin{align}
  \nonumber  f_0^s(q^2_{\text{max}})|_{\text{fit}} = & \,\,\left( 1 + \rho \text{log}\left(M_{\eta_h}\over M_{\eta_c}\right)\right) \times \\ \nonumber
  &\hspace{-5em}\sum_{i,j,k = 0}^{2,2,2} d_{ijk} \left({2\Lambda_{\text{QCD}}\over M_{\eta_h} }\right)^{i} \left({ am^{\text{val}}_{h0} \over \pi }\right)^{2j} \left({ am^{\text{val}}_{c0} \over \pi }\right)^{2k} \\ & \times \left(\, 1 + \mathcal{N}_{\text{mistuning}} \,\right) \, .
      \label{eq:mhfitqmax}
\end{align}
This is the same fit function as described earlier for the individual 
$z$-space coefficients in Equation~(\ref{eq:mhfitfun}) and we take the same 
priors for the corresponding coefficients as discussed there. 

The fit has $\chi^2/N_{\text{dof}} =0.21$, for 12 degrees of freedom.
Evaluating the result at $a=0$ and physical $b$ quark 
mass, we find
\begin{align}
\label{eq:f0qmax}
  f_0^s(q^2_{\text{max}}) = 0.907(16)\,.
\end{align}
We show the dependence on $M_{\eta_h}$ of our results and the 
fit in Figure~\ref{fig:directq2max}. 
The error budget corresponding to Equation~\eqref{eq:f0qmax} is given 
in Table~\ref{directq2max_budget}. 
Note that we do not impose the constraint that $f_0^s(q^2_{\text{max}}) = 1.0$
when $m_h=m_c$. If we do this, we reduce the 
uncertainty in Equation~(\ref{eq:f0qmax}) by 25\%. 

We include in Figure~\ref{fig:directq2max} a previous lattice 
determination of $f_0^s(q^2_{\text{max}})$~\cite{Monahan:2017uby}, 
shown as a red triangle. Our result, containing independent 
uncertainties, is in agreement with this earlier value but much more 
accurate. The older study used the $n_f=2+1$ MILC asqtad gluon 
ensembles, with HISQ $s$ and $c$ valence quarks, and an NRQCD 
$b$ quark. Using NRQCD meant that the calculation could be 
performed directly at the physical $b$ mass. 
However, the matching of lattice NRQCD currents to continuum 
QCD, performed at $\mathcal{O}(\alpha_s)$, is a significant source 
of systematic error absent in our calculation.

We perform a number of tests of the fit at zero recoil, 
and present results in Figure~\ref{fig:directtests}.
The tests show that the fits are robust. 

\begin{figure*}[ht]
  \begin{center}
  \includegraphics[width=0.9\textwidth]{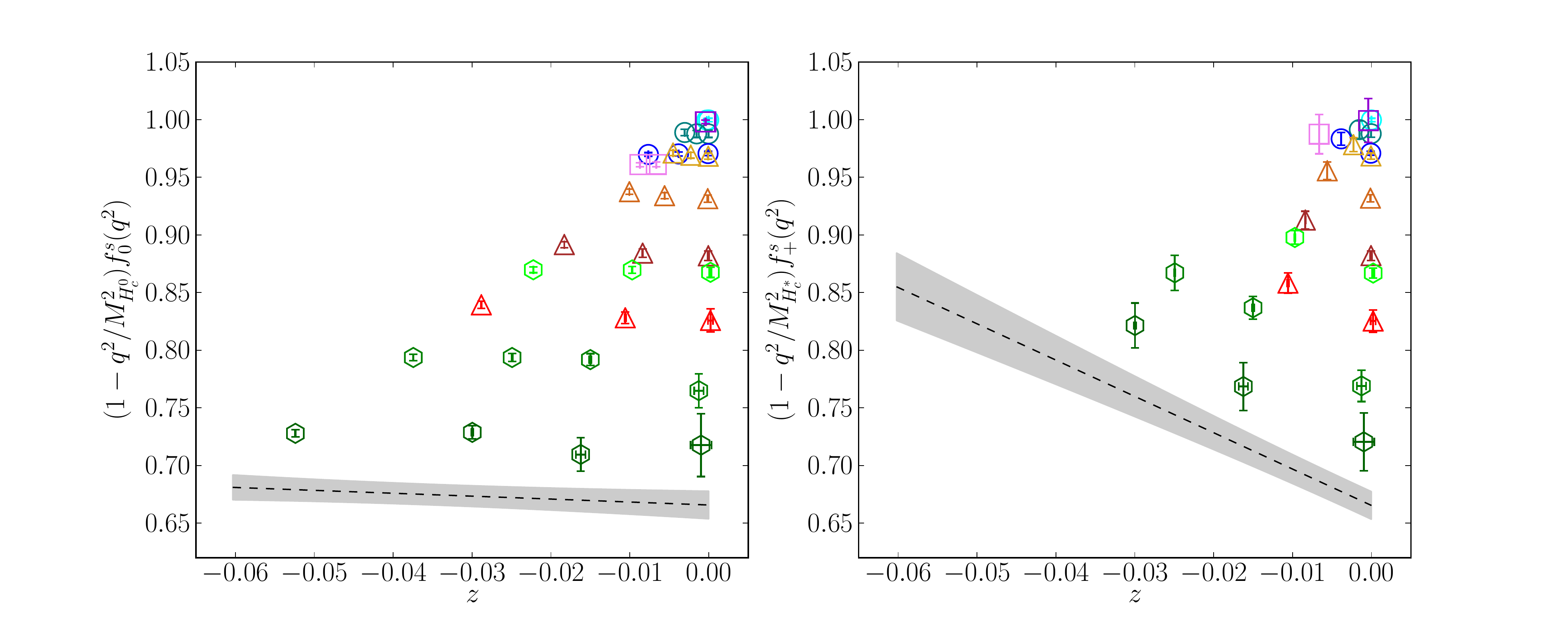}
  \caption{ $Pf_{0,+}^s$ in $z$-space, where $P$ is the appropriate 
pole function for each form factor given in Equation~(\ref{eq:BCL}). 
The grey band shows the 
result of our fit at $a=0$ and physical $l$,$s$, $c$ {\it{and}} 
$b$ masses. Sets listed in the legend follow the order of sets in 
Table~\ref{tab:ensembles}. \label{fig:directz}}
    \end{center}
\end{figure*}

\subsubsection{Full $q^2$ range}
\label{sec:directfullq2}

\begin{figure*}[ht]
  \begin{center}
  \includegraphics[width=0.9\textwidth]{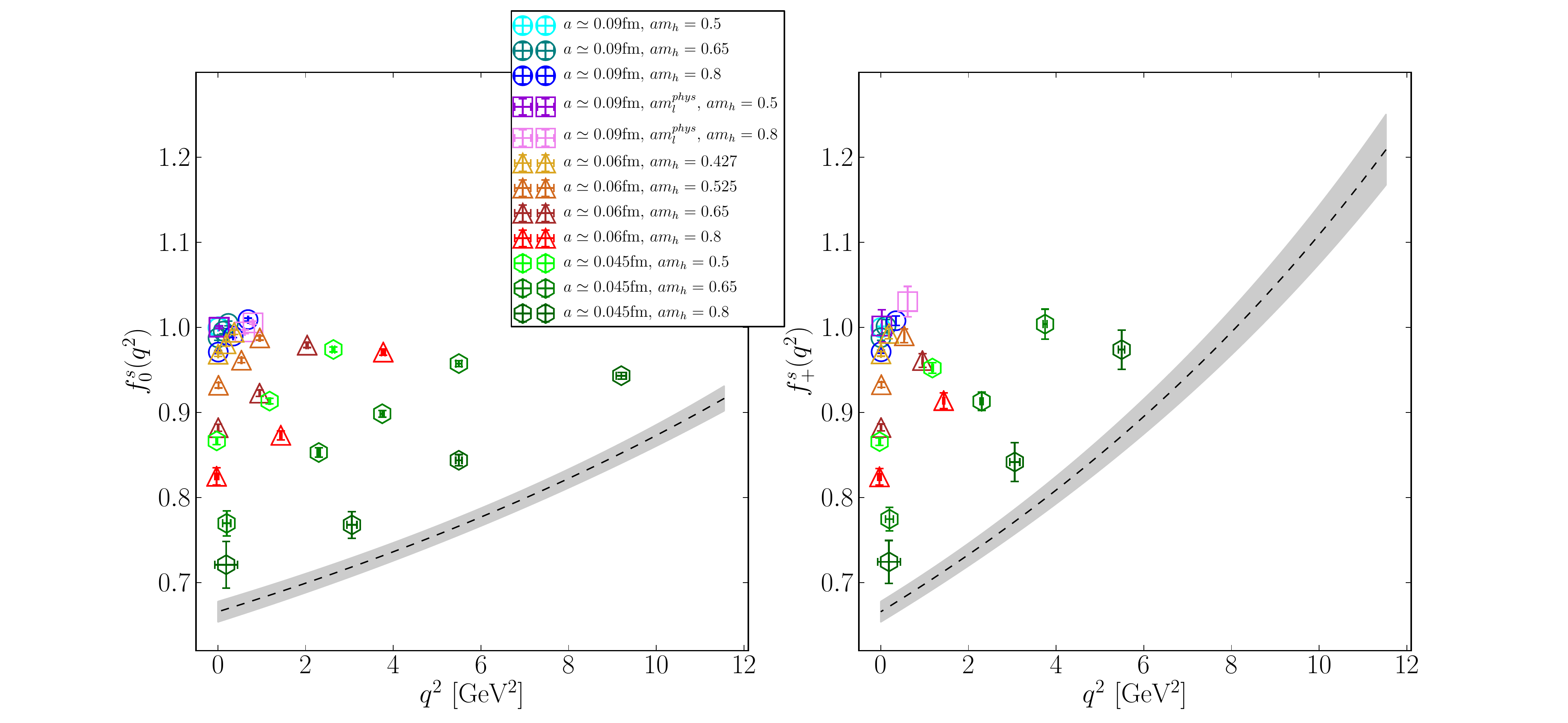}
  \caption{ $f_{0,+}^s(q^2)$ against $q^2$. The grey band shows the 
result of our fit at $a=0$ and physical $l$,$s$, $c$ {\it{and}} 
$b$ masses. Sets listed in the legend follow the order of sets in 
Table~\ref{tab:ensembles}. \label{fig:directdata}}
    \end{center}
\end{figure*}

\begin{figure}[ht]
  \begin{center}
  \includegraphics[width=0.50\textwidth]{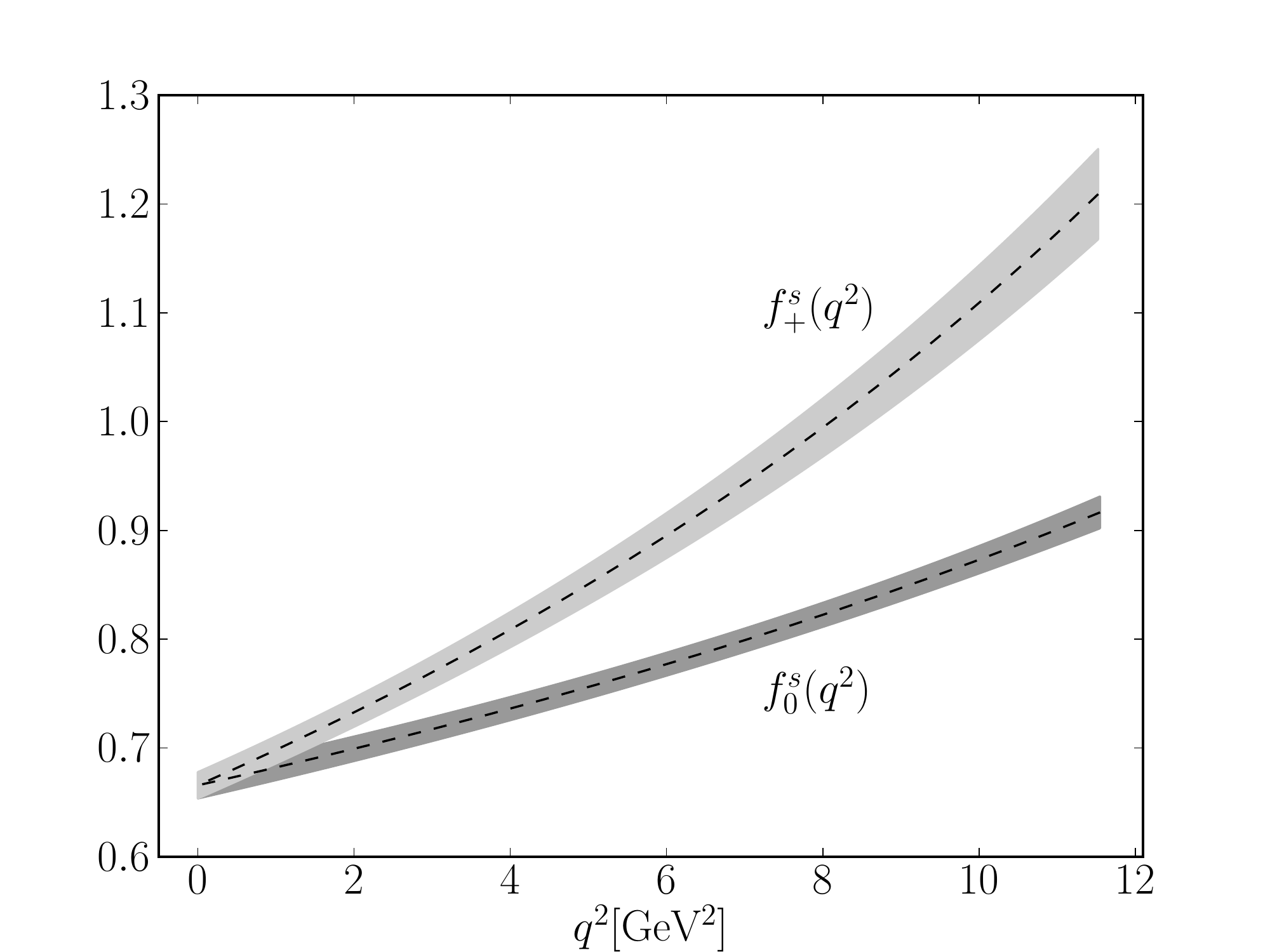}
  \caption{ Final result for $f_{0,+}^s(q^2)$ against $q^2$ at the physical point \label{fig:directfinal}.}
    \end{center}
\end{figure}

\begin{figure}[ht]
  \begin{center}
  \includegraphics[width=0.45\textwidth]{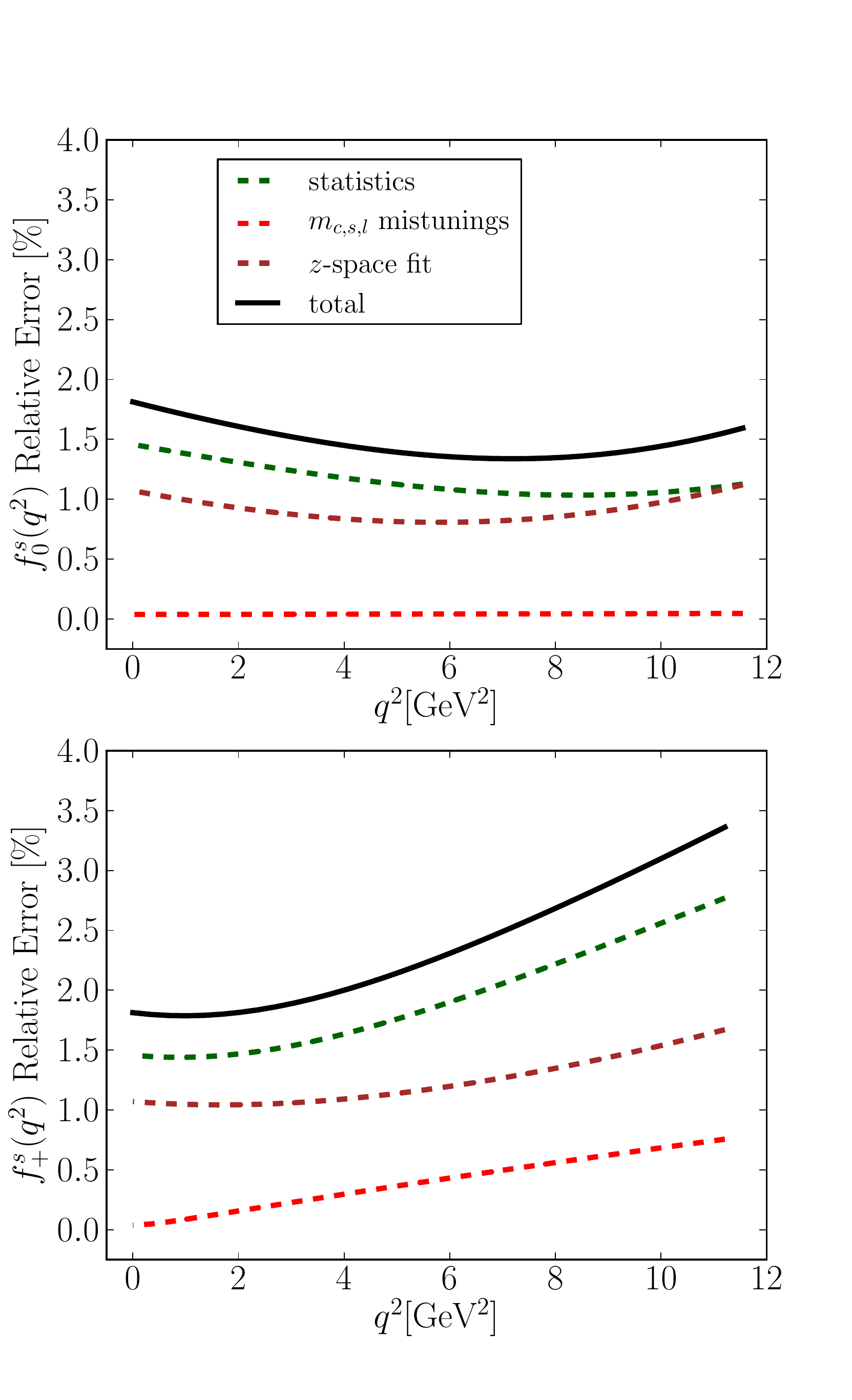}
  \caption{ Error budget for $f_{0,+}^s(q^2)$ as a function of $q^2$ \label{fig:directerrorbudget}.}
    \end{center}
\end{figure}

\begin{figure}[ht]
  \begin{center}
  \hspace{-10pt}
  \includegraphics[width=0.5\textwidth]{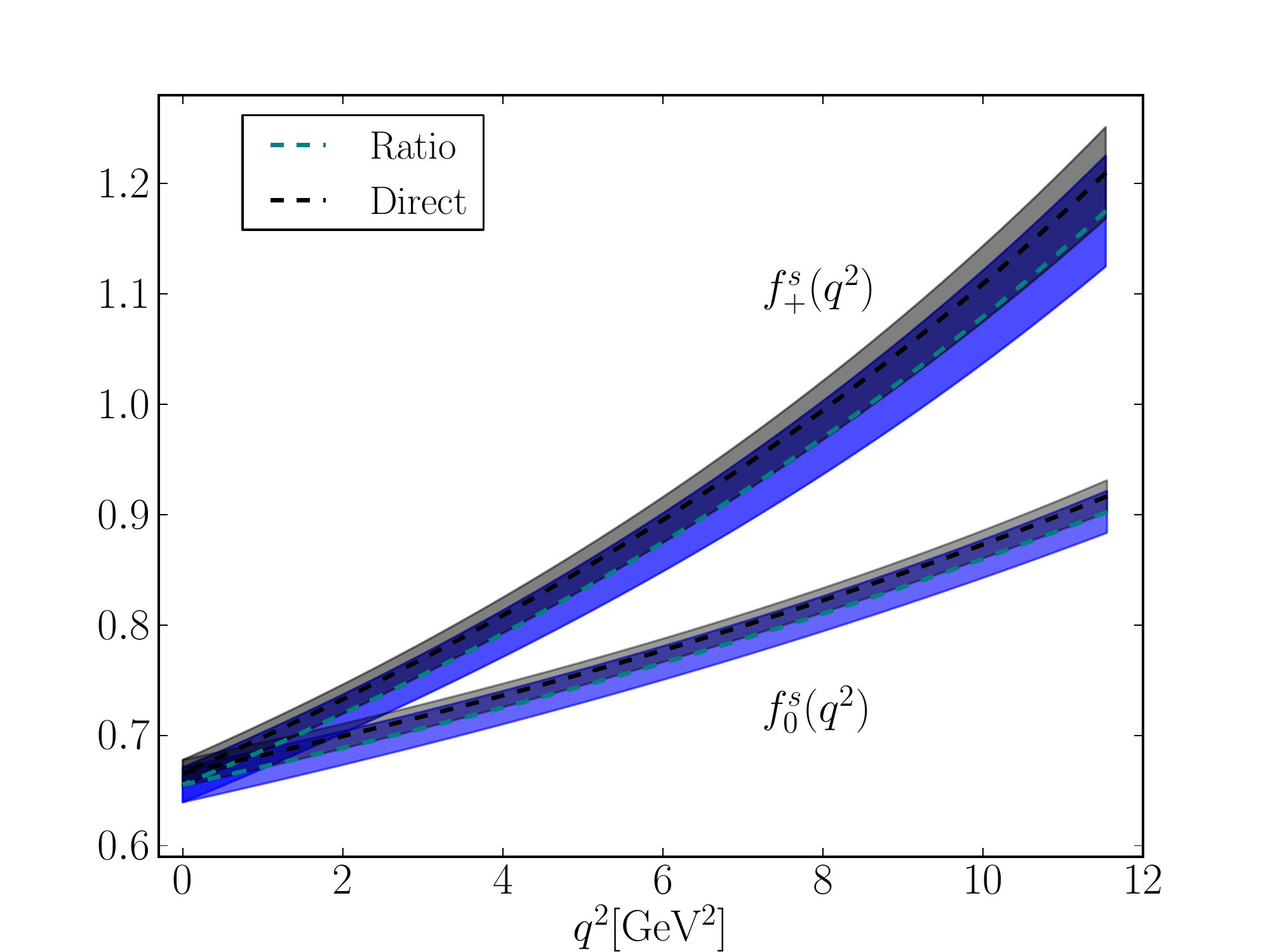}
  \caption{ Results for $f_{0,+}^s(q^2)$ against $q^2$ at the physical point, 
comparing the ratio method (from Appendix~\ref{appendix:ratio}) 
and the direct method (from Section~\ref{sec:directfullq2}). \label{fig:ratiovsdirect}}
    \end{center}
\end{figure}

\begin{figure}[ht]
  \hspace{-10pt}
  \includegraphics[width=0.52\textwidth]{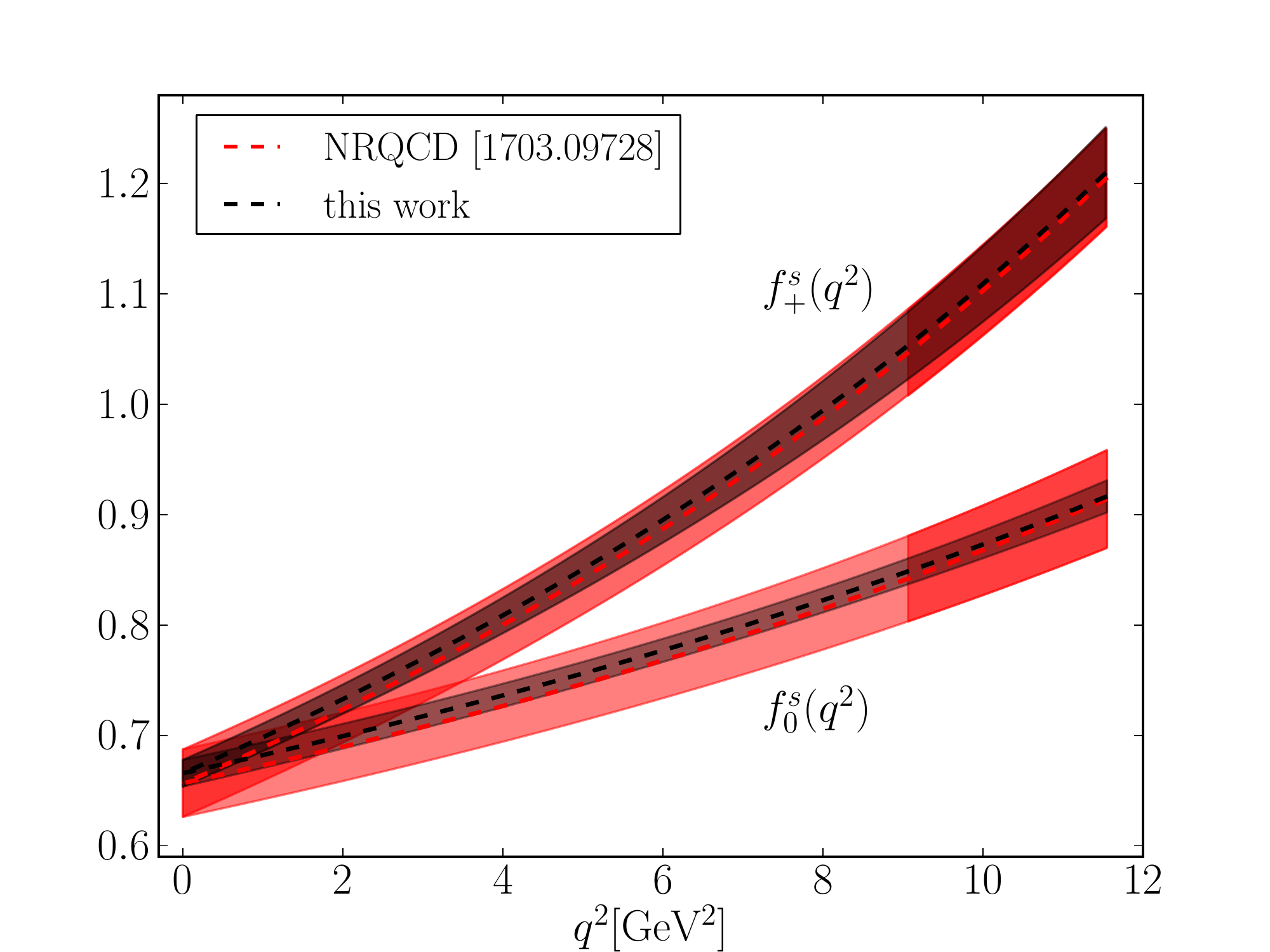}
  \caption{ Our final result for $f_{0,+}^s(q^2)$ compared to form 
factors calculated using an NRQCD action for the $b$ 
quark~\cite{Monahan:2017uby}. 
Part of the NRQCD band is shaded darker than the rest 
($q^2 \gtrapprox 9.5$GeV$^2$) to signify the region where lattice results 
were directly calculated. The NRQCD form factors in the rest of 
the $q^2$ range are the result of an extrapolation 
using a BCL parameterization.\label{fig:nrqcd}}
\end{figure}

We now proceed to fit our full set of data including zero and 
non-zero recoil points to the fit form given in Equations~(\ref{eq:BCL}) 
and~(\ref{eq:mhfitfun}) and discussed in Section~\ref{sec:extrapolation}. 
We include the covariance matrix between 
the form factor values obtained from correlator fits on a given ensemble 
along with correlated lattice spacing uncertainties.  
The goodness of fit obtained is 
$\chi^2/N_{\text{dof}} = 0.51$, with 58 degrees of freedom.

In Figure~\ref{fig:directz} we show our results and fit function in 
$z$-space for the form factors multiplied by their appropriate 
pole factors, $P$, given by $1-q^2/M_{H_c^{0,*}}$ in Equation~(\ref{eq:BCL}). 
This shows that the $z$-dependence is relatively benign for 
both form factors and the main $m_h$-dependent effect is the 
smooth reduction in value of $Pf$ as $m_h$ increases. The final 
result at the physical $b$ quark mass is given by the grey band.  

In Figure~\ref{fig:directdata}, we show the results and fit 
function in $q^2$-space. The form factors for the physical 
$b$ quark mass (i.e. those corresponding to $B_s \to D_s$ decay) 
are given by the grey band.   

Figure~\ref{fig:directfinal} shows the physical $f_+$ and $f_0$ 
form factors on the same plot and covering the full $q^2$ range 
for the $B_s \to D_s$ decay. 
Figure~\ref{fig:directerrorbudget} plots the 
associated error budget for the two form factors throughout 
the $q^2$ range. The dominant uncertainty comes from statistical 
errors. There are also significant uncertainties from the $q^2$ and 
$m_h$ dependence for $f_+$ at larger values of $q^2$. This is because 
there are no lattice QCD results at $q^2_{\text{max}}$ for $f_+$. 
The impact of uncertainties in the lattice spacing 
(both in $w_0/a$ and in $w_0$) are smaller than the errors shown 
in the Figure and so not plotted there. This is because the form 
factors themselves are dimensionless and lattice spacing effects in 
the determination of $q^2$ largely cancel as, for example, the pole masses 
are given in terms of lattice masses.  

As discussed in Section~\ref{sec:extrapolation} an alternative approach 
to the fit is to take ratios of the form factors to the $H_c$ decay 
constant and fit the ratios to the fit form of Eqs.~(\ref{eq:BCL}) 
and~(\ref{eq:mhfitfun}). This fit is described in 
Appendix~\ref{appendix:ratio}. It has the advantage of smaller discretisation 
effects but the disadvantage of larger lattice spacing uncertainties
because the ratios being fit are dimensionful. In the end the ratio 
method has larger uncertainty for the final physical form factors.  
We therefore take the results from the direct method as our final 
result, and use the ratio method results as a consistency test. 
Since the two approaches have quite different systematic 
errors, their comparison supplies a strong consistency check. 
In Figure~\ref{fig:ratiovsdirect}, we plot the form factors from 
the two methods on top of each other. As is clear from this plot, 
the results are in good agreement. The direct method gives a more accurate 
result for both form factors and at all $q^2$.  

We compare the coefficients from our fits to 
unitarity bounds in Appendix~\ref{appendix:unitarity} as a further test. 

In Figure~\ref{fig:nrqcd}, we compare our final form factors to those 
determined from the lattice QCD calculation using the NRQCD approach 
for the $b$ quark already used as a comparison at $q^2_{\text{max}}$ 
in Figure~\ref{fig:directq2max}~\cite{Monahan:2017uby}. The 
NRQCD calculation works directly at the $b$ quark mass but on 
relatively coarse lattices and hence is unable to obtain results at 
large physical momenta for the $D_s$ meson. The results close to zero-recoil 
are extrapolated to $q^2=0$ using a $z$-space parameterisation.  
As the Figure shows, our results are in excellent agreement with 
the NRQCD calculation but are more precise for 
both $f^s_0(q^2)$ and $f^s_+(q^2)$ throughout all $q^2$. This is because we 
can avoid the significant systematic uncertainty that the NRQCD calculation 
has from the perturbative matching to continuum QCD 
of the NRQCD current that couples 
to the $W$. 

\subsection{$R({D_s})$}
\label{sec:rds}

\begin{figure}[ht]
  \hspace{-20pt}
  \includegraphics[width=0.53\textwidth]{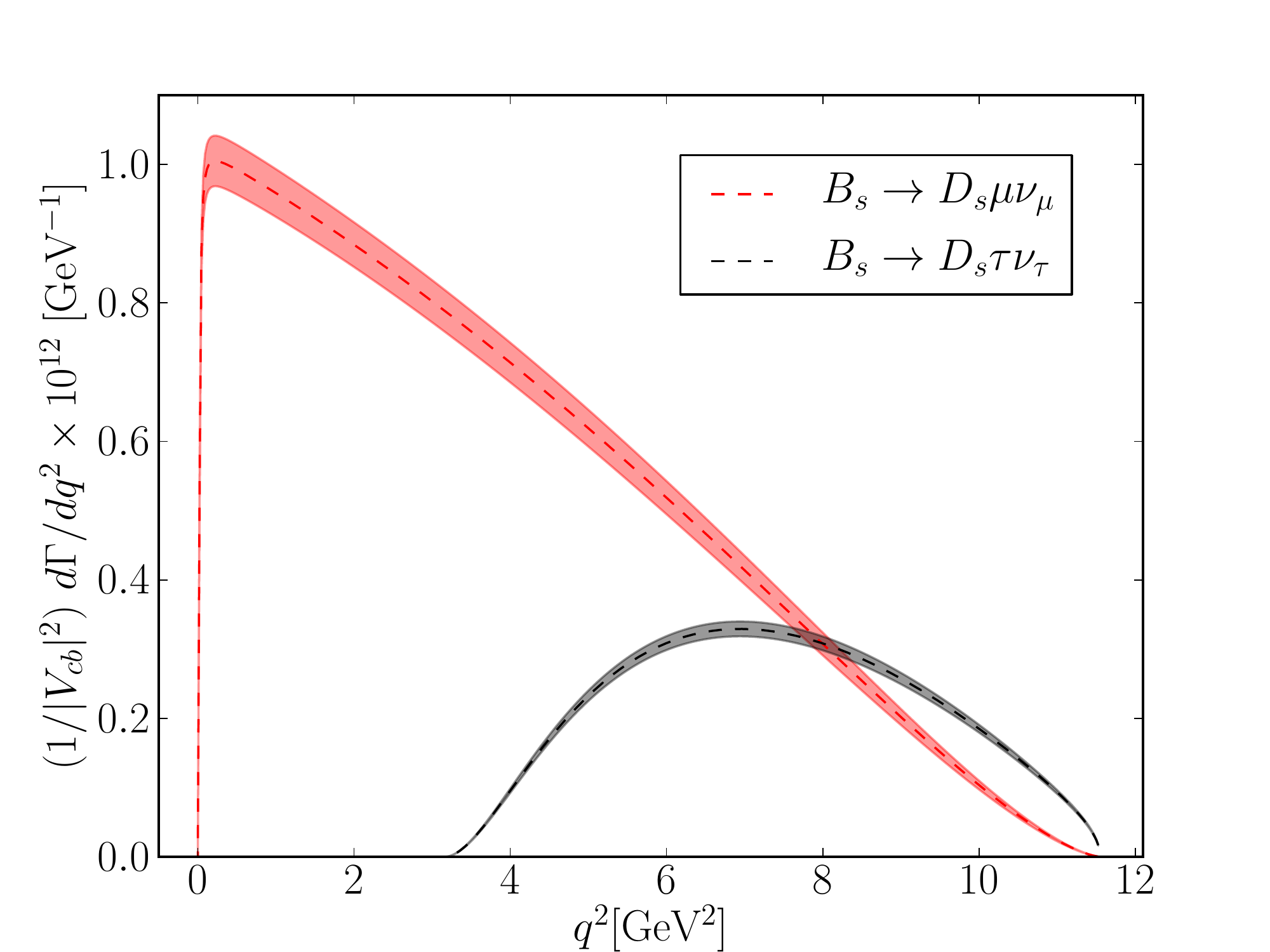}
  \caption{Differential decay rates for 
the $B_s\to D_s\mu\nu_{\mu}$ and 
$B_s\to D_s\tau\nu_{\tau}$ decays, calculated using the form factors 
determined in this work.\label{fig:diffrate}}
\end{figure}

\begin{table}
  \begin{center}
    \begin{tabular}{c c}
      \hline
      Source & \% Fractional Error \\ [0.5ex]
      \hline
      Statistics & 1.11  \\ [1ex]
      $z$-space fit & 1.05  \\ [1ex]
      Quark Mass Mistuning & 0.12 \\ [1ex]
      \hline
      Total & 1.54 \\ [1ex]
      \hline
    \end{tabular}
  \end{center}
  \caption{Error budget for our result for $R({D_s})$ in the SM. $z$-space fit refers to the error associated with the fit of the dependence on heavy quark mass and lattice 
spacing and interpolation in $q^2$. \label{RDs_budget}}
\end{table}

Using our calculated form factors $f^s_{0,+}(q^2)$, we can 
calculate the differential rate for $B_s \to D_s \ell \nu$ decay
from Equation~(\ref{eq:diffrate}). This is a function of the lepton 
mass and so differs between the heavy $\tau$ and the light 
$e$, $\mu$ leptons. The differential rate for $\mu$ and $\tau$ 
is compared in Figure~\ref{fig:diffrate}. We take the meson 
and lepton masses needed for Equation~(\ref{eq:diffrate}) from~\cite{Tanabashi:2018oca} 
and $\eta_{\text{EW}}$ = 1.011(5)~\cite{Na:2015kha}. The distribution in the $\tau$ 
case is cut off at $q^2=m_{\tau}^2$ and so, although there is enhancement 
from $m_{\ell}^2/q^2$ terms in Equation~(\ref{eq:diffrate}) that reflect reduced helicity 
suppression, the integrated branching fraction for the $\tau$ case 
is smaller than for the $\mu$.   

The ratio of branching fractions for semileptonic $B$ decays to $\tau$ 
and to $e$/$\mu$ is being used as a probe of lepton universality 
with an interesting picture emerging~\cite{Amhis:2016xyh, CariaMoriond}.  
Here we provide a new SM prediction for the quantity
\begin{align}
  R({D_s}) = {\mathcal{B}(B_s\to D_s \tau \nu_{\tau})\over
    \mathcal{B}(B_s\to D_s l \nu_{l})}\,,
\end{align}
where $l=e$ or $\mu$ (the difference between $e$ and $\mu$ is negligible 
in comparison to our precision on $R(D_s)$). 
Our result is 
\begin{align}
\left.  R(D_s)\right|_{\text{SM}} = 0.2987(46),
  \label{eq:RDs}
\end{align}
in which we averaged over the $l=e$ and $l=\mu$ cases. Note that $|V_{cb}|$ and 
$\eta_{\text{EW}}$ cancel in this ratio.  
We give an error budget for this result in terms of the uncertainties from 
our lattice QCD calculation in Table~\ref{RDs_budget}.
Our result agrees with, but is more accurate than, the previous lattice 
QCD value of $R(D_s)$ (0.301(6)) 
from~\cite{Monahan:2017uby}. An experimental result for $R(D_s)$ would 
allow a new test of lepton universality.   

We expect very little difference between $R(D_s)$ and the analogous 
quantity $R(D)$ because the mass of the spectator quark has little 
effect on the form factors~\cite{Bailey:2012rr}.  
Lattice QCD calculations that involve light spectator quarks have 
larger statistical errors, however, which is why the process $B_s \to D_s$ 
is under better control. Previous lattice QCD results for 
$R(D)$ are 0.300(8)~\cite{Na:2015kha} and 0.299(11)~\cite{Lattice:2015rga}, 
in which any difference with our result for $R(D_s)$ is too small to 
be visible with these uncertainties.  

\section{Comparison to HQET}
\label{sec:HQET}

\begin{figure}[ht]
  \includegraphics[width=0.53\textwidth]{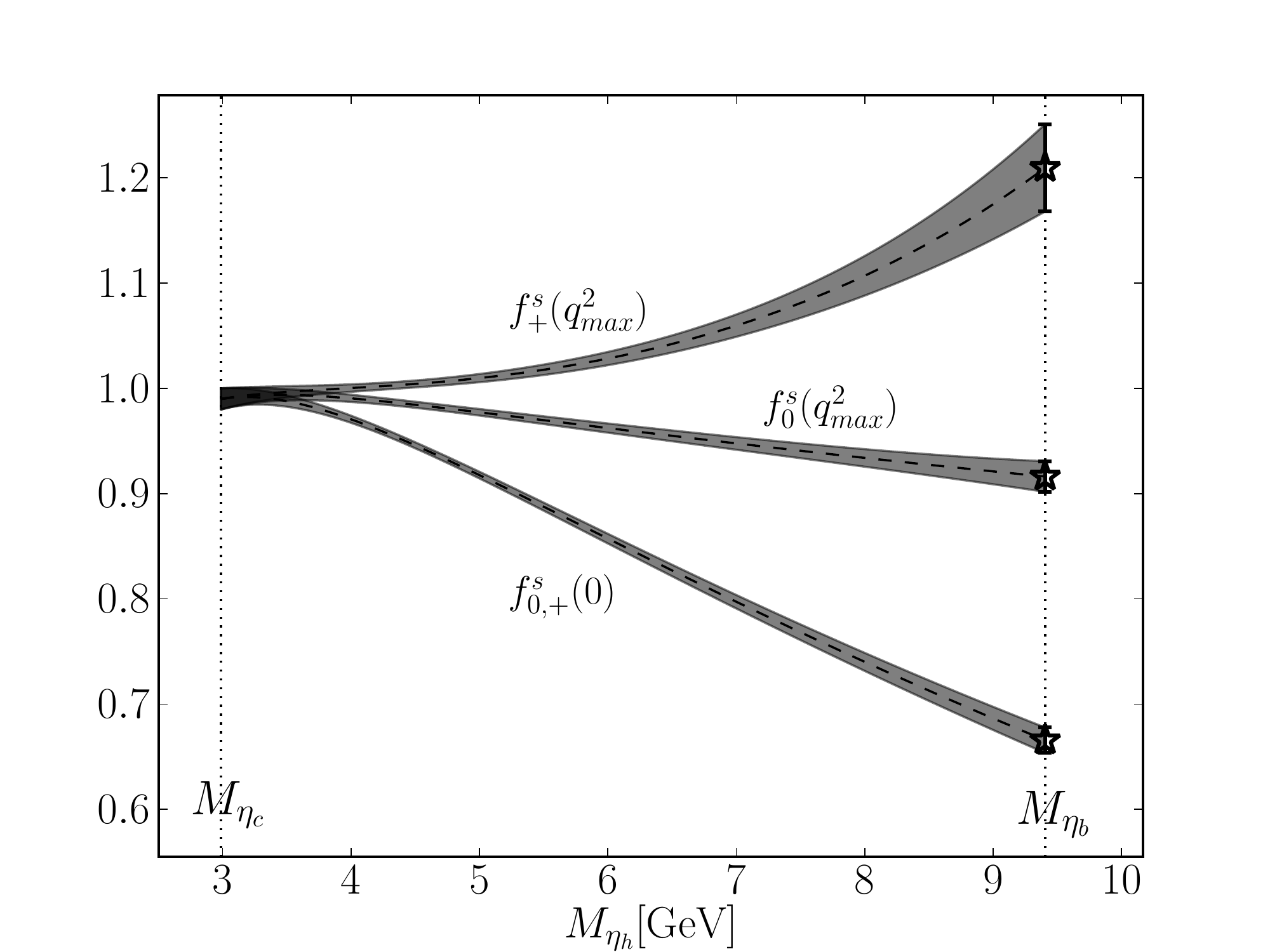}
  \caption{Form factor values at $q^2_{\text{max}}$ and $q^2=0$ plotted
against $M_{\eta_h}$, a proxy for the heavy quark mass. \label{fig:q0qmaxvmh}}
\end{figure}

\begin{figure}[ht]
  \includegraphics[width=0.53\textwidth]{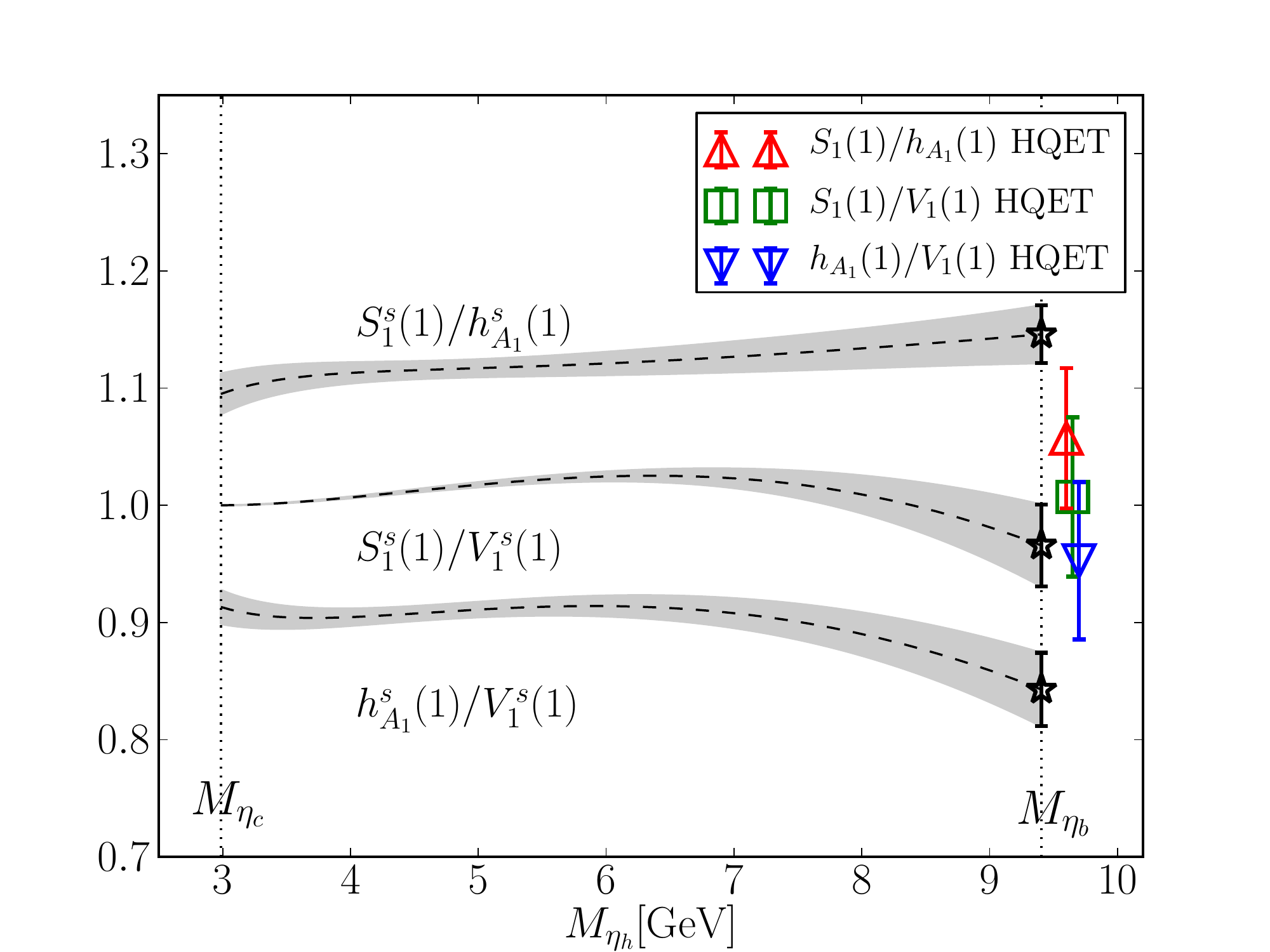}
  \caption{Form factor ratios against $M_{\eta_h}$, a proxy for the heavy quark mass. 
$S_1^s$ and $V_1^s$ are defined in Eqs.~\eqref{eq:Shqet} and \eqref{eq:Vhqet}. 
Note that only two of these ratios are independent.
The colourful points are NLO HQET expectations from \cite{Bernlochner:2017jka}, 
derived with input from QCD sum rules. The HQET error bars each include a 
6\% uncertainty to allow for missing higher-order terms. \label{fig:HQSB}}
\end{figure}

\begin{figure}[ht]
  \includegraphics[width=0.53\textwidth]{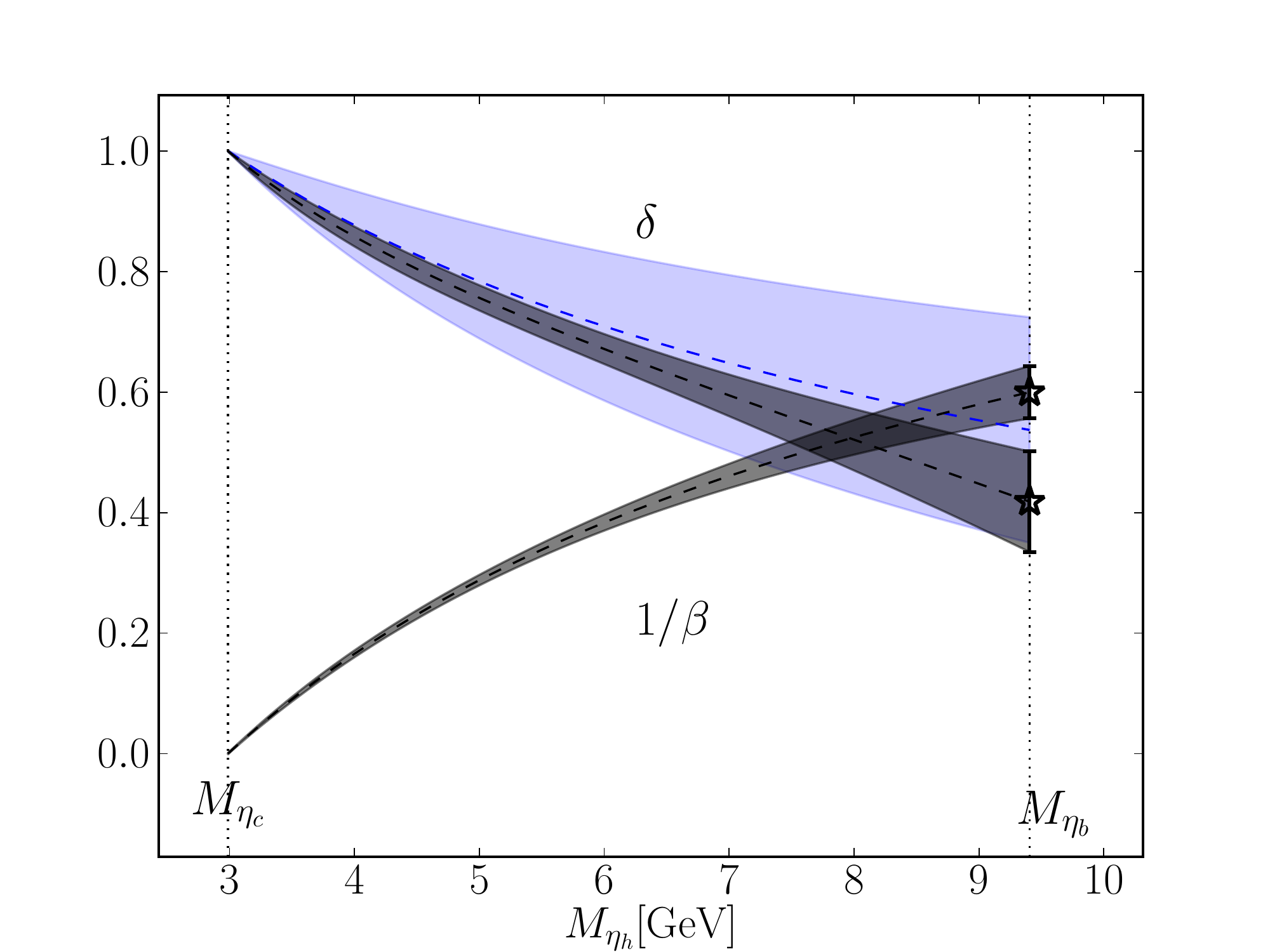}
  \caption{We show two quantities derived from the form 
factor slopes as a function of $M_{\eta_h}$. 
$1/\beta(m_h)$ is defined in Equation~\eqref{eq:beta}) and $\delta$ 
in Equation~\eqref{eq:delta}. Our results are shown by the grey bands. 
The blue band shows the leading order HQET expectation for $\delta$ given 
in Equation~\eqref{eq:delta_hqet}. \label{fig:delta}}
\end{figure}

In Figure~\ref{fig:q0qmaxvmh} we show our form factor results 
at two key values of $q^2$, the zero recoil point and $q^2=0$, 
as a function of heavy quark mass, given by $M_{\eta_h}$. 
The plot demonstrates how $f_+$ at zero-recoil increases as the 
heavy quark mass increases, but $f_0$ changes very little. 
The value at $q^2=0$, where the form factors are equal, falls 
with growing heavy quark mass, as the $q^2$ range opens up.  

Knowledge of the functional form of $f_0^s(q^2)$ and $f_+^s(q^2)$, 
along with that of the $B_s \rightarrow D_s^* \ell \nu$ form 
factor at zero recoil, $h_{A_1}^s(1)$, from~\cite{McLean:2019sds}, against $m_h$ 
gives us access to the functional form in $m_h$ of a number of 
quantities of interest in HQET.

In HQET the vector current matrix element is parameterized with a different set of form factors, $h^s_+(w)$ and $h^s_-(w)$ according to
\begin{align}
  {\langle D_s | V^{\mu} | H_s \rangle \over \sqrt{M_{D_s}M_{H_s}} } =
  h^s_+(w) ( v + v' )^{\mu} + h^s_-(w) ( v - v' )^{\mu}\,,
\end{align}
where $v^{\mu} = p_{H_s}^{\mu}/M_{H_s}$ and $v'^{\mu} = p_{D_s}^{\mu}/M_{D_s}$ 
are the 4-velocities of the initial and final state mesons, 
and $w=v\cdot v'$ is an alternative parameter to $q^2$ used in the context of HQET.

As a test of HQET, one can construct ratios of form factors that should 
become unity in the $m_c,m_h\to \infty$ limit. 
Following~\cite{Caprini:1997mu}, one can redefine the form factors such that each of them reduce to the Isgur-Wise function $\xi(w)$ in the $m_c,m_h\to \infty$ limit. In the $B_s\to D_s$ case these new form factors are
\begin{align}
  \label{eq:Shqet}
  S^s_1(w) &= h^s_+(w) - {1+r\over 1-r} { w - 1 \over w + 1 } h^s_-(w)\,, \\
  V^s_1(w) &= h^s_+(w) - { 1 - r \over 1 + r } h^s_-(w)\,,
  \label{eq:Vhqet}
\end{align}
where $r = M_{D_s}/M_{H_s}$. $h_{A_1}^s$ also reduces 
to $\xi$ in the infinite mass limit. Hence any ratio 
between $S^s_1$,$V^s_1$ and $h_{A_1}^s$ should become unity in 
this limit. From our results at zero recoil and for $m_h=m_b$ we find
\begin{align}
  {S^s_1(1)\over h^s_{A_1}(1)}\bigg|_{\text{lat}} &=  1.146(25) \\
  {S^s_1(1)\over V^s_1(1)}\bigg|_{\text{lat}} &=  0.966(35) \\
  {h^s_{A_1}(1)\over V^s_1(1)}\bigg|_{\text{lat}} &=  0.843(31)
\end{align}

Fig. \ref{fig:HQSB} illustrates how these ratios vary with $m_h$, and gives 
the NLO HQET expectation for these values for 
comparison~\cite{Bernlochner:2017jka}. The HQET results include a 6\% uncertainty 
to allow for missing higher order terms in $\alpha_s^2$, $\alpha_s\overline{\Lambda}/m_c$ 
and $(\overline{\Lambda}/m_c)^2$ as suggested for these ratios 
in~\cite{Bigi:2017jbd}. 
Our results show some tension with the HQET expectations that might 
indicate that the missing higher order contributions are bigger than 6\%. 
This is discussed in~\cite{Bigi:2017jbd} 
in the context of earlier lattice QCD results. 
Preliminary 
results from the JLQCD collaboration~\cite{Kaneko:2018mcr} 
show similar values to ours. 

Another set of quantities of interest in HQET are the slopes of the 
form factors at $q^2=0$~\cite{Hill:2005ju,Hill:2006ub}:
\begin{align}
  \label{eq:beta}
  {1\over \beta(m_h)} \equiv&\,\, {M_{H_s}^2-M_{D_s}^2 \over f^s_+(0) } \,{df^s_+\over dq^2}\bigg|_{q^2=0}\,, \\
  \delta(m_h) \equiv& \,\,1 - {M_{H_s}^2 - M_{D_s}^2 \over f^s_+(0) } \left( {df^s_+\over dq^2} \bigg|_{q^2=0} - {df^s_0\over dq^2} \bigg|_{q^2=0} \right)\,.
  \label{eq:delta}
\end{align}
To obtain these values from our results for the form factors, 
we take the derivative of the fit function 
(Equation~(\ref{eq:mhfitfun})) evaluated at 
continuum and physical $l,s$ and $c$ masses. At $m_h=m_b$ we find
\begin{align}
  {1\over \beta(m_b)} = 0.600(43) \,,\quad \delta(m_b) = 0.405(84)\,.
\end{align}
In Figure~\ref{fig:delta} we show how these 
quantities vary with $m_h$. By rewriting Equation~\eqref{eq:delta} in terms 
of $h^s_{+,-}$, and recognising that in the heavy quark 
limit $h^s_+ \approx \xi$ and $h^s_- \approx 0$, one can find a 
leading order HQET expectation for $\delta$~\cite{Hill:2006ub}:
\begin{align}
  \label{eq:delta_hqet}
  \delta(m_h) &= { 2M_{D_s} \over M_{H_s} + M_{D_s} } { 1 + {h_-\over h_+ } \over 1 - {M_{H_s}-M_{D_s}\over M_{H_s}+M_{D_s}} {h_-\over h_+} } \\
  &= { 2M_{D_s} \over M_{H_s} + M_{D_s} } \left[ 1 + \right. \nonumber \\
&\left. \left(\frac{M_{H_s}-M_{D_s}}{M_{H_s}+M_{D_s}}\right)\order{\alpha_s,\Lambda_{\text{QCD}}/m_h,\Lambda_{\text{QCD}}/m_c } \right]\,. \nonumber
\end{align}
This leading order expression, along with an uncertainty from 
missing higher-orders as indicated above, is shown in Figure~\ref{fig:delta} as a 
blue band. Note that by definition $\delta(m_c)=1$. 
Our results (grey band) are in good agreement with this, up to the 
uncertainties from higher order terms shown in Equation~(\ref{eq:delta_hqet}).

\section{Conclusions}
\label{sec:conclusions}

We have calculated the scalar and vector form factors for the $B_s \to D_s \ell \nu$ 
decay for the full $q^2$ range using lattice QCD with a fully nonperturbative 
normalisation of the current operators for the first time.  
Our calculation used correlation functions at three values of the 
lattice spacing, including an ensemble with an approximately physical light quark mass. 
We used the relativistic HISQ action with a range of values for the heavy valence 
quark and by fitting this dependence, along with the lattice spacing dependence, we 
are able to determine results at the $b$ quark mass. The valence $c$ and $s$ quark 
masses are accurately tuned to their physical values. By working on very fine lattices 
we are able both to reach a heavy quark mass close to the $b$ but also to cover 
the full $q^2$ range of the decay.   

Our results for the form factors are given in Figure~\ref{fig:directfinal} and 
the differential rate that this implies for $B_s \to D_s \ell \nu$ in 
Figure~\ref{fig:diffrate}. This will allow a determination of $|V_{cb}|$ 
from future experimental data from this semileptonic process. 
Instructions on how to reproduce our form factors 
are given in Appendix~\ref{sec:reconstructing_formfactors}. 
Our error budget is given in Figure~\ref{fig:directerrorbudget}. 

Our results are more accurate than previous lattice QCD determinations 
using a nonrelativistic $b$ quark formalism because we 
do not have a systematic uncertainty from the perturbative matching of the 
lattice current to continuum QCD. 

From our results we can predict the ratio $R(D_s)$ of the branching fraction 
to a $\tau$ lepton compared to that to $e$/$\mu$ (see Section~\ref{sec:rds}).  
We are also able to compare functions of the form factors and their slopes to 
HQET expectations (see Section~\ref{sec:HQET}). 

Our calculation shows that a heavy-HISQ determination 
of the $B\to D\ell\nu$ form factors is feasible. 
This would allow direct comparison to existing experimental data. 
Such a calculation could use an identical strategy to the one demonstrated here, 
with the strange valence quark replaced with a light one 
and additional calculations on ensembles spanning a range of 
light quark masses. Higher statistics would be needed since statistical 
uncertainties will be larger than in this calculation, and the issue of 
topology freezing on fine lattices will be more significant. 
Our calculation demonstrates, however, that lattice QCD is no longer
limited for these form factors by the systematic uncertainties coming from 
current matching and, with sufficient computer time, a ~1\% accurate 
result for $B \to D\ell \nu$ form factors is achieveable. 

\subsection*{\bf{Acknowledgements}}

We are grateful to the MILC collaboration for the use
of their configurations and their code. Computing was
done on the Cambridge Service for Data Driven Discovery (CSD3) supercomputer, part of which is operated by the University of Cambridge Research Computing Service on behalf of the UK Science and Technology Facilities Council (STFC) DiRAC HPC Facility. The DiRAC component of CSD3 was funded by BEIS via STFC capital grants and is operated by STFC operations grants.
We are grateful to the CSD3 support staff for assistance. Funding for this work 
came from STFC. We would also like to thank C. Bouchard, B. Colquhoun, 
D. Hatton, J. Harrison, P. Lepage, Z. Ligeti and M. Wingate for useful discussions

\appendix

\section{Reconstructing Form Factors}
\label{sec:reconstructing_formfactors}

This appendix gives the necessary information to reproduce the functional form of the form factors through $q^2$ reproduced in this work. We here express the form factors in terms of the BCL parameterisation \cite{Bourrely:2008za}:
\begin{align}
  \label{eq:BCLappendix}
  &f^s_0(q^2) = {1\over 1 - {q^2\over M_{B_{c0}}^2} } \sum_{n=0}^{2} a_n^0 z^n(q^2), \\
  \nonumber
  &f^s_+(q^2) = {1\over 1 - {q^2\over M_{B_c^*}^2} } \sum_{n=0}^{2} a_n^+ \left( z^n(q^2) - {n\over 3} (-1)^{n-3} z^3(q^2) \right),  \nonumber
\end{align}
where the function $z(q^2)$ is defined by defined by
\begin{align}
  z(q^2) = {\sqrt{t_+ - q^2} - \sqrt{t_+} \over \sqrt{t_+ - q^2} + \sqrt{t_+} },
\end{align}
and $t_+ = (M_{B_s}+M_{D_s})^2$. We take the PDG 2018 values for these masses, 
5.3669~GeV for the $B_s$ and 1.9683~GeV for the $D_s$~\cite{Tanabashi:2018oca}.
For the position of the poles, we use $M_{B_{c0}} = 6.704$GeV 
and $M_{B_c^*} = 6.329$GeV. The coefficients $a_n^{0,+}$ found 
from our fit, along with their covariance, is given in Table \ref{tab:coeffs}. 
The form factor values at the two ends of the $q^2$ range are: 
$f^s_+(q^2_{\text{max}})=1.209(42)$; $f^s_0(q^2_{\text{max}})=0.917(15)$ 
(note that this value differs slightly from that in Eq.~(\ref{eq:f0qmax}) 
because this value comes from a fit that includes all $q^2$ values); 
$f^s_+(q^2=0)=f_0(q^2=0)=0.666(12)$. 

\begin{table}
\begin{center}
\begin{tabular}{ c c c c c c c }
\hline
 & $a^0_0$ & $a^0_1$ & $a^0_2$ & $a^+_0$ & $a^+_1$ & $a^+_2$\\ [0.5ex]
\hline
 & 0.66574 & -0.25944 & -0.10636 & 0.66574 & -3.23599 & -0.07478\\ [1ex]
\hline
 & 0.00015 & 0.00188 & 0.00070 & 0.00015 & 0.00022 & 0.00003\\ [1ex]
 &  & 0.06129 & 0.16556 & 0.00188 & 0.01449 & 0.00001\\ [1ex]
 &  &  & 3.29493 & 0.00070 & 0.18757 & -0.00614\\ [1ex]
 &  &  &  & 0.00015 & 0.00022 & 0.00003\\ [1ex]
 &  &  &  &  & 0.20443 & 0.10080\\ [1ex]
 &  &  &  &  &  & 4.04413\\ [1ex]
\hline
\end{tabular}
    \caption{Our results for $z$-coefficients in the BCL parameterization Equation~\eqref{eq:BCLappendix}. The first row gives mean values, and the rest of the table gives the covarance matrix associated with these parameters. \label{tab:coeffs}}
\end{center}
\end{table}

\section{The ratio method for obtaining the form factors}
\label{appendix:ratio}

Here we show results from the {\textit{ratio}} approach to determining the 
form factors. In this approach we fit the ratio of the form factors to 
decay constant of the pseudoscalar $H_c$ meson~\cite{McLean:2019sds}:
\begin{align}
  R^s_{0,+}(q^2) \equiv {f^s_{0,+}(q^2)\over f_{H_c}\sqrt{M_{H_c}}}.
  \label{eq:ratio2}
\end{align}
Two-point correlation functions for the $H_c$ meson are calculated 
along with the other two- and three-point correlation functions that 
we need (see Section~\ref{sec:correlationfunctions}) and 
included in the simultaneous fits to these correlation functions 
described in Section~\ref{sec:analysiscorrs}. 
This enables us to determine the heavy-charm meson decay constant 
(and in fact the combination $f_{H_c}\sqrt{M_{H_c}}$ needed 
for Equation~(\ref{eq:ratio2})) from the amplitude for the ground-state in the two-point 
correlation functions as given in Equation~(\ref{eq:decayconstantfit}). 
Our results for $R^s_{0,+}(q^2)$ are given in Table~\ref{tab:kinetic}.  

We fit $R^s_{0,+}(q^2)$ using an identical fit function to that of the direct approach, given in Equations \eqref{eq:BCL} and \eqref{eq:mhfitfun}.

Discretisation effects change in the ratio given 
in Equation~(\ref{eq:ratio2}), compared to those from the form factors 
themselves, changing the continuum extrapolation.  
The dependence on heavy quark mass of the ratio is also very 
different. 
The value of the ratio at the physical point (where $m_h=m_b$ and $a=0$) 
can then multiplied by $f_{B_c}\sqrt{M_{B_c}}$ to obtain the form factors. 
We find $f_{B_c}$ via a separate calculation (detailed in 
Appendix A of~\cite{McLean:2019sds}) and take the experimental 
value for the $B_c$ meson mass, $M_{B_c} = 6.2756(11)$GeV~\cite{Tanabashi:2018oca}. 
This approach has the disadvantage 
of introducing larger uncertainties from scale-setting since $R_{0,+}^s(q^2)$ 
are dimensionful quantities (as opposed to $f^s_{0,+}(q^2)$ which are dimensionless).
Hence we do not use it for our final value. It provides a useful test of 
our uncertainties, however. 

\begin{figure}[ht]
  \hspace{-10pt}
  \includegraphics[width=0.53\textwidth]{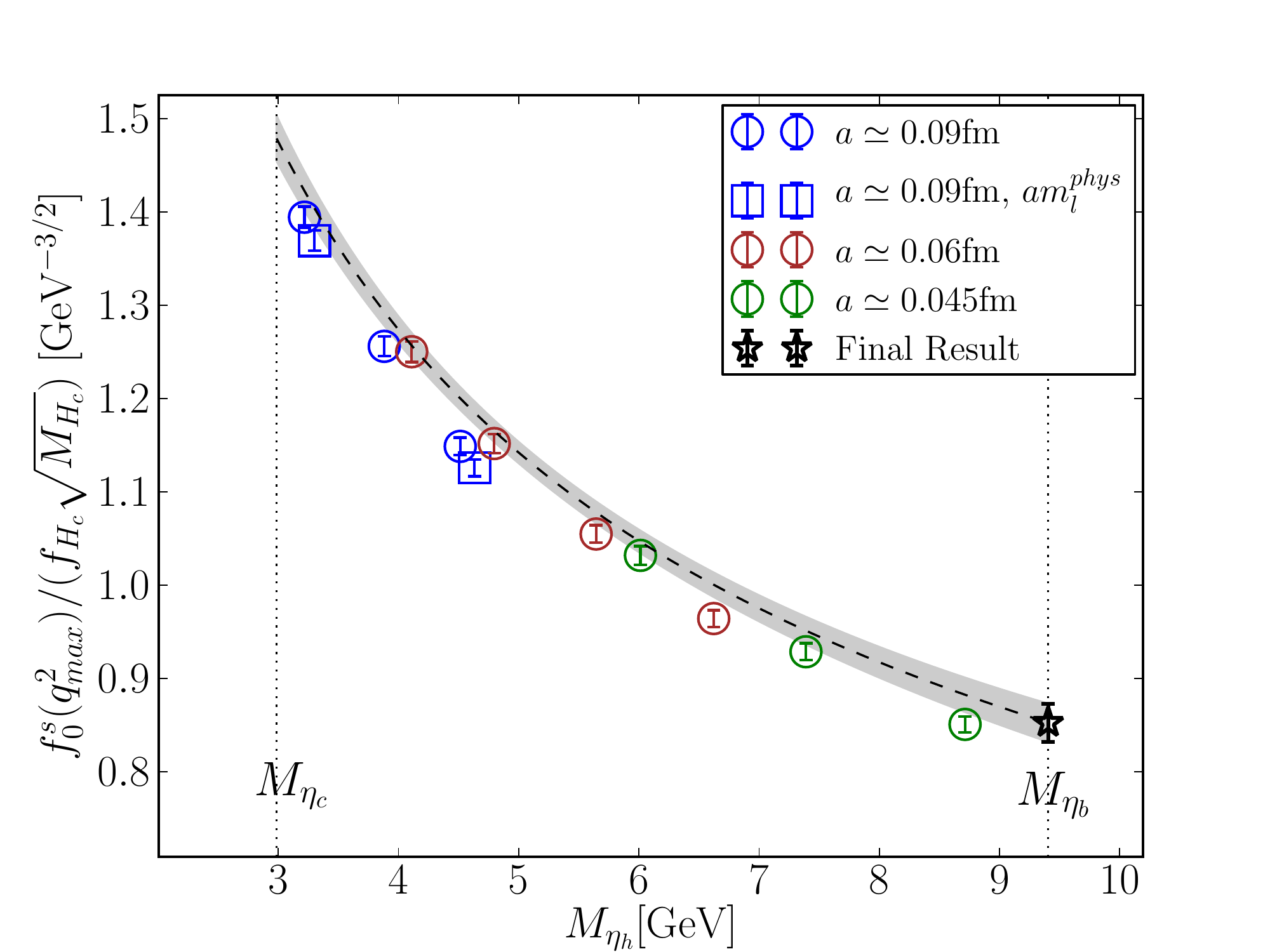}
  \caption{ $R_0^s(q^2_{\text{max}}) = f_0^s(q^2_{\text{max}})/(f_{H_c}\sqrt{M_{H_c}})$ against $M_{\eta_h}$ (a proxy for the heavy quark mass). The grey band shows the result of the extrapolation at $a=0$ and physical $l$,$s$ and $c$ masses. Sets listed in the legend follow the order of sets in table \ref{tab:ensembles}. \label{fig:ratioq2max}}
\end{figure}

\begin{figure*}[ht]
  \begin{center}
  \includegraphics[width=1.00\textwidth]{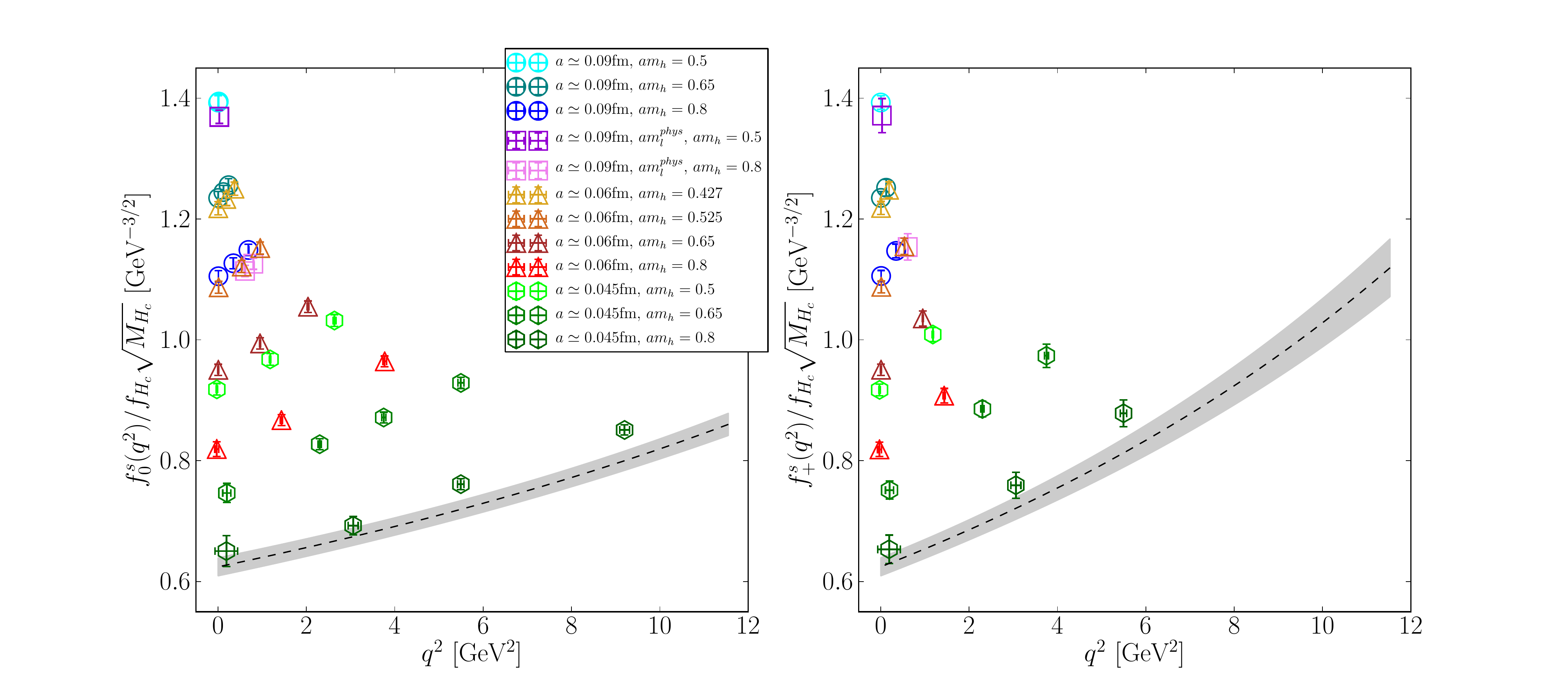}
  \caption{ $R_{0,+}^s(q^2) = f_{0,+}^s(q^2)/(f_{H_c}\sqrt{M_{H_c}})$ against 
$q^2$. The grey band shows the result of our fit at $a=0$ and physical 
$l$,$s$, $c$ {\it{and}} $b$ masses. Sets listed in the legend follow 
the order of sets in table \ref{tab:ensembles}. \label{fig:ratiodata}}
    \end{center}
\end{figure*}

\subsubsection{Zero Recoil}

\begin{table}
  \begin{center}
    \begin{tabular}{c c}
      \hline
      Source & \% Fractional Error \\ [0.5ex]
      \hline
      Statistics & 1.10  \\ [1ex]
      Scale Setting & 1.30  \\ [1ex]
      $m_h \to m_b$ and $a\to 0$ & 1.44 \\ [1ex]
      Quark mistuning & 0.87 \\ [1ex]
      \hline
      Total & 2.39 \\ [1ex]
      \hline
    \end{tabular}
  \end{center}
  \caption{Error budget for $R_0^s(q^2_{\text{max}})$.\label{ratioq2max_budget}}
\end{table}

We first show results from a fit at the zero recoil 
point, $R^s_0(q^2_{\text{max}})$, as a function of 
heavy quark mass and lattice spacing. 
To do this we use the same fit form as for our fits to $f^s_{0,+}(q^2_{\text{max}})$, 
given in Equation~(\ref{eq:mhfitqmax}), with the same priors.  
We find, with physical parameters for all quark 
masses and $a=0$
\begin{align}
\label{eq:r0phys}
  R_0^s(q^2_{\text{max}}) = {f_0^s(q^2_{\text{max}})\over f_{B_c}\sqrt{M_{B_c}}} = 0.853(20) \text{GeV}^{-3/2}\,.
\end{align}
The fit has $\chi^2/N_{\text{dof}} = 0.63$ for $N_{\text{dof}}=16$.
The lattice QCD results and fit are shown in a plot against $M_{\eta_h}$ 
in Figure~\ref{fig:ratioq2max}. As can be seen from this plot, the lattice results 
have stronger dependence on the heavy quark mass and somewhat 
less on the lattice spacing compared to that seen 
in Figure~\ref{fig:directq2max}. 
The error budget for our physical value in Equation~(\ref{eq:r0phys}) 
is given in Table~\ref{ratioq2max_budget}. Notice that, compared to 
Table~\ref{directq2max_budget} it now has a significant scale-setting 
uncertainty.  

\subsubsection{Full $q^2$ range}

In Figure~\ref{fig:ratiodata}, we show the complete set of lattice results along 
with the results of the full fit given by the fit form in Equation~(\ref{eq:mhfitfun}).  
As for Figure~\ref{fig:directdata} we see that as the lattice spacing 
decreases, the range of heavy quark masses increases and the 
$q^2$ range, $0 < (M_{H_s}-M_{D_s})^2$ expands. The goodness of fit here was $\chi^2/N_{\text{dof}} = 0.57$, $N_{\text{dof}}=58$.

To obtain the form factors, the resulting functions $R_{0,+}^s(q^2)$ 
were multiplied by $\sqrt{M_{B_c}}$ (using the experimental value) 
and $f_{B_c}$ from our determination detailed in appendix A of~\cite{McLean:2019sds}. 
The resulting form factors are shown in Figure~\ref{fig:ratiovsdirect} in 
a comparison to those found by our direct method of Section~\ref{sec:directfullq2}. 

\section{Tests of Unitarity Bounds}
\label{appendix:unitarity}

Unitarity and crossing symmetry imposes bounds on the coefficients 
of the BCL parameterization of 
$f_{0,+}(q^2)$, $\{a_n\}$~\cite{Okubo:1971my,Okubo:1971jf}. 
As another consistency test, we show here that the coefficients found in our fit satisfy these bounds.

To obtain bounds on the BCL coefficients~\cite{Bourrely:2008za}, one must relate them to those 
of a different parameterisation, that of Boyd, Grinstein and 
Lebed (BGL) \cite{Boyd:1995sq}:
\begin{align}
  f_{0,+}^s(q^2) = {1\over B(z)\phi(z)} \sum^N_{n\geq 0} b_n z^n.
\end{align}
$B(z)$ is known as the Blashke factor:
\begin{align}
  B(z) = { z - z_* \over 1 - z z_* }\,,
\end{align}
where $z_* = z(M^2_{B_c^0})$ for $f_0^s$, or $z(M^2_{B_c^*})$ for $f_+^s\,$. $\phi(z)$ is the outer function;
\begin{align}
  \phi(z) &= M_{B_s}^{2-s} 2^{2+p} \sqrt{\kappa n_f}
  \left[ {M_{D_s}\over M_{B_s}} (1+z)\right]^{s-3/2} \nonumber \\
  \times &\left[ (1-z)\left(1+{M_{D_s}\over M_{B_s}}\right)+2\sqrt{M_{D_s}\over M_{B_s}}(1+z)\right]^{-s-p}.
\end{align}
In the $f^s_0$ case, $\kappa=12\pi M^2_{B_s}\chi_A$, $p=1$,\,$s=3$. 
In the $f^s_+$ case, $\kappa=6\pi M^2_{B_s} \chi_V$, $p=3$,\,$s=2$. 
The quantities $\chi_{V,A}$ are the once-subtracted dispersion relations 
at $q^2=0$ for vector and axial $b\to c$ currents respectively, 
computed in \cite{Boyd:1995sq} to 
be $\chi_V = 5.7\times 10^{-3}/m_b^2$ and $\chi_A = 9.6\times 10^{-3}/m_b^2$.

The BGL coefficients, $\{ b_n \}$, obey the unitarity constaint
\begin{align}
  \sum_{m=0}^{\infty} |b_m|^2 \leq 1
  \label{eq:unitarity1}
\end{align}
by construction of the parameterisation. To see how this applies to the 
BCL coefficients $\{ a_n \}$, one must relate them to $\{ b_m \}$ by 
equating the two parameterisations to find~\cite{Bourrely:2008za}
\begin{align}
  \label{eq:BCL-BGL}
  \sum^M_{m=0} b_m z^m = \psi(z) \sum_{n=0}^N a_n z^n,
\end{align}
where $\psi(z)$ is given by
\begin{align}
  \psi(z) = { M_{\text{pole}}^2 \over 4(t_+ - t_0) } \phi(z) {(1-z)^2(1-z_*)^2\over (1-zz_*)^2 },
\end{align}
and $M_{\text{pole}}=M_{B_c^0}$ in the $f^s_0$ case 
and $M_{B_c^*}$ in the $f^s_+$ case. Expanding $\psi(z)$ around $z=0$, 
comparing coefficients of $z$ in Equation~\eqref{eq:BCL-BGL}, and imposing the 
constraint of Equation~\eqref{eq:unitarity1}, we arrive at a constraint 
for the BCL coefficients
\begin{align}
    \label{eq:unitarity2}
    &\mathcal{B} \equiv \sum_{j,k=0}^{L,L} B_{jk} a_j a_k \leq 1\,, \\
    &B_{jk} = \sum_{n=0}^{\infty} \eta_n \eta_{n+|j-k|}\,,
        \label{eq:Bmatrix}
\end{align}
where $\{\eta_n\}$ are the Taylor coefficients of $\psi(z)$.

\begin{table}
  \begin{center}
    \begin{tabular}{c c c c}
      \hline
      & $B_{00}$ & $B_{01}$ & $B_{02}$ \\ [0.5ex]
      \hline
      $f^s_0$ & 0.00186 & -0.000258 & -0.000703 \\ [0.5ex]
      $f^s_+$ & 0.00179 & -0.000367 & 0.00108 \\ [0.5ex]
      \hline
    \end{tabular}
  \end{center}
  \caption{Numerical values for $B_{jk}$ appearing in the unitarity bound for BCL coefficients, defined in \eqref{eq:Bmatrix}, for the $f^s_0$ and $f^s_+$ cases. The rest of the elements can be obtained from these using the properties $B_{j(j+k)}=B_{0k}$ and $B_{jk}=B_{kj}$. \label{tab:Bmatrix}}
\end{table}

$\psi(z)$ is bounded on the closed disk $|z|<1$, so its Taylor coefficients 
are rapidly decreasing. We computed values for $B_{jk}$ by truncating 
the sum in its definition (Equation~\eqref{eq:Bmatrix}) at 100. 
These values are given in Table~\ref{tab:Bmatrix}. 
With these $B_{jk}$ values, and the $a_n$ coefficients 
from our fit (via the direct method), we find
\begin{align}
  \nonumber
  \mathcal{B}_0 = 0.00073(89)\,,
  \\ \mathcal{B}_+ = 0.0210(54)\,.
  \nonumber
\end{align}
These comfortably satisfy the unitarity bound. Additionally, 
as discussed in~\cite{Becher:2005bg}, the leading contributions 
to $\mathcal{B}_{0,+}$ are of order $(\Lambda_{\text{QCD}}/m_b)^3 \simeq 10^{-3}$ 
in the heavy quark expansion. This expectation is 
approximately fulfilled by our result.

\bibliography{paper}

\begin{thebibliography}{76}%
\makeatletter
\providecommand \@ifxundefined [1]{%
 \@ifx{#1\undefined}
}%
\providecommand \@ifnum [1]{%
 \ifnum #1\expandafter \@firstoftwo
 \else \expandafter \@secondoftwo
 \fi
}%
\providecommand \@ifx [1]{%
 \ifx #1\expandafter \@firstoftwo
 \else \expandafter \@secondoftwo
 \fi
}%
\providecommand \natexlab [1]{#1}%
\providecommand \enquote  [1]{``#1''}%
\providecommand \bibnamefont  [1]{#1}%
\providecommand \bibfnamefont [1]{#1}%
\providecommand \citenamefont [1]{#1}%
\providecommand \href@noop [0]{\@secondoftwo}%
\providecommand \href [0]{\begingroup \@sanitize@url \@href}%
\providecommand \@href[1]{\@@startlink{#1}\@@href}%
\providecommand \@@href[1]{\endgroup#1\@@endlink}%
\providecommand \@sanitize@url [0]{\catcode `\\12\catcode `\$12\catcode
  `\&12\catcode `\#12\catcode `\^12\catcode `\_12\catcode `\%12\relax}%
\providecommand \@@startlink[1]{}%
\providecommand \@@endlink[0]{}%
\providecommand \url  [0]{\begingroup\@sanitize@url \@url }%
\providecommand \@url [1]{\endgroup\@href {#1}{\urlprefix }}%
\providecommand \urlprefix  [0]{URL }%
\providecommand \Eprint [0]{\href }%
\providecommand \doibase [0]{http://dx.doi.org/}%
\providecommand \selectlanguage [0]{\@gobble}%
\providecommand \bibinfo  [0]{\@secondoftwo}%
\providecommand \bibfield  [0]{\@secondoftwo}%
\providecommand \translation [1]{[#1]}%
\providecommand \BibitemOpen [0]{}%
\providecommand \bibitemStop [0]{}%
\providecommand \bibitemNoStop [0]{.\EOS\space}%
\providecommand \EOS [0]{\spacefactor3000\relax}%
\providecommand \BibitemShut  [1]{\csname bibitem#1\endcsname}%
\let\auto@bib@innerbib\@empty
\bibitem [{\citenamefont {Wei}\ \emph {et~al.}(2009)\citenamefont {Wei} \emph
  {et~al.}}]{Wei:2009zv}%
  \BibitemOpen
  \bibfield  {author} {\bibinfo {author} {\bibfnamefont {J.~T.}\ \bibnamefont
  {Wei}} \emph {et~al.} (\bibinfo {collaboration} {Belle}),\ }\href {\doibase
  10.1103/PhysRevLett.103.171801} {\bibfield  {journal} {\bibinfo  {journal}
  {Phys. Rev. Lett.}\ }\textbf {\bibinfo {volume} {103}},\ \bibinfo {pages}
  {171801} (\bibinfo {year} {2009})},\ \Eprint {http://arxiv.org/abs/0904.0770}
  {arXiv:0904.0770 [hep-ex]} \BibitemShut {NoStop}%
\bibitem [{\citenamefont {Lees}\ \emph
  {et~al.}(2012{\natexlab{a}})\citenamefont {Lees} \emph
  {et~al.}}]{Lees:2012xj}%
  \BibitemOpen
  \bibfield  {author} {\bibinfo {author} {\bibfnamefont {J.~P.}\ \bibnamefont
  {Lees}} \emph {et~al.} (\bibinfo {collaboration} {BaBar}),\ }\href {\doibase
  10.1103/PhysRevLett.109.101802} {\bibfield  {journal} {\bibinfo  {journal}
  {Phys. Rev. Lett.}\ }\textbf {\bibinfo {volume} {109}},\ \bibinfo {pages}
  {101802} (\bibinfo {year} {2012}{\natexlab{a}})},\ \Eprint
  {http://arxiv.org/abs/1205.5442} {arXiv:1205.5442 [hep-ex]} \BibitemShut
  {NoStop}%
\bibitem [{\citenamefont {Lees}\ \emph
  {et~al.}(2012{\natexlab{b}})\citenamefont {Lees} \emph
  {et~al.}}]{Lees:2012tva}%
  \BibitemOpen
  \bibfield  {author} {\bibinfo {author} {\bibfnamefont {J.~P.}\ \bibnamefont
  {Lees}} \emph {et~al.} (\bibinfo {collaboration} {BaBar}),\ }\href {\doibase
  10.1103/PhysRevD.86.032012} {\bibfield  {journal} {\bibinfo  {journal} {Phys.
  Rev.}\ }\textbf {\bibinfo {volume} {D86}},\ \bibinfo {pages} {032012}
  (\bibinfo {year} {2012}{\natexlab{b}})},\ \Eprint
  {http://arxiv.org/abs/1204.3933} {arXiv:1204.3933 [hep-ex]} \BibitemShut
  {NoStop}%
\bibitem [{\citenamefont {Lees}\ \emph {et~al.}(2013)\citenamefont {Lees} \emph
  {et~al.}}]{Lees:2013uzd}%
  \BibitemOpen
  \bibfield  {author} {\bibinfo {author} {\bibfnamefont {J.~P.}\ \bibnamefont
  {Lees}} \emph {et~al.} (\bibinfo {collaboration} {BaBar}),\ }\href {\doibase
  10.1103/PhysRevD.88.072012} {\bibfield  {journal} {\bibinfo  {journal} {Phys.
  Rev.}\ }\textbf {\bibinfo {volume} {D88}},\ \bibinfo {pages} {072012}
  (\bibinfo {year} {2013})},\ \Eprint {http://arxiv.org/abs/1303.0571}
  {arXiv:1303.0571 [hep-ex]} \BibitemShut {NoStop}%
\bibitem [{\citenamefont {Aaij}\ \emph
  {et~al.}(2014{\natexlab{a}})\citenamefont {Aaij} \emph
  {et~al.}}]{Aaij:2014pli}%
  \BibitemOpen
  \bibfield  {author} {\bibinfo {author} {\bibfnamefont {R.}~\bibnamefont
  {Aaij}} \emph {et~al.} (\bibinfo {collaboration} {LHCb}),\ }\href {\doibase
  10.1007/JHEP06(2014)133} {\bibfield  {journal} {\bibinfo  {journal} {JHEP}\
  }\textbf {\bibinfo {volume} {06}},\ \bibinfo {pages} {133} (\bibinfo {year}
  {2014}{\natexlab{a}})},\ \Eprint {http://arxiv.org/abs/1403.8044}
  {arXiv:1403.8044 [hep-ex]} \BibitemShut {NoStop}%
\bibitem [{\citenamefont {Aaij}\ \emph
  {et~al.}(2014{\natexlab{b}})\citenamefont {Aaij} \emph
  {et~al.}}]{Aaij:2014ora}%
  \BibitemOpen
  \bibfield  {author} {\bibinfo {author} {\bibfnamefont {R.}~\bibnamefont
  {Aaij}} \emph {et~al.} (\bibinfo {collaboration} {LHCb}),\ }\href {\doibase
  10.1103/PhysRevLett.113.151601} {\bibfield  {journal} {\bibinfo  {journal}
  {Phys. Rev. Lett.}\ }\textbf {\bibinfo {volume} {113}},\ \bibinfo {pages}
  {151601} (\bibinfo {year} {2014}{\natexlab{b}})},\ \Eprint
  {http://arxiv.org/abs/1406.6482} {arXiv:1406.6482 [hep-ex]} \BibitemShut
  {NoStop}%
\bibitem [{\citenamefont {Huschle}\ \emph {et~al.}(2015)\citenamefont {Huschle}
  \emph {et~al.}}]{Huschle:2015rga}%
  \BibitemOpen
  \bibfield  {author} {\bibinfo {author} {\bibfnamefont {M.}~\bibnamefont
  {Huschle}} \emph {et~al.} (\bibinfo {collaboration} {Belle}),\ }\href
  {\doibase 10.1103/PhysRevD.92.072014} {\bibfield  {journal} {\bibinfo
  {journal} {Phys. Rev.}\ }\textbf {\bibinfo {volume} {D92}},\ \bibinfo {pages}
  {072014} (\bibinfo {year} {2015})},\ \Eprint
  {http://arxiv.org/abs/1507.03233} {arXiv:1507.03233 [hep-ex]} \BibitemShut
  {NoStop}%
\bibitem [{\citenamefont {Aaij}\ \emph
  {et~al.}(2016{\natexlab{a}})\citenamefont {Aaij} \emph
  {et~al.}}]{Aaij:2015oid}%
  \BibitemOpen
  \bibfield  {author} {\bibinfo {author} {\bibfnamefont {R.}~\bibnamefont
  {Aaij}} \emph {et~al.} (\bibinfo {collaboration} {LHCb}),\ }\href {\doibase
  10.1007/JHEP02(2016)104} {\bibfield  {journal} {\bibinfo  {journal} {JHEP}\
  }\textbf {\bibinfo {volume} {02}},\ \bibinfo {pages} {104} (\bibinfo {year}
  {2016}{\natexlab{a}})},\ \Eprint {http://arxiv.org/abs/1512.04442}
  {arXiv:1512.04442 [hep-ex]} \BibitemShut {NoStop}%
\bibitem [{\citenamefont {Aaij}\ \emph
  {et~al.}(2015{\natexlab{a}})\citenamefont {Aaij} \emph
  {et~al.}}]{Aaij:2015yra}%
  \BibitemOpen
  \bibfield  {author} {\bibinfo {author} {\bibfnamefont {R.}~\bibnamefont
  {Aaij}} \emph {et~al.} (\bibinfo {collaboration} {LHCb}),\ }\href {\doibase
  10.1103/PhysRevLett.115.159901, 10.1103/PhysRevLett.115.111803} {\bibfield
  {journal} {\bibinfo  {journal} {Phys. Rev. Lett.}\ }\textbf {\bibinfo
  {volume} {115}},\ \bibinfo {pages} {111803} (\bibinfo {year}
  {2015}{\natexlab{a}})},\ \bibinfo {note} {[Erratum: Phys. Rev.
  Lett.115,159901(2015)]},\ \Eprint {http://arxiv.org/abs/1506.08614}
  {arXiv:1506.08614 [hep-ex]} \BibitemShut {NoStop}%
\bibitem [{\citenamefont {Aaij}\ \emph
  {et~al.}(2015{\natexlab{b}})\citenamefont {Aaij} \emph
  {et~al.}}]{Aaij:2015xza}%
  \BibitemOpen
  \bibfield  {author} {\bibinfo {author} {\bibfnamefont {R.}~\bibnamefont
  {Aaij}} \emph {et~al.} (\bibinfo {collaboration} {LHCb}),\ }\href {\doibase
  10.1007/JHEP09(2018)145, 10.1007/JHEP06(2015)115} {\bibfield  {journal}
  {\bibinfo  {journal} {JHEP}\ }\textbf {\bibinfo {volume} {06}},\ \bibinfo
  {pages} {115} (\bibinfo {year} {2015}{\natexlab{b}})},\ \bibinfo {note}
  {[Erratum: JHEP09,145(2018)]},\ \Eprint {http://arxiv.org/abs/1503.07138}
  {arXiv:1503.07138 [hep-ex]} \BibitemShut {NoStop}%
\bibitem [{\citenamefont {Aaij}\ \emph
  {et~al.}(2016{\natexlab{b}})\citenamefont {Aaij} \emph
  {et~al.}}]{Aaij:2016flj}%
  \BibitemOpen
  \bibfield  {author} {\bibinfo {author} {\bibfnamefont {R.}~\bibnamefont
  {Aaij}} \emph {et~al.} (\bibinfo {collaboration} {LHCb}),\ }\href {\doibase
  10.1007/JHEP11(2016)047, 10.1007/JHEP04(2017)142} {\bibfield  {journal}
  {\bibinfo  {journal} {JHEP}\ }\textbf {\bibinfo {volume} {11}},\ \bibinfo
  {pages} {047} (\bibinfo {year} {2016}{\natexlab{b}})},\ \bibinfo {note}
  {[Erratum: JHEP04,142(2017)]},\ \Eprint {http://arxiv.org/abs/1606.04731}
  {arXiv:1606.04731 [hep-ex]} \BibitemShut {NoStop}%
\bibitem [{\citenamefont {Wehle}\ \emph {et~al.}(2017)\citenamefont {Wehle}
  \emph {et~al.}}]{Wehle:2016yoi}%
  \BibitemOpen
  \bibfield  {author} {\bibinfo {author} {\bibfnamefont {S.}~\bibnamefont
  {Wehle}} \emph {et~al.} (\bibinfo {collaboration} {Belle}),\ }\href {\doibase
  10.1103/PhysRevLett.118.111801} {\bibfield  {journal} {\bibinfo  {journal}
  {Phys. Rev. Lett.}\ }\textbf {\bibinfo {volume} {118}},\ \bibinfo {pages}
  {111801} (\bibinfo {year} {2017})},\ \Eprint
  {http://arxiv.org/abs/1612.05014} {arXiv:1612.05014 [hep-ex]} \BibitemShut
  {NoStop}%
\bibitem [{\citenamefont {Sato}\ \emph {et~al.}(2016)\citenamefont {Sato} \emph
  {et~al.}}]{Sato:2016svk}%
  \BibitemOpen
  \bibfield  {author} {\bibinfo {author} {\bibfnamefont {Y.}~\bibnamefont
  {Sato}} \emph {et~al.} (\bibinfo {collaboration} {Belle}),\ }\href {\doibase
  10.1103/PhysRevD.94.072007} {\bibfield  {journal} {\bibinfo  {journal} {Phys.
  Rev.}\ }\textbf {\bibinfo {volume} {D94}},\ \bibinfo {pages} {072007}
  (\bibinfo {year} {2016})},\ \Eprint {http://arxiv.org/abs/1607.07923}
  {arXiv:1607.07923 [hep-ex]} \BibitemShut {NoStop}%
\bibitem [{\citenamefont {Hirose}\ \emph {et~al.}(2018)\citenamefont {Hirose}
  \emph {et~al.}}]{Hirose:2017dxl}%
  \BibitemOpen
  \bibfield  {author} {\bibinfo {author} {\bibfnamefont {S.}~\bibnamefont
  {Hirose}} \emph {et~al.} (\bibinfo {collaboration} {Belle}),\ }\href
  {\doibase 10.1103/PhysRevD.97.012004} {\bibfield  {journal} {\bibinfo
  {journal} {Phys. Rev.}\ }\textbf {\bibinfo {volume} {D97}},\ \bibinfo {pages}
  {012004} (\bibinfo {year} {2018})},\ \Eprint
  {http://arxiv.org/abs/1709.00129} {arXiv:1709.00129 [hep-ex]} \BibitemShut
  {NoStop}%
\bibitem [{\citenamefont {Aaij}\ \emph
  {et~al.}(2018{\natexlab{a}})\citenamefont {Aaij} \emph
  {et~al.}}]{Aaij:2017deq}%
  \BibitemOpen
  \bibfield  {author} {\bibinfo {author} {\bibfnamefont {R.}~\bibnamefont
  {Aaij}} \emph {et~al.} (\bibinfo {collaboration} {LHCb}),\ }\href {\doibase
  10.1103/PhysRevD.97.072013} {\bibfield  {journal} {\bibinfo  {journal} {Phys.
  Rev.}\ }\textbf {\bibinfo {volume} {D97}},\ \bibinfo {pages} {072013}
  (\bibinfo {year} {2018}{\natexlab{a}})},\ \Eprint
  {http://arxiv.org/abs/1711.02505} {arXiv:1711.02505 [hep-ex]} \BibitemShut
  {NoStop}%
\bibitem [{\citenamefont {Aaij}\ \emph
  {et~al.}(2018{\natexlab{b}})\citenamefont {Aaij} \emph
  {et~al.}}]{Aaij:2017uff}%
  \BibitemOpen
  \bibfield  {author} {\bibinfo {author} {\bibfnamefont {R.}~\bibnamefont
  {Aaij}} \emph {et~al.} (\bibinfo {collaboration} {LHCb}),\ }\href {\doibase
  10.1103/PhysRevLett.120.171802} {\bibfield  {journal} {\bibinfo  {journal}
  {Phys. Rev. Lett.}\ }\textbf {\bibinfo {volume} {120}},\ \bibinfo {pages}
  {171802} (\bibinfo {year} {2018}{\natexlab{b}})},\ \Eprint
  {http://arxiv.org/abs/1708.08856} {arXiv:1708.08856 [hep-ex]} \BibitemShut
  {NoStop}%
\bibitem [{\citenamefont {Aaij}\ \emph {et~al.}(2017)\citenamefont {Aaij} \emph
  {et~al.}}]{Aaij:2017vbb}%
  \BibitemOpen
  \bibfield  {author} {\bibinfo {author} {\bibfnamefont {R.}~\bibnamefont
  {Aaij}} \emph {et~al.} (\bibinfo {collaboration} {LHCb}),\ }\href {\doibase
  10.1007/JHEP08(2017)055} {\bibfield  {journal} {\bibinfo  {journal} {JHEP}\
  }\textbf {\bibinfo {volume} {08}},\ \bibinfo {pages} {055} (\bibinfo {year}
  {2017})},\ \Eprint {http://arxiv.org/abs/1705.05802} {arXiv:1705.05802
  [hep-ex]} \BibitemShut {NoStop}%
\bibitem [{\citenamefont {Sirunyan}\ \emph {et~al.}(2018)\citenamefont
  {Sirunyan} \emph {et~al.}}]{Sirunyan:2017dhj}%
  \BibitemOpen
  \bibfield  {author} {\bibinfo {author} {\bibfnamefont {A.~M.}\ \bibnamefont
  {Sirunyan}} \emph {et~al.} (\bibinfo {collaboration} {CMS}),\ }\href
  {\doibase 10.1016/j.physletb.2018.04.030} {\bibfield  {journal} {\bibinfo
  {journal} {Phys. Lett.}\ }\textbf {\bibinfo {volume} {B781}},\ \bibinfo
  {pages} {517} (\bibinfo {year} {2018})},\ \Eprint
  {http://arxiv.org/abs/1710.02846} {arXiv:1710.02846 [hep-ex]} \BibitemShut
  {NoStop}%
\bibitem [{\citenamefont {Aaboud}\ \emph {et~al.}(2018)\citenamefont {Aaboud}
  \emph {et~al.}}]{Aaboud:2018krd}%
  \BibitemOpen
  \bibfield  {author} {\bibinfo {author} {\bibfnamefont {M.}~\bibnamefont
  {Aaboud}} \emph {et~al.} (\bibinfo {collaboration} {ATLAS}),\ }\href
  {\doibase 10.1007/JHEP10(2018)047} {\bibfield  {journal} {\bibinfo  {journal}
  {JHEP}\ }\textbf {\bibinfo {volume} {10}},\ \bibinfo {pages} {047} (\bibinfo
  {year} {2018})},\ \Eprint {http://arxiv.org/abs/1805.04000} {arXiv:1805.04000
  [hep-ex]} \BibitemShut {NoStop}%
\bibitem [{\citenamefont {Aubert}\ \emph {et~al.}(2009)\citenamefont {Aubert}
  \emph {et~al.}}]{Aubert:2008yv}%
  \BibitemOpen
  \bibfield  {author} {\bibinfo {author} {\bibfnamefont {B.}~\bibnamefont
  {Aubert}} \emph {et~al.} (\bibinfo {collaboration} {BaBar}),\ }\href
  {\doibase 10.1103/PhysRevD.79.012002} {\bibfield  {journal} {\bibinfo
  {journal} {Phys. Rev.}\ }\textbf {\bibinfo {volume} {D79}},\ \bibinfo {pages}
  {012002} (\bibinfo {year} {2009})},\ \Eprint {http://arxiv.org/abs/0809.0828}
  {arXiv:0809.0828 [hep-ex]} \BibitemShut {NoStop}%
\bibitem [{\citenamefont {Aubert}\ \emph {et~al.}(2010)\citenamefont {Aubert}
  \emph {et~al.}}]{Aubert:2009ac}%
  \BibitemOpen
  \bibfield  {author} {\bibinfo {author} {\bibfnamefont {B.}~\bibnamefont
  {Aubert}} \emph {et~al.} (\bibinfo {collaboration} {BaBar}),\ }\href
  {\doibase 10.1103/PhysRevLett.104.011802} {\bibfield  {journal} {\bibinfo
  {journal} {Phys. Rev. Lett.}\ }\textbf {\bibinfo {volume} {104}},\ \bibinfo
  {pages} {011802} (\bibinfo {year} {2010})},\ \Eprint
  {http://arxiv.org/abs/0904.4063} {arXiv:0904.4063 [hep-ex]} \BibitemShut
  {NoStop}%
\bibitem [{\citenamefont {Bailey}\ \emph {et~al.}(2015)\citenamefont {Bailey}
  \emph {et~al.}}]{Lattice:2015rga}%
  \BibitemOpen
  \bibfield  {author} {\bibinfo {author} {\bibfnamefont {J.~A.}\ \bibnamefont
  {Bailey}} \emph {et~al.} (\bibinfo {collaboration} {MILC}),\ }\href {\doibase
  10.1103/PhysRevD.92.034506} {\bibfield  {journal} {\bibinfo  {journal} {Phys.
  Rev.}\ }\textbf {\bibinfo {volume} {D92}},\ \bibinfo {pages} {034506}
  (\bibinfo {year} {2015})},\ \Eprint {http://arxiv.org/abs/1503.07237}
  {arXiv:1503.07237 [hep-lat]} \BibitemShut {NoStop}%
\bibitem [{\citenamefont {Na}\ \emph {et~al.}(2015)\citenamefont {Na},
  \citenamefont {Bouchard}, \citenamefont {Lepage}, \citenamefont {Monahan},\
  and\ \citenamefont {Shigemitsu}}]{Na:2015kha}%
  \BibitemOpen
  \bibfield  {author} {\bibinfo {author} {\bibfnamefont {H.}~\bibnamefont
  {Na}}, \bibinfo {author} {\bibfnamefont {C.~M.}\ \bibnamefont {Bouchard}},
  \bibinfo {author} {\bibfnamefont {G.~P.}\ \bibnamefont {Lepage}}, \bibinfo
  {author} {\bibfnamefont {C.}~\bibnamefont {Monahan}}, \ and\ \bibinfo
  {author} {\bibfnamefont {J.}~\bibnamefont {Shigemitsu}} (\bibinfo
  {collaboration} {HPQCD}),\ }\href {\doibase 10.1103/PhysRevD.93.119906,
  10.1103/PhysRevD.92.054510} {\bibfield  {journal} {\bibinfo  {journal} {Phys.
  Rev.}\ }\textbf {\bibinfo {volume} {D92}},\ \bibinfo {pages} {054510}
  (\bibinfo {year} {2015})},\ \bibinfo {note} {[Erratum: Phys.
  Rev.D93,no.11,119906(2016)]},\ \Eprint {http://arxiv.org/abs/1505.03925}
  {arXiv:1505.03925 [hep-lat]} \BibitemShut {NoStop}%
\bibitem [{\citenamefont {Glattauer}\ \emph {et~al.}(2016)\citenamefont
  {Glattauer} \emph {et~al.}}]{Glattauer:2015teq}%
  \BibitemOpen
  \bibfield  {author} {\bibinfo {author} {\bibfnamefont {R.}~\bibnamefont
  {Glattauer}} \emph {et~al.} (\bibinfo {collaboration} {Belle}),\ }\href
  {\doibase 10.1103/PhysRevD.93.032006} {\bibfield  {journal} {\bibinfo
  {journal} {Phys. Rev.}\ }\textbf {\bibinfo {volume} {D93}},\ \bibinfo {pages}
  {032006} (\bibinfo {year} {2016})},\ \Eprint
  {http://arxiv.org/abs/1510.03657} {arXiv:1510.03657 [hep-ex]} \BibitemShut
  {NoStop}%
\bibitem [{\citenamefont {Bernlochner}\ \emph {et~al.}(2017)\citenamefont
  {Bernlochner}, \citenamefont {Ligeti}, \citenamefont {Papucci},\ and\
  \citenamefont {Robinson}}]{Bernlochner:2017jka}%
  \BibitemOpen
  \bibfield  {author} {\bibinfo {author} {\bibfnamefont {F.~U.}\ \bibnamefont
  {Bernlochner}}, \bibinfo {author} {\bibfnamefont {Z.}~\bibnamefont {Ligeti}},
  \bibinfo {author} {\bibfnamefont {M.}~\bibnamefont {Papucci}}, \ and\
  \bibinfo {author} {\bibfnamefont {D.~J.}\ \bibnamefont {Robinson}},\ }\href
  {\doibase 10.1103/PhysRevD.95.115008, 10.1103/PhysRevD.97.059902} {\bibfield
  {journal} {\bibinfo  {journal} {Phys. Rev.}\ }\textbf {\bibinfo {volume}
  {D95}},\ \bibinfo {pages} {115008} (\bibinfo {year} {2017})},\ \bibinfo
  {note} {[Erratum: Phys. Rev.D97,no.5,059902(2018)]},\ \Eprint
  {http://arxiv.org/abs/1703.05330} {arXiv:1703.05330 [hep-ph]} \BibitemShut
  {NoStop}%
\bibitem [{\citenamefont {Bigi}\ \emph
  {et~al.}(2017{\natexlab{a}})\citenamefont {Bigi}, \citenamefont {Gambino},\
  and\ \citenamefont {Schacht}}]{Bigi:2017njr}%
  \BibitemOpen
  \bibfield  {author} {\bibinfo {author} {\bibfnamefont {D.}~\bibnamefont
  {Bigi}}, \bibinfo {author} {\bibfnamefont {P.}~\bibnamefont {Gambino}}, \
  and\ \bibinfo {author} {\bibfnamefont {S.}~\bibnamefont {Schacht}},\ }\href
  {\doibase 10.1016/j.physletb.2017.04.022} {\bibfield  {journal} {\bibinfo
  {journal} {Phys. Lett.}\ }\textbf {\bibinfo {volume} {B769}},\ \bibinfo
  {pages} {441} (\bibinfo {year} {2017}{\natexlab{a}})},\ \Eprint
  {http://arxiv.org/abs/1703.06124} {arXiv:1703.06124 [hep-ph]} \BibitemShut
  {NoStop}%
\bibitem [{\citenamefont {Grinstein}\ and\ \citenamefont
  {Kobach}(2017)}]{Grinstein:2017nlq}%
  \BibitemOpen
  \bibfield  {author} {\bibinfo {author} {\bibfnamefont {B.}~\bibnamefont
  {Grinstein}}\ and\ \bibinfo {author} {\bibfnamefont {A.}~\bibnamefont
  {Kobach}},\ }\href {\doibase 10.1016/j.physletb.2017.05.078} {\bibfield
  {journal} {\bibinfo  {journal} {Phys. Lett.}\ }\textbf {\bibinfo {volume}
  {B771}},\ \bibinfo {pages} {359} (\bibinfo {year} {2017})},\ \Eprint
  {http://arxiv.org/abs/1703.08170} {arXiv:1703.08170 [hep-ph]} \BibitemShut
  {NoStop}%
\bibitem [{\citenamefont {Tanabashi}\ \emph {et~al.}(2018)\citenamefont
  {Tanabashi} \emph {et~al.}}]{Tanabashi:2018oca}%
  \BibitemOpen
  \bibfield  {author} {\bibinfo {author} {\bibfnamefont {M.}~\bibnamefont
  {Tanabashi}} \emph {et~al.} (\bibinfo {collaboration} {Particle Data
  Group}),\ }\href {\doibase 10.1103/PhysRevD.98.030001} {\bibfield  {journal}
  {\bibinfo  {journal} {Phys. Rev.}\ }\textbf {\bibinfo {volume} {D98}},\
  \bibinfo {pages} {030001} (\bibinfo {year} {2018})}\BibitemShut {NoStop}%
\bibitem [{\citenamefont {Lees}\ \emph {et~al.}(2019)\citenamefont {Lees} \emph
  {et~al.}}]{Dey:2019bgc}%
  \BibitemOpen
  \bibfield  {author} {\bibinfo {author} {\bibfnamefont {J.~P.}\ \bibnamefont
  {Lees}} \emph {et~al.} (\bibinfo {collaboration} {BaBar}),\ }\href@noop {} {\
   (\bibinfo {year} {2019})},\ \Eprint {http://arxiv.org/abs/1903.10002}
  {arXiv:1903.10002 [hep-ex]} \BibitemShut {NoStop}%
\bibitem [{\citenamefont {Gambino}\ \emph {et~al.}(2019)\citenamefont
  {Gambino}, \citenamefont {Jung},\ and\ \citenamefont
  {Schacht}}]{Gambino:2019sif}%
  \BibitemOpen
  \bibfield  {author} {\bibinfo {author} {\bibfnamefont {P.}~\bibnamefont
  {Gambino}}, \bibinfo {author} {\bibfnamefont {M.}~\bibnamefont {Jung}}, \
  and\ \bibinfo {author} {\bibfnamefont {S.}~\bibnamefont {Schacht}},\
  }\href@noop {} {\  (\bibinfo {year} {2019})},\ \Eprint
  {http://arxiv.org/abs/1905.08209} {arXiv:1905.08209 [hep-ph]} \BibitemShut
  {NoStop}%
\bibitem [{\citenamefont {Koponen}\ \emph {et~al.}(2013)\citenamefont
  {Koponen}, \citenamefont {Davies}, \citenamefont {Donald}, \citenamefont
  {Follana}, \citenamefont {Lepage}, \citenamefont {Na},\ and\ \citenamefont
  {Shigemitsu}}]{Koponen:2013tua}%
  \BibitemOpen
  \bibfield  {author} {\bibinfo {author} {\bibfnamefont {J.}~\bibnamefont
  {Koponen}}, \bibinfo {author} {\bibfnamefont {C.~T.~H.}\ \bibnamefont
  {Davies}}, \bibinfo {author} {\bibfnamefont {G.~C.}\ \bibnamefont {Donald}},
  \bibinfo {author} {\bibfnamefont {E.}~\bibnamefont {Follana}}, \bibinfo
  {author} {\bibfnamefont {G.~P.}\ \bibnamefont {Lepage}}, \bibinfo {author}
  {\bibfnamefont {H.}~\bibnamefont {Na}}, \ and\ \bibinfo {author}
  {\bibfnamefont {J.}~\bibnamefont {Shigemitsu}} (\bibinfo {collaboration}
  {HPQCD}),\ }\href@noop {} {\  (\bibinfo {year} {2013})},\ \Eprint
  {http://arxiv.org/abs/1305.1462} {arXiv:1305.1462 [hep-lat]} \BibitemShut
  {NoStop}%
\bibitem [{\citenamefont {McLean}\ \emph {et~al.}(2018)\citenamefont {McLean},
  \citenamefont {Davies}, \citenamefont {Lytle},\ and\ \citenamefont
  {Koponen}}]{McLean:2019jll}%
  \BibitemOpen
  \bibfield  {author} {\bibinfo {author} {\bibfnamefont {E.}~\bibnamefont
  {McLean}}, \bibinfo {author} {\bibfnamefont {C.~T.~H.}\ \bibnamefont
  {Davies}}, \bibinfo {author} {\bibfnamefont {A.~T.}\ \bibnamefont {Lytle}}, \
  and\ \bibinfo {author} {\bibfnamefont {J.}~\bibnamefont {Koponen}} (\bibinfo
  {collaboration} {HPQCD}),\ }in\ \href@noop {} {\emph {\bibinfo {booktitle}
  {{Lattice(2018)}}}}\ (\bibinfo {year} {2018})\ \Eprint
  {http://arxiv.org/abs/1901.04979} {arXiv:1901.04979 [hep-lat]} \BibitemShut
  {NoStop}%
\bibitem [{\citenamefont {Laiho}\ and\ \citenamefont {Van~de
  Water}(2006)}]{Laiho:2005ue}%
  \BibitemOpen
  \bibfield  {author} {\bibinfo {author} {\bibfnamefont {J.}~\bibnamefont
  {Laiho}}\ and\ \bibinfo {author} {\bibfnamefont {R.~S.}\ \bibnamefont {Van~de
  Water}},\ }\href {\doibase 10.1103/PhysRevD.73.054501} {\bibfield  {journal}
  {\bibinfo  {journal} {Phys. Rev.}\ }\textbf {\bibinfo {volume} {D73}},\
  \bibinfo {pages} {054501} (\bibinfo {year} {2006})},\ \Eprint
  {http://arxiv.org/abs/hep-lat/0512007} {arXiv:hep-lat/0512007 [hep-lat]}
  \BibitemShut {NoStop}%
\bibitem [{\citenamefont {Bailey}\ \emph {et~al.}(2012)\citenamefont {Bailey}
  \emph {et~al.}}]{Bailey:2012rr}%
  \BibitemOpen
  \bibfield  {author} {\bibinfo {author} {\bibfnamefont {J.~A.}\ \bibnamefont
  {Bailey}} \emph {et~al.} (\bibinfo {collaboration} {Fermilab/MILC}),\ }\href
  {\doibase 10.1103/PhysRevD.85.114502, 10.1103/PhysRevD.86.039904} {\bibfield
  {journal} {\bibinfo  {journal} {Phys. Rev.}\ }\textbf {\bibinfo {volume}
  {D85}},\ \bibinfo {pages} {114502} (\bibinfo {year} {2012})},\ \bibinfo
  {note} {[Erratum: Phys. Rev.D86,039904(2012)]},\ \Eprint
  {http://arxiv.org/abs/1202.6346} {arXiv:1202.6346 [hep-lat]} \BibitemShut
  {NoStop}%
\bibitem [{\citenamefont {Monahan}\ \emph {et~al.}(2017)\citenamefont
  {Monahan}, \citenamefont {Na}, \citenamefont {Bouchard}, \citenamefont
  {Lepage},\ and\ \citenamefont {Shigemitsu}}]{Monahan:2017uby}%
  \BibitemOpen
  \bibfield  {author} {\bibinfo {author} {\bibfnamefont {C.~J.}\ \bibnamefont
  {Monahan}}, \bibinfo {author} {\bibfnamefont {H.}~\bibnamefont {Na}},
  \bibinfo {author} {\bibfnamefont {C.~M.}\ \bibnamefont {Bouchard}}, \bibinfo
  {author} {\bibfnamefont {G.~P.}\ \bibnamefont {Lepage}}, \ and\ \bibinfo
  {author} {\bibfnamefont {J.}~\bibnamefont {Shigemitsu}} (\bibinfo
  {collaboration} {HPQCD}),\ }\href {\doibase 10.1103/PhysRevD.95.114506}
  {\bibfield  {journal} {\bibinfo  {journal} {Phys. Rev.}\ }\textbf {\bibinfo
  {volume} {D95}},\ \bibinfo {pages} {114506} (\bibinfo {year} {2017})},\
  \Eprint {http://arxiv.org/abs/1703.09728} {arXiv:1703.09728 [hep-lat]}
  \BibitemShut {NoStop}%
\bibitem [{\citenamefont {Amhis}\ \emph {et~al.}(2017)\citenamefont {Amhis}
  \emph {et~al.}}]{Amhis:2016xyh}%
  \BibitemOpen
  \bibfield  {author} {\bibinfo {author} {\bibfnamefont {Y.}~\bibnamefont
  {Amhis}} \emph {et~al.} (\bibinfo {collaboration} {HFLAV}),\ }\href {\doibase
  10.1140/epjc/s10052-017-5058-4} {\bibfield  {journal} {\bibinfo  {journal}
  {Eur. Phys. J.}\ }\textbf {\bibinfo {volume} {C77}},\ \bibinfo {pages} {895}
  (\bibinfo {year} {2017})},\ \Eprint {http://arxiv.org/abs/1612.07233}
  {arXiv:1612.07233 [hep-ex]} \BibitemShut {NoStop}%
\bibitem [{\citenamefont {Caria}(2019)}]{CariaMoriond}%
  \BibitemOpen
  \bibfield  {author} {\bibinfo {author} {\bibfnamefont {G.}~\bibnamefont
  {Caria}} (\bibinfo {collaboration} {Belle}),\ }\href
  {https://moriond.in2p3.fr/2019/EW/Program.html} {\enquote {\bibinfo {title}
  {{Measurement of $R(D)$ and $R(D^*)$ with a Semileptonic tag at Belle}},}\ }
  (\bibinfo {year} {2019}),\ \bibinfo {note} {54th Rencontres de Moriond on
  Electroweak Interactions, March 16-23, 2019, La Thuile, Italy}\BibitemShut
  {NoStop}%
\bibitem [{\citenamefont {Atoui}\ \emph {et~al.}(2014)\citenamefont {Atoui},
  \citenamefont {Morénas}, \citenamefont {Bečirevic},\ and\ \citenamefont
  {Sanfilippo}}]{Atoui:2013zza}%
  \BibitemOpen
  \bibfield  {author} {\bibinfo {author} {\bibfnamefont {M.}~\bibnamefont
  {Atoui}}, \bibinfo {author} {\bibfnamefont {V.}~\bibnamefont {Morénas}},
  \bibinfo {author} {\bibfnamefont {D.}~\bibnamefont {Bečirevic}}, \ and\
  \bibinfo {author} {\bibfnamefont {F.}~\bibnamefont {Sanfilippo}},\ }\href
  {\doibase 10.1140/epjc/s10052-014-2861-z} {\bibfield  {journal} {\bibinfo
  {journal} {Eur. Phys. J.}\ }\textbf {\bibinfo {volume} {C74}},\ \bibinfo
  {pages} {2861} (\bibinfo {year} {2014})},\ \Eprint
  {http://arxiv.org/abs/1310.5238} {arXiv:1310.5238 [hep-lat]} \BibitemShut
  {NoStop}%
\bibitem [{\citenamefont {Flynn}\ \emph {et~al.}(2018)\citenamefont {Flynn},
  \citenamefont {Hill}, \citenamefont {Jüttner}, \citenamefont {Soni},
  \citenamefont {Tsang},\ and\ \citenamefont {Witzel}}]{Flynn:2019any}%
  \BibitemOpen
  \bibfield  {author} {\bibinfo {author} {\bibfnamefont {J.~M.}\ \bibnamefont
  {Flynn}}, \bibinfo {author} {\bibfnamefont {R.~C.}\ \bibnamefont {Hill}},
  \bibinfo {author} {\bibfnamefont {A.}~\bibnamefont {Jüttner}}, \bibinfo
  {author} {\bibfnamefont {A.}~\bibnamefont {Soni}}, \bibinfo {author}
  {\bibfnamefont {J.~T.}\ \bibnamefont {Tsang}}, \ and\ \bibinfo {author}
  {\bibfnamefont {O.}~\bibnamefont {Witzel}},\ }in\ \href@noop {} {\emph
  {\bibinfo {booktitle} {{Lattice(2018)}}}}\ (\bibinfo {year} {2018})\ \Eprint
  {http://arxiv.org/abs/1903.02100} {arXiv:1903.02100 [hep-lat]} \BibitemShut
  {NoStop}%
\bibitem [{\citenamefont {Follana}\ \emph {et~al.}(2007)\citenamefont
  {Follana}, \citenamefont {Mason}, \citenamefont {Davies}, \citenamefont
  {Hornbostel}, \citenamefont {Lepage}, \citenamefont {Shigemitsu},
  \citenamefont {Trottier},\ and\ \citenamefont {Wong}}]{Follana:2006rc}%
  \BibitemOpen
  \bibfield  {author} {\bibinfo {author} {\bibfnamefont {E.}~\bibnamefont
  {Follana}}, \bibinfo {author} {\bibfnamefont {Q.}~\bibnamefont {Mason}},
  \bibinfo {author} {\bibfnamefont {C.}~\bibnamefont {Davies}}, \bibinfo
  {author} {\bibfnamefont {K.}~\bibnamefont {Hornbostel}}, \bibinfo {author}
  {\bibfnamefont {G.~P.}\ \bibnamefont {Lepage}}, \bibinfo {author}
  {\bibfnamefont {J.}~\bibnamefont {Shigemitsu}}, \bibinfo {author}
  {\bibfnamefont {H.}~\bibnamefont {Trottier}}, \ and\ \bibinfo {author}
  {\bibfnamefont {K.}~\bibnamefont {Wong}} (\bibinfo {collaboration} {HPQCD,
  UKQCD}),\ }\href {\doibase 10.1103/PhysRevD.75.054502} {\bibfield  {journal}
  {\bibinfo  {journal} {Phys. Rev.}\ }\textbf {\bibinfo {volume} {D75}},\
  \bibinfo {pages} {054502} (\bibinfo {year} {2007})},\ \Eprint
  {http://arxiv.org/abs/hep-lat/0610092} {arXiv:hep-lat/0610092 [hep-lat]}
  \BibitemShut {NoStop}%
\bibitem [{\citenamefont {Bazavov}\ \emph {et~al.}(2013)\citenamefont {Bazavov}
  \emph {et~al.}}]{Bazavov:2012xda}%
  \BibitemOpen
  \bibfield  {author} {\bibinfo {author} {\bibfnamefont {A.}~\bibnamefont
  {Bazavov}} \emph {et~al.} (\bibinfo {collaboration} {MILC}),\ }\href
  {\doibase 10.1103/PhysRevD.87.054505} {\bibfield  {journal} {\bibinfo
  {journal} {Phys. Rev.}\ }\textbf {\bibinfo {volume} {D87}},\ \bibinfo {pages}
  {054505} (\bibinfo {year} {2013})},\ \Eprint {http://arxiv.org/abs/1212.4768}
  {arXiv:1212.4768 [hep-lat]} \BibitemShut {NoStop}%
\bibitem [{\citenamefont {McNeile}\ \emph {et~al.}(2010)\citenamefont
  {McNeile}, \citenamefont {Davies}, \citenamefont {Follana}, \citenamefont
  {Hornbostel},\ and\ \citenamefont {Lepage}}]{McNeile:2010ji}%
  \BibitemOpen
  \bibfield  {author} {\bibinfo {author} {\bibfnamefont {C.}~\bibnamefont
  {McNeile}}, \bibinfo {author} {\bibfnamefont {C.~T.~H.}\ \bibnamefont
  {Davies}}, \bibinfo {author} {\bibfnamefont {E.}~\bibnamefont {Follana}},
  \bibinfo {author} {\bibfnamefont {K.}~\bibnamefont {Hornbostel}}, \ and\
  \bibinfo {author} {\bibfnamefont {G.~P.}\ \bibnamefont {Lepage}},\ }\href
  {\doibase 10.1103/PhysRevD.82.034512} {\bibfield  {journal} {\bibinfo
  {journal} {Phys. Rev.}\ }\textbf {\bibinfo {volume} {D82}},\ \bibinfo {pages}
  {034512} (\bibinfo {year} {2010})},\ \Eprint {http://arxiv.org/abs/1004.4285}
  {arXiv:1004.4285 [hep-lat]} \BibitemShut {NoStop}%
\bibitem [{\citenamefont {McNeile}\ \emph
  {et~al.}(2012{\natexlab{a}})\citenamefont {McNeile}, \citenamefont {Davies},
  \citenamefont {Follana}, \citenamefont {Hornbostel},\ and\ \citenamefont
  {Lepage}}]{McNeile:2011ng}%
  \BibitemOpen
  \bibfield  {author} {\bibinfo {author} {\bibfnamefont {C.}~\bibnamefont
  {McNeile}}, \bibinfo {author} {\bibfnamefont {C.~T.~H.}\ \bibnamefont
  {Davies}}, \bibinfo {author} {\bibfnamefont {E.}~\bibnamefont {Follana}},
  \bibinfo {author} {\bibfnamefont {K.}~\bibnamefont {Hornbostel}}, \ and\
  \bibinfo {author} {\bibfnamefont {G.~P.}\ \bibnamefont {Lepage}} (\bibinfo
  {collaboration} {HPQCD}),\ }\href {\doibase 10.1103/PhysRevD.85.031503}
  {\bibfield  {journal} {\bibinfo  {journal} {Phys. Rev.}\ }\textbf {\bibinfo
  {volume} {D85}},\ \bibinfo {pages} {031503} (\bibinfo {year}
  {2012}{\natexlab{a}})},\ \Eprint {http://arxiv.org/abs/1110.4510}
  {arXiv:1110.4510 [hep-lat]} \BibitemShut {NoStop}%
\bibitem [{\citenamefont {McNeile}\ \emph
  {et~al.}(2012{\natexlab{b}})\citenamefont {McNeile}, \citenamefont {Davies},
  \citenamefont {Follana}, \citenamefont {Hornbostel},\ and\ \citenamefont
  {Lepage}}]{McNeile:2012qf}%
  \BibitemOpen
  \bibfield  {author} {\bibinfo {author} {\bibfnamefont {C.}~\bibnamefont
  {McNeile}}, \bibinfo {author} {\bibfnamefont {C.~T.~H.}\ \bibnamefont
  {Davies}}, \bibinfo {author} {\bibfnamefont {E.}~\bibnamefont {Follana}},
  \bibinfo {author} {\bibfnamefont {K.}~\bibnamefont {Hornbostel}}, \ and\
  \bibinfo {author} {\bibfnamefont {G.~P.}\ \bibnamefont {Lepage}},\ }\href
  {\doibase 10.1103/PhysRevD.86.074503} {\bibfield  {journal} {\bibinfo
  {journal} {Phys. Rev.}\ }\textbf {\bibinfo {volume} {D86}},\ \bibinfo {pages}
  {074503} (\bibinfo {year} {2012}{\natexlab{b}})},\ \Eprint
  {http://arxiv.org/abs/1207.0994} {arXiv:1207.0994 [hep-lat]} \BibitemShut
  {NoStop}%
\bibitem [{\citenamefont {Bazavov}\ \emph {et~al.}(2018)\citenamefont {Bazavov}
  \emph {et~al.}}]{Bazavov:2017lyh}%
  \BibitemOpen
  \bibfield  {author} {\bibinfo {author} {\bibfnamefont {A.}~\bibnamefont
  {Bazavov}} \emph {et~al.} (\bibinfo {collaboration} {Fermilab/MILC}),\ }\href
  {\doibase 10.1103/PhysRevD.98.074512} {\bibfield  {journal} {\bibinfo
  {journal} {Phys. Rev.}\ }\textbf {\bibinfo {volume} {D98}},\ \bibinfo {pages}
  {074512} (\bibinfo {year} {2018})},\ \Eprint
  {http://arxiv.org/abs/1712.09262} {arXiv:1712.09262 [hep-lat]} \BibitemShut
  {NoStop}%
\bibitem [{\citenamefont {Petreczky}\ and\ \citenamefont
  {Weber}(2019)}]{Petreczky:2019ozv}%
  \BibitemOpen
  \bibfield  {author} {\bibinfo {author} {\bibfnamefont {P.}~\bibnamefont
  {Petreczky}}\ and\ \bibinfo {author} {\bibfnamefont {J.~H.}\ \bibnamefont
  {Weber}},\ }\href@noop {} {\  (\bibinfo {year} {2019})},\ \Eprint
  {http://arxiv.org/abs/1901.06424} {arXiv:1901.06424 [hep-lat]} \BibitemShut
  {NoStop}%
\bibitem [{\citenamefont {Lytle}\ \emph {et~al.}(2016)\citenamefont {Lytle},
  \citenamefont {Colquhoun}, \citenamefont {Davies}, \citenamefont {Koponen},\
  and\ \citenamefont {McNeile}}]{Lytle:2016ixw}%
  \BibitemOpen
  \bibfield  {author} {\bibinfo {author} {\bibfnamefont {A.}~\bibnamefont
  {Lytle}}, \bibinfo {author} {\bibfnamefont {B.}~\bibnamefont {Colquhoun}},
  \bibinfo {author} {\bibfnamefont {C.}~\bibnamefont {Davies}}, \bibinfo
  {author} {\bibfnamefont {J.}~\bibnamefont {Koponen}}, \ and\ \bibinfo
  {author} {\bibfnamefont {C.}~\bibnamefont {McNeile}} (\bibinfo
  {collaboration} {HPQCD}),\ }\href {\doibase 10.22323/1.273.0069} {\bibfield
  {journal} {\bibinfo  {journal} {PoS}\ }\textbf {\bibinfo {volume}
  {BEAUTY2016}},\ \bibinfo {pages} {069} (\bibinfo {year} {2016})},\ \Eprint
  {http://arxiv.org/abs/1605.05645} {arXiv:1605.05645 [hep-lat]} \BibitemShut
  {NoStop}%
\bibitem [{\citenamefont {Colquhoun}\ \emph {et~al.}(2016)\citenamefont
  {Colquhoun}, \citenamefont {Davies}, \citenamefont {Koponen}, \citenamefont
  {Lytle},\ and\ \citenamefont {McNeile}}]{Colquhoun:2016osw}%
  \BibitemOpen
  \bibfield  {author} {\bibinfo {author} {\bibfnamefont {B.}~\bibnamefont
  {Colquhoun}}, \bibinfo {author} {\bibfnamefont {C.}~\bibnamefont {Davies}},
  \bibinfo {author} {\bibfnamefont {J.}~\bibnamefont {Koponen}}, \bibinfo
  {author} {\bibfnamefont {A.}~\bibnamefont {Lytle}}, \ and\ \bibinfo {author}
  {\bibfnamefont {C.}~\bibnamefont {McNeile}} (\bibinfo {collaboration}
  {HPQCD}),\ }\href@noop {} {\bibfield  {journal} {\bibinfo  {journal} {PoS}\
  }\textbf {\bibinfo {volume} {LATTICE2016}},\ \bibinfo {pages} {281} (\bibinfo
  {year} {2016})},\ \Eprint {http://arxiv.org/abs/1611.01987} {arXiv:1611.01987
  [hep-lat]} \BibitemShut {NoStop}%
\bibitem [{\citenamefont {McLean}\ \emph {et~al.}(2019)\citenamefont {McLean},
  \citenamefont {Davies}, \citenamefont {Lytle},\ and\ \citenamefont
  {Koponen}}]{McLean:2019sds}%
  \BibitemOpen
  \bibfield  {author} {\bibinfo {author} {\bibfnamefont {E.}~\bibnamefont
  {McLean}}, \bibinfo {author} {\bibfnamefont {C.~T.~H.}\ \bibnamefont
  {Davies}}, \bibinfo {author} {\bibfnamefont {A.~T.}\ \bibnamefont {Lytle}}, \
  and\ \bibinfo {author} {\bibfnamefont {J.}~\bibnamefont {Koponen}},\
  }\href@noop {} {\  (\bibinfo {year} {2019})},\ \Eprint
  {http://arxiv.org/abs/1904.02046} {arXiv:1904.02046 [hep-lat]} \BibitemShut
  {NoStop}%
\bibitem [{\citenamefont {Bazavov}\ \emph {et~al.}(2010)\citenamefont {Bazavov}
  \emph {et~al.}}]{Bazavov:2010ru}%
  \BibitemOpen
  \bibfield  {author} {\bibinfo {author} {\bibfnamefont {A.}~\bibnamefont
  {Bazavov}} \emph {et~al.} (\bibinfo {collaboration} {MILC}),\ }\href
  {\doibase 10.1103/PhysRevD.82.074501} {\bibfield  {journal} {\bibinfo
  {journal} {Phys. Rev.}\ }\textbf {\bibinfo {volume} {D82}},\ \bibinfo {pages}
  {074501} (\bibinfo {year} {2010})},\ \Eprint {http://arxiv.org/abs/1004.0342}
  {arXiv:1004.0342 [hep-lat]} \BibitemShut {NoStop}%
\bibitem [{\citenamefont {Borsanyi}\ \emph {et~al.}(2012)\citenamefont
  {Borsanyi} \emph {et~al.}}]{Borsanyi:2012zs}%
  \BibitemOpen
  \bibfield  {author} {\bibinfo {author} {\bibfnamefont {S.}~\bibnamefont
  {Borsanyi}} \emph {et~al.},\ }\href {\doibase 10.1007/JHEP09(2012)010}
  {\bibfield  {journal} {\bibinfo  {journal} {JHEP}\ }\textbf {\bibinfo
  {volume} {09}},\ \bibinfo {pages} {010} (\bibinfo {year} {2012})},\ \Eprint
  {http://arxiv.org/abs/1203.4469} {arXiv:1203.4469 [hep-lat]} \BibitemShut
  {NoStop}%
\bibitem [{\citenamefont {Chakraborty}\ \emph {et~al.}(2017)\citenamefont
  {Chakraborty}, \citenamefont {Davies}, \citenamefont {de~Oliviera},
  \citenamefont {Koponen}, \citenamefont {Lepage},\ and\ \citenamefont {Van~de
  Water}}]{Chakraborty:2016mwy}%
  \BibitemOpen
  \bibfield  {author} {\bibinfo {author} {\bibfnamefont {B.}~\bibnamefont
  {Chakraborty}}, \bibinfo {author} {\bibfnamefont {C.~T.~H.}\ \bibnamefont
  {Davies}}, \bibinfo {author} {\bibfnamefont {P.~G.}\ \bibnamefont
  {de~Oliviera}}, \bibinfo {author} {\bibfnamefont {J.}~\bibnamefont
  {Koponen}}, \bibinfo {author} {\bibfnamefont {G.~P.}\ \bibnamefont {Lepage}},
  \ and\ \bibinfo {author} {\bibfnamefont {R.~S.}\ \bibnamefont {Van~de Water}}
  (\bibinfo {collaboration} {HPQCD}),\ }\href {\doibase
  10.1103/PhysRevD.96.034516} {\bibfield  {journal} {\bibinfo  {journal} {Phys.
  Rev.}\ }\textbf {\bibinfo {volume} {D96}},\ \bibinfo {pages} {034516}
  (\bibinfo {year} {2017})},\ \Eprint {http://arxiv.org/abs/1601.03071}
  {arXiv:1601.03071 [hep-lat]} \BibitemShut {NoStop}%
\bibitem [{\citenamefont {Chakraborty}\ \emph {et~al.}(2015)\citenamefont
  {Chakraborty}, \citenamefont {Davies}, \citenamefont {Galloway},
  \citenamefont {Knecht}, \citenamefont {Koponen}, \citenamefont {Donald},
  \citenamefont {Dowdall}, \citenamefont {Lepage},\ and\ \citenamefont
  {McNeile}}]{Chakraborty:2014aca}%
  \BibitemOpen
  \bibfield  {author} {\bibinfo {author} {\bibfnamefont {B.}~\bibnamefont
  {Chakraborty}}, \bibinfo {author} {\bibfnamefont {C.~T.~H.}\ \bibnamefont
  {Davies}}, \bibinfo {author} {\bibfnamefont {B.}~\bibnamefont {Galloway}},
  \bibinfo {author} {\bibfnamefont {P.}~\bibnamefont {Knecht}}, \bibinfo
  {author} {\bibfnamefont {J.}~\bibnamefont {Koponen}}, \bibinfo {author}
  {\bibfnamefont {G.~C.}\ \bibnamefont {Donald}}, \bibinfo {author}
  {\bibfnamefont {R.~J.}\ \bibnamefont {Dowdall}}, \bibinfo {author}
  {\bibfnamefont {G.~P.}\ \bibnamefont {Lepage}}, \ and\ \bibinfo {author}
  {\bibfnamefont {C.}~\bibnamefont {McNeile}} (\bibinfo {collaboration}
  {HPQCD}),\ }\href {\doibase 10.1103/PhysRevD.91.054508} {\bibfield  {journal}
  {\bibinfo  {journal} {Phys. Rev.}\ }\textbf {\bibinfo {volume} {D91}},\
  \bibinfo {pages} {054508} (\bibinfo {year} {2015})},\ \Eprint
  {http://arxiv.org/abs/1408.4169} {arXiv:1408.4169 [hep-lat]} \BibitemShut
  {NoStop}%
\bibitem [{\citenamefont {McNeile}(2015)}]{mcneile:private}%
  \BibitemOpen
  \bibfield  {author} {\bibinfo {author} {\bibfnamefont {C.}~\bibnamefont
  {McNeile}},\ }\href@noop {} {}\bibinfo {howpublished} {private communication}
  (\bibinfo {year} {2015})\BibitemShut {NoStop}%
\bibitem [{\citenamefont {Dowdall}\ \emph {et~al.}(2013)\citenamefont
  {Dowdall}, \citenamefont {Davies}, \citenamefont {Lepage},\ and\
  \citenamefont {McNeile}}]{Dowdall:2013rya}%
  \BibitemOpen
  \bibfield  {author} {\bibinfo {author} {\bibfnamefont {R.~J.}\ \bibnamefont
  {Dowdall}}, \bibinfo {author} {\bibfnamefont {C.~T.~H.}\ \bibnamefont
  {Davies}}, \bibinfo {author} {\bibfnamefont {G.~P.}\ \bibnamefont {Lepage}},
  \ and\ \bibinfo {author} {\bibfnamefont {C.}~\bibnamefont {McNeile}}
  (\bibinfo {collaboration} {HPQCD}),\ }\href {\doibase
  10.1103/PhysRevD.88.074504} {\bibfield  {journal} {\bibinfo  {journal} {Phys.
  Rev.}\ }\textbf {\bibinfo {volume} {D88}},\ \bibinfo {pages} {074504}
  (\bibinfo {year} {2013})},\ \Eprint {http://arxiv.org/abs/1303.1670}
  {arXiv:1303.1670 [hep-lat]} \BibitemShut {NoStop}%
\bibitem [{\citenamefont {Hart}\ \emph {et~al.}(2009)\citenamefont {Hart},
  \citenamefont {von Hippel},\ and\ \citenamefont {Horgan}}]{Hart:2008sq}%
  \BibitemOpen
  \bibfield  {author} {\bibinfo {author} {\bibfnamefont {A.}~\bibnamefont
  {Hart}}, \bibinfo {author} {\bibfnamefont {G.~M.}\ \bibnamefont {von
  Hippel}}, \ and\ \bibinfo {author} {\bibfnamefont {R.~R.}\ \bibnamefont
  {Horgan}} (\bibinfo {collaboration} {HPQCD}),\ }\href {\doibase
  10.1103/PhysRevD.79.074008} {\bibfield  {journal} {\bibinfo  {journal} {Phys.
  Rev.}\ }\textbf {\bibinfo {volume} {D79}},\ \bibinfo {pages} {074008}
  (\bibinfo {year} {2009})},\ \Eprint {http://arxiv.org/abs/0812.0503}
  {arXiv:0812.0503 [hep-lat]} \BibitemShut {NoStop}%
\bibitem [{\citenamefont {Davies}\ \emph
  {et~al.}(2010{\natexlab{a}})\citenamefont {Davies}, \citenamefont {McNeile},
  \citenamefont {Follana}, \citenamefont {Lepage}, \citenamefont {Na},\ and\
  \citenamefont {Shigemitsu}}]{Davies:2010ip}%
  \BibitemOpen
  \bibfield  {author} {\bibinfo {author} {\bibfnamefont {C.~T.~H.}\
  \bibnamefont {Davies}}, \bibinfo {author} {\bibfnamefont {C.}~\bibnamefont
  {McNeile}}, \bibinfo {author} {\bibfnamefont {E.}~\bibnamefont {Follana}},
  \bibinfo {author} {\bibfnamefont {G.~P.}\ \bibnamefont {Lepage}}, \bibinfo
  {author} {\bibfnamefont {H.}~\bibnamefont {Na}}, \ and\ \bibinfo {author}
  {\bibfnamefont {J.}~\bibnamefont {Shigemitsu}} (\bibinfo {collaboration}
  {HPQCD}),\ }\href {\doibase 10.1103/PhysRevD.82.114504} {\bibfield  {journal}
  {\bibinfo  {journal} {Phys. Rev.}\ }\textbf {\bibinfo {volume} {D82}},\
  \bibinfo {pages} {114504} (\bibinfo {year} {2010}{\natexlab{a}})},\ \Eprint
  {http://arxiv.org/abs/1008.4018} {arXiv:1008.4018 [hep-lat]} \BibitemShut
  {NoStop}%
\bibitem [{\citenamefont {Guadagnoli}\ \emph {et~al.}(2006)\citenamefont
  {Guadagnoli}, \citenamefont {Mescia},\ and\ \citenamefont
  {Simula}}]{Guadagnoli:2005be}%
  \BibitemOpen
  \bibfield  {author} {\bibinfo {author} {\bibfnamefont {D.}~\bibnamefont
  {Guadagnoli}}, \bibinfo {author} {\bibfnamefont {F.}~\bibnamefont {Mescia}},
  \ and\ \bibinfo {author} {\bibfnamefont {S.}~\bibnamefont {Simula}},\ }\href
  {\doibase 10.1103/PhysRevD.73.114504} {\bibfield  {journal} {\bibinfo
  {journal} {Phys. Rev.}\ }\textbf {\bibinfo {volume} {D73}},\ \bibinfo {pages}
  {114504} (\bibinfo {year} {2006})},\ \Eprint
  {http://arxiv.org/abs/hep-lat/0512020} {arXiv:hep-lat/0512020 [hep-lat]}
  \BibitemShut {NoStop}%
\bibitem [{\citenamefont {Davies}\ \emph
  {et~al.}(2010{\natexlab{b}})\citenamefont {Davies}, \citenamefont {Follana},
  \citenamefont {Kendall}, \citenamefont {Lepage},\ and\ \citenamefont
  {McNeile}}]{Davies:2009tsa}%
  \BibitemOpen
  \bibfield  {author} {\bibinfo {author} {\bibfnamefont {C.~T.~H.}\
  \bibnamefont {Davies}}, \bibinfo {author} {\bibfnamefont {E.}~\bibnamefont
  {Follana}}, \bibinfo {author} {\bibfnamefont {I.~D.}\ \bibnamefont
  {Kendall}}, \bibinfo {author} {\bibfnamefont {G.~P.}\ \bibnamefont {Lepage}},
  \ and\ \bibinfo {author} {\bibfnamefont {C.}~\bibnamefont {McNeile}}
  (\bibinfo {collaboration} {HPQCD}),\ }\href {\doibase
  10.1103/PhysRevD.81.034506} {\bibfield  {journal} {\bibinfo  {journal} {Phys.
  Rev.}\ }\textbf {\bibinfo {volume} {D81}},\ \bibinfo {pages} {034506}
  (\bibinfo {year} {2010}{\natexlab{b}})},\ \Eprint
  {http://arxiv.org/abs/0910.1229} {arXiv:0910.1229 [hep-lat]} \BibitemShut
  {NoStop}%
\bibitem [{\citenamefont {Lepage}\ \emph {et~al.}(2002)\citenamefont {Lepage},
  \citenamefont {Clark}, \citenamefont {Davies}, \citenamefont {Hornbostel},
  \citenamefont {Mackenzie}, \citenamefont {Morningstar},\ and\ \citenamefont
  {Trottier}}]{Lepage:2001ym}%
  \BibitemOpen
  \bibfield  {author} {\bibinfo {author} {\bibfnamefont {G.~P.}\ \bibnamefont
  {Lepage}}, \bibinfo {author} {\bibfnamefont {B.}~\bibnamefont {Clark}},
  \bibinfo {author} {\bibfnamefont {C.~T.~H.}\ \bibnamefont {Davies}}, \bibinfo
  {author} {\bibfnamefont {K.}~\bibnamefont {Hornbostel}}, \bibinfo {author}
  {\bibfnamefont {P.~B.}\ \bibnamefont {Mackenzie}}, \bibinfo {author}
  {\bibfnamefont {C.}~\bibnamefont {Morningstar}}, \ and\ \bibinfo {author}
  {\bibfnamefont {H.}~\bibnamefont {Trottier}},\ }\href {\doibase
  10.1016/S0920-5632(01)01638-3} {\bibfield  {journal} {\bibinfo  {journal}
  {Nucl. Phys. Proc. Suppl.}\ }\textbf {\bibinfo {volume} {106}},\ \bibinfo
  {pages} {12} (\bibinfo {year} {2002})},\ \Eprint
  {http://arxiv.org/abs/hep-lat/0110175} {arXiv:hep-lat/0110175 [hep-lat]}
  \BibitemShut {NoStop}%
\bibitem [{cor(2018)}]{corrfitter}%
  \BibitemOpen
  \href@noop {} {\enquote {\bibinfo {title} {Corrfitter},}\ }\bibinfo
  {howpublished} {\url{https:/github.com/gplepage/corrfitter}} (\bibinfo {year}
  {2018})\BibitemShut {NoStop}%
\bibitem [{\citenamefont {Donald}\ \emph {et~al.}(2012)\citenamefont {Donald},
  \citenamefont {Davies}, \citenamefont {Dowdall}, \citenamefont {Follana},
  \citenamefont {Hornbostel}, \citenamefont {Koponen}, \citenamefont {Lepage},\
  and\ \citenamefont {McNeile}}]{Donald:2012ga}%
  \BibitemOpen
  \bibfield  {author} {\bibinfo {author} {\bibfnamefont {G.~C.}\ \bibnamefont
  {Donald}}, \bibinfo {author} {\bibfnamefont {C.~T.~H.}\ \bibnamefont
  {Davies}}, \bibinfo {author} {\bibfnamefont {R.~J.}\ \bibnamefont {Dowdall}},
  \bibinfo {author} {\bibfnamefont {E.}~\bibnamefont {Follana}}, \bibinfo
  {author} {\bibfnamefont {K.}~\bibnamefont {Hornbostel}}, \bibinfo {author}
  {\bibfnamefont {J.}~\bibnamefont {Koponen}}, \bibinfo {author} {\bibfnamefont
  {G.~P.}\ \bibnamefont {Lepage}}, \ and\ \bibinfo {author} {\bibfnamefont
  {C.}~\bibnamefont {McNeile}},\ }\href {\doibase 10.1103/PhysRevD.86.094501}
  {\bibfield  {journal} {\bibinfo  {journal} {Phys. Rev.}\ }\textbf {\bibinfo
  {volume} {D86}},\ \bibinfo {pages} {094501} (\bibinfo {year} {2012})},\
  \Eprint {http://arxiv.org/abs/1208.2855} {arXiv:1208.2855 [hep-lat]}
  \BibitemShut {NoStop}%
\bibitem [{\citenamefont {Donald}\ \emph {et~al.}(2014)\citenamefont {Donald},
  \citenamefont {Davies}, \citenamefont {Koponen},\ and\ \citenamefont
  {Lepage}}]{Donald:2013pea}%
  \BibitemOpen
  \bibfield  {author} {\bibinfo {author} {\bibfnamefont {G.~C.}\ \bibnamefont
  {Donald}}, \bibinfo {author} {\bibfnamefont {C.~T.~H.}\ \bibnamefont
  {Davies}}, \bibinfo {author} {\bibfnamefont {J.}~\bibnamefont {Koponen}}, \
  and\ \bibinfo {author} {\bibfnamefont {G.~P.}\ \bibnamefont {Lepage}}
  (\bibinfo {collaboration} {HPQCD}),\ }\href {\doibase
  10.1103/PhysRevD.90.074506} {\bibfield  {journal} {\bibinfo  {journal} {Phys.
  Rev.}\ }\textbf {\bibinfo {volume} {D90}},\ \bibinfo {pages} {074506}
  (\bibinfo {year} {2014})},\ \Eprint {http://arxiv.org/abs/1311.6669}
  {arXiv:1311.6669 [hep-lat]} \BibitemShut {NoStop}%
\bibitem [{\citenamefont {Monahan}\ \emph {et~al.}(2013)\citenamefont
  {Monahan}, \citenamefont {Shigemitsu},\ and\ \citenamefont
  {Horgan}}]{Monahan:2012dq}%
  \BibitemOpen
  \bibfield  {author} {\bibinfo {author} {\bibfnamefont {C.}~\bibnamefont
  {Monahan}}, \bibinfo {author} {\bibfnamefont {J.}~\bibnamefont {Shigemitsu}},
  \ and\ \bibinfo {author} {\bibfnamefont {R.}~\bibnamefont {Horgan}},\ }\href
  {\doibase 10.1103/PhysRevD.87.034017} {\bibfield  {journal} {\bibinfo
  {journal} {Phys. Rev.}\ }\textbf {\bibinfo {volume} {D87}},\ \bibinfo {pages}
  {034017} (\bibinfo {year} {2013})},\ \Eprint {http://arxiv.org/abs/1211.6966}
  {arXiv:1211.6966 [hep-lat]} \BibitemShut {NoStop}%
\bibitem [{\citenamefont {Bourrely}\ \emph {et~al.}(2009)\citenamefont
  {Bourrely}, \citenamefont {Caprini},\ and\ \citenamefont
  {Lellouch}}]{Bourrely:2008za}%
  \BibitemOpen
  \bibfield  {author} {\bibinfo {author} {\bibfnamefont {C.}~\bibnamefont
  {Bourrely}}, \bibinfo {author} {\bibfnamefont {I.}~\bibnamefont {Caprini}}, \
  and\ \bibinfo {author} {\bibfnamefont {L.}~\bibnamefont {Lellouch}},\ }\href
  {\doibase 10.1103/PhysRevD.82.099902, 10.1103/PhysRevD.79.013008} {\bibfield
  {journal} {\bibinfo  {journal} {Phys. Rev.}\ }\textbf {\bibinfo {volume}
  {D79}},\ \bibinfo {pages} {013008} (\bibinfo {year} {2009})},\ \bibinfo
  {note} {[Erratum: Phys. Rev.D82,099902(2010)]},\ \Eprint
  {http://arxiv.org/abs/0807.2722} {arXiv:0807.2722 [hep-ph]} \BibitemShut
  {NoStop}%
\bibitem [{\citenamefont {Dowdall}\ \emph {et~al.}(2012)\citenamefont
  {Dowdall}, \citenamefont {Davies}, \citenamefont {Hammant},\ and\
  \citenamefont {Horgan}}]{Dowdall:2012ab}%
  \BibitemOpen
  \bibfield  {author} {\bibinfo {author} {\bibfnamefont {R.~J.}\ \bibnamefont
  {Dowdall}}, \bibinfo {author} {\bibfnamefont {C.~T.~H.}\ \bibnamefont
  {Davies}}, \bibinfo {author} {\bibfnamefont {T.~C.}\ \bibnamefont {Hammant}},
  \ and\ \bibinfo {author} {\bibfnamefont {R.~R.}\ \bibnamefont {Horgan}},\
  }\href {\doibase 10.1103/PhysRevD.86.094510} {\bibfield  {journal} {\bibinfo
  {journal} {Phys. Rev.}\ }\textbf {\bibinfo {volume} {D86}},\ \bibinfo {pages}
  {094510} (\bibinfo {year} {2012})},\ \Eprint {http://arxiv.org/abs/1207.5149}
  {arXiv:1207.5149 [hep-lat]} \BibitemShut {NoStop}%
\bibitem [{\citenamefont {Bernard}\ and\ \citenamefont
  {Toussaint}(2018)}]{Bernard:2017npd}%
  \BibitemOpen
  \bibfield  {author} {\bibinfo {author} {\bibfnamefont {C.}~\bibnamefont
  {Bernard}}\ and\ \bibinfo {author} {\bibfnamefont {D.}~\bibnamefont
  {Toussaint}} (\bibinfo {collaboration} {MILC}),\ }\href {\doibase
  10.1103/PhysRevD.97.074502} {\bibfield  {journal} {\bibinfo  {journal} {Phys.
  Rev.}\ }\textbf {\bibinfo {volume} {D97}},\ \bibinfo {pages} {074502}
  (\bibinfo {year} {2018})},\ \Eprint {http://arxiv.org/abs/1707.05430}
  {arXiv:1707.05430 [hep-lat]} \BibitemShut {NoStop}%
\bibitem [{\citenamefont {Caprini}\ \emph {et~al.}(1998)\citenamefont
  {Caprini}, \citenamefont {Lellouch},\ and\ \citenamefont
  {Neubert}}]{Caprini:1997mu}%
  \BibitemOpen
  \bibfield  {author} {\bibinfo {author} {\bibfnamefont {I.}~\bibnamefont
  {Caprini}}, \bibinfo {author} {\bibfnamefont {L.}~\bibnamefont {Lellouch}}, \
  and\ \bibinfo {author} {\bibfnamefont {M.}~\bibnamefont {Neubert}},\ }\href
  {\doibase 10.1016/S0550-3213(98)00350-2} {\bibfield  {journal} {\bibinfo
  {journal} {Nucl. Phys.}\ }\textbf {\bibinfo {volume} {B530}},\ \bibinfo
  {pages} {153} (\bibinfo {year} {1998})},\ \Eprint
  {http://arxiv.org/abs/hep-ph/9712417} {arXiv:hep-ph/9712417 [hep-ph]}
  \BibitemShut {NoStop}%
\bibitem [{\citenamefont {Bigi}\ \emph
  {et~al.}(2017{\natexlab{b}})\citenamefont {Bigi}, \citenamefont {Gambino},\
  and\ \citenamefont {Schacht}}]{Bigi:2017jbd}%
  \BibitemOpen
  \bibfield  {author} {\bibinfo {author} {\bibfnamefont {D.}~\bibnamefont
  {Bigi}}, \bibinfo {author} {\bibfnamefont {P.}~\bibnamefont {Gambino}}, \
  and\ \bibinfo {author} {\bibfnamefont {S.}~\bibnamefont {Schacht}},\ }\href
  {\doibase 10.1007/JHEP11(2017)061} {\bibfield  {journal} {\bibinfo  {journal}
  {JHEP}\ }\textbf {\bibinfo {volume} {11}},\ \bibinfo {pages} {061} (\bibinfo
  {year} {2017}{\natexlab{b}})},\ \Eprint {http://arxiv.org/abs/1707.09509}
  {arXiv:1707.09509 [hep-ph]} \BibitemShut {NoStop}%
\bibitem [{\citenamefont {Kaneko}\ \emph {et~al.}(2018)\citenamefont {Kaneko},
  \citenamefont {Aoki}, \citenamefont {Colquhoun}, \citenamefont {Fukaya},\
  and\ \citenamefont {Hashimoto}}]{Kaneko:2018mcr}%
  \BibitemOpen
  \bibfield  {author} {\bibinfo {author} {\bibfnamefont {T.}~\bibnamefont
  {Kaneko}}, \bibinfo {author} {\bibfnamefont {Y.}~\bibnamefont {Aoki}},
  \bibinfo {author} {\bibfnamefont {B.}~\bibnamefont {Colquhoun}}, \bibinfo
  {author} {\bibfnamefont {H.}~\bibnamefont {Fukaya}}, \ and\ \bibinfo {author}
  {\bibfnamefont {S.}~\bibnamefont {Hashimoto}} (\bibinfo {collaboration}
  {JLQCD}),\ }in\ \href@noop {} {\emph {\bibinfo {booktitle} {{36th
  International Symposium on Lattice Field Theory (Lattice 2018) East Lansing,
  MI, United States, July 22-28, 2018}}}}\ (\bibinfo {year} {2018})\ \Eprint
  {http://arxiv.org/abs/1811.00794} {arXiv:1811.00794 [hep-lat]} \BibitemShut
  {NoStop}%
\bibitem [{\citenamefont {Hill}(2006{\natexlab{a}})}]{Hill:2005ju}%
  \BibitemOpen
  \bibfield  {author} {\bibinfo {author} {\bibfnamefont {R.~J.}\ \bibnamefont
  {Hill}},\ }\href {\doibase 10.1103/PhysRevD.73.014012} {\bibfield  {journal}
  {\bibinfo  {journal} {Phys. Rev.}\ }\textbf {\bibinfo {volume} {D73}},\
  \bibinfo {pages} {014012} (\bibinfo {year} {2006}{\natexlab{a}})},\ \Eprint
  {http://arxiv.org/abs/hep-ph/0505129} {arXiv:hep-ph/0505129 [hep-ph]}
  \BibitemShut {NoStop}%
\bibitem [{\citenamefont {Hill}(2006{\natexlab{b}})}]{Hill:2006ub}%
  \BibitemOpen
  \bibfield  {author} {\bibinfo {author} {\bibfnamefont {R.~J.}\ \bibnamefont
  {Hill}},\ }\bibfield  {booktitle} {\emph {\bibinfo {booktitle} {{Proceedings,
  4th Conference on Flavor Physics and CP Violation (FPCP 2006): Vancouver,
  British Columbia, Canada, April 9-12, 2006}}},\ }\href@noop {} {\bibfield
  {journal} {\bibinfo  {journal} {eConf}\ }\textbf {\bibinfo {volume}
  {C060409}},\ \bibinfo {pages} {027} (\bibinfo {year} {2006}{\natexlab{b}})},\
  \Eprint {http://arxiv.org/abs/hep-ph/0606023} {arXiv:hep-ph/0606023 [hep-ph]}
  \BibitemShut {NoStop}%
\bibitem [{\citenamefont {Okubo}(1971{\natexlab{a}})}]{Okubo:1971my}%
  \BibitemOpen
  \bibfield  {author} {\bibinfo {author} {\bibfnamefont {S.}~\bibnamefont
  {Okubo}},\ }\href {\doibase 10.1103/PhysRevD.4.725} {\bibfield  {journal}
  {\bibinfo  {journal} {Phys. Rev.}\ }\textbf {\bibinfo {volume} {D4}},\
  \bibinfo {pages} {725} (\bibinfo {year} {1971}{\natexlab{a}})}\BibitemShut
  {NoStop}%
\bibitem [{\citenamefont {Okubo}(1971{\natexlab{b}})}]{Okubo:1971jf}%
  \BibitemOpen
  \bibfield  {author} {\bibinfo {author} {\bibfnamefont {S.}~\bibnamefont
  {Okubo}},\ }\href {\doibase 10.1103/PhysRevD.3.2807} {\bibfield  {journal}
  {\bibinfo  {journal} {Phys. Rev.}\ }\textbf {\bibinfo {volume} {D3}},\
  \bibinfo {pages} {2807} (\bibinfo {year} {1971}{\natexlab{b}})}\BibitemShut
  {NoStop}%
\bibitem [{\citenamefont {Boyd}\ \emph {et~al.}(1996)\citenamefont {Boyd},
  \citenamefont {Grinstein},\ and\ \citenamefont {Lebed}}]{Boyd:1995sq}%
  \BibitemOpen
  \bibfield  {author} {\bibinfo {author} {\bibfnamefont {C.~G.}\ \bibnamefont
  {Boyd}}, \bibinfo {author} {\bibfnamefont {B.}~\bibnamefont {Grinstein}}, \
  and\ \bibinfo {author} {\bibfnamefont {R.~F.}\ \bibnamefont {Lebed}},\ }\href
  {\doibase 10.1016/0550-3213(95)00653-2} {\bibfield  {journal} {\bibinfo
  {journal} {Nucl. Phys.}\ }\textbf {\bibinfo {volume} {B461}},\ \bibinfo
  {pages} {493} (\bibinfo {year} {1996})},\ \Eprint
  {http://arxiv.org/abs/hep-ph/9508211} {arXiv:hep-ph/9508211 [hep-ph]}
  \BibitemShut {NoStop}%
\bibitem [{\citenamefont {Becher}\ and\ \citenamefont
  {Hill}(2006)}]{Becher:2005bg}%
  \BibitemOpen
  \bibfield  {author} {\bibinfo {author} {\bibfnamefont {T.}~\bibnamefont
  {Becher}}\ and\ \bibinfo {author} {\bibfnamefont {R.~J.}\ \bibnamefont
  {Hill}},\ }\href {\doibase 10.1016/j.physletb.2005.11.063} {\bibfield
  {journal} {\bibinfo  {journal} {Phys. Lett.}\ }\textbf {\bibinfo {volume}
  {B633}},\ \bibinfo {pages} {61} (\bibinfo {year} {2006})},\ \Eprint
  {http://arxiv.org/abs/hep-ph/0509090} {arXiv:hep-ph/0509090 [hep-ph]}
  \BibitemShut {NoStop}%
\end{thebibliography}%

\end{document}